\documentclass[english,3p, final]{elsarticle}
\usepackage[T1]{fontenc}
\usepackage[utf8]{luainputenc}
\usepackage{amsmath}
\usepackage{graphicx}
\usepackage{esint}

\makeatletter
\usepackage[english]{babel}

\addto\captionsenglish{}

\biboptions{numbers,sort&compress}
\usepackage{mciteplus}

\def\ps@pprintTitle{%
  \let\@oddhead\@empty
  \let\@evenhead\@empty
  \def\@oddfoot{}%
  \let\@evenfoot\@oddfoot}

\makeatother

\usepackage{babel}
\begin{document}

\title{Contour-time approach to the Bose-Hubbard model in the strong coupling
regime: Studying two-point spatio-temporal correlations at the Hartree-Fock-Bogoliubov
level}

\author[sfu]{Matthew R. C. Fitzpatrick\corref{cor1}}

\ead{mrfitzpa@sfu.ca}

\author[sfu]{Malcolm P. Kennett}

\ead{malcolmk@sfu.ca}

\cortext[cor1]{Corresponding author}

\address[sfu]{Department of Physics, Simon Fraser University, 8888 University Drive,
Burnaby, British Columbia V5A 1S6, Canada}
\begin{abstract}
We develop a formalism that allows the study of correlations in space
and time in both the superfluid and Mott insulating phases of the Bose-Hubbard Model.
Specifically, we obtain a two particle irreducible effective
action within the contour-time formalism that allows for both
equilibrium and out of equilibrium phenomena. We derive equations
of motion for both the superfluid order parameter and two-point correlation
functions. To assess the accuracy of this formalism, we study the
equilibrium solution of the equations of motion and compare our results
to existing strong coupling methods as well as exact methods where
possible. We discuss applications of this formalism to out
of equilibrium situations. 
\end{abstract}
\maketitle

\section{Introduction\label{sec:introduction - 1}}

The out of equilibrium dynamics of cold atoms trapped in optical lattices
has received considerable attention in recent years \citep{BlochRev1,BlochRev2,Morsch,Lewenstein,Bloch,KennettRev}.
The ability to tune experimental parameters over a wide range of values
in real time makes these systems very versatile and gives the opportunity
to study quantum systems out of equilibrium in a controlled fashion.
Quantum quenches, in which parameters in the Hamiltonian are varied
in time faster than the system can respond adiabatically, e.g. when
a system is driven through a quantum critical point, are a protocol
that is natural to study in this context and have been studied intensely
both theoretically and experimentally.

The Bose-Hubbard model (BHM) \citep{Fisher} has been shown to describe
interacting ultracold bosons in an optical lattice \citep{Jaksch},
allowing the opportunity for experiments to probe the out of equilibrium
dynamics of the model \citep{Jaksch,Greiner,Chen,Bakr,Jimenez,Spielman,Sherson,Stoferle,Kohl,Schori,Greiner2,Gerbier4,Gerbier3,Will,Trotzky,Spielman08,Trotzky1d,Mahmud}.
The BHM is a particularly convenient context for studying quantum
quenches as it displays a quantum phase transition between the superfluid
and Mott-insulator phases (or vice versa) as the ratio of intersite
hopping $J$ to the on-site repulsion $U$ is varied, as observed
by Greiner \emph{et al}. \citep{Greiner}. Theoretical studies of
the BHM suggest that whether equilibration occurs or not after a quantum
quench depends sensitively on the initial and final values of $J/U$
and the chemical potential \citep{Kollath,Sciolla,Sciolla2,Fischer0,Fischer1,Kennett,Werner,Nessi}.
In the case of quenches from superfluid (large $J/U$) to Mott insulator
(small $J/U$) there have been suggestions that there may be aging
behaviour and glassiness that might be experimentally observable in
two time correlations or in violations of the fluctuation dissipation
theorem \citep{Kollath,Sciolla,Sciolla2,Kennett,Nessi,KennettRev}.
In the alternative quench from Mott insulator to superfluid, it has
been suggested that Kibble-Zurek \citep{Kibble,Zurek,Zurek05} scaling
of defects should be observed \citep{Polkovnikov_nex,Gritsev}, which
has recently been tested experimentally \citep{Chen}.

In experiments, the combination of a harmonic trap and small $J/U$
leads to a wedding cake structure of the equilibrium density, with
alternating Mott insulating and superfluid regions \citep{Lannert,Batrouni02}.
The presence of Mott insulating regions has been predicted to retard
relaxation to equilibrium after a quench to small $J/U$ by impeding
mass transport of bosons through these regions \citep{Natu,Bernier}
which has also been observed experimentally \citep{Hung}. This gives
a picture in which relaxation after a quench takes place in two steps
-- fast relaxation to local equilibrium followed by slower relaxation
via mass transport \citep{Natu,Dutta}.

In addition to slow dynamics, several analytical and numerical studies
have also shown a Lieb-Robinson-like \citep{LiebRobinson} bound of
a maximal velocity which leads to a light-cone like spreading of density
correlations in one dimensional systems for quenches from the superfluid
to Mott-insulating regime as well as quenches within the superfluid
\citep{Carleo} or Mott-insulating phases \citep{Fischer0,Lauchli,Bernier,Barmettler}.
 The latter case
was recently observed experimentally by Cheneau \emph{et al}. \citep{Cheneau}.
Similar predictions have been made for higher dimensional systems
\citep{Navez,Natu2,Carleo}. The results summarized above motivate
the study of the temporal and spatial correlations of the BHM after
a quantum quench in order to elucidate the dynamics observed after
quenches.

A generic problem in the theoretical description of quantum quenches
is that it is necessary to have a formalism that is able to describe
the physics in the phases on both sides of a quantum critical point.
In the case of the Bose Hubbard model, numerical approaches such as
exact diagonalization and the time-dependent density matrix renormalization
group (t-DMRG) \citep{Kollath,Lauchli,Bernier,Bernier2,Clark,Cheneau,Trotzky1d}
can be essentially exact in all parts of parameter space but are limited
by system size and usually are practical only in one dimension. For
dimensions higher than one, methods such as time-dependent Gutzwiller
mean field theory \citep{Lewenstein,Zakrzewski,Amico,Natu} and dynamical
mean field theory \citep{Werner} have been used which can capture
the presence of a quantum phase transition, but in their simplest
form do not capture spatial correlations, although there has been
work on including perturbative corrections \citep{Trefzger,TrefzgerDutta,Navez,Schroll,Yanay,Navez2,Krutitsky}.
An analytical approach based on using two Hubbard Stratonovich transformations
to capture both weak-coupling and strong-coupling physics in the same
formalism was developed by Sengupta and Dupuis \citep{SenguptaDupuis}.
Within their effective theory, they performed a mean-field calculation
of the superfluid order parameter and a Bogoliubov (1-loop) approximation
to the two-point Green's function to study the excitation spectrum.
Their work was generalized by one of us from an equilibrium theory
to out of equilibrium by using the Schwinger-Keldysh formalism to
obtain a one-particle irreducible (1PI) effective action which was
then used to study the superfluid order parameter after a quench \citep{Kennett}.

Here, we extend the approach developed in Ref. \citep{Kennett}
to obtain a two-particle irreducible (2PI) effective action using
the contour-time formalism, which is a generalisation of the Schwinger-Keldysh
formalism. In the 2PI approach, the evolution of the order parameter
and the two-point Green's functions are treated on the same footing
\citep{Rey1} which allows us to describe correlations both in the broken
symmetry (superfluid) phase and the Mott phase. Moreover, the method
provides a systematic way to go beyond the mean-field or the 1-loop
approximation. We obtain two main results. First, we develop the 2PI strong coupling
formalism for the BHM. Second, we derive equations of motion within a
Hartree-Fock-Bogoliubov-Popov approximation suitable for both equilibrium
and out of equilibrium calculations.  We obtain equilibrium solutions of these
equations that allow us to obtain phase boundaries and excitation spectra that
we compare to previous equilibrium results obtained in a 1-loop calculation \citep{SenguptaDupuis} 
and numerically exact results where possible. 

This paper is structured as follows. In Section \ref{sec:model and formalism - 1},
we describe the model that we study and derive the 2PI effective action
for the BHM. In Section \ref{sec:eqns of motion - 1}, we obtain the
equations of motion for both the order parameter and the two-particle
Green's function by taking appropriate variations of the 2PI effective
action. In Section \ref{sec:equilibrium solution - 1}, we study the
equilibrium solution of the equations of motion at the HFBP level.
Finally in Section \ref{sec:discussion and conclusions - 1} we discuss
our results and present our conclusions.

\section{Model and formalism\label{sec:model and formalism - 1}}

In this section we introduce the Bose Hubbard model and discuss the
generalization of the 1PI approach developed in Ref.~\citep{Kennett}
to a 2PI effective action within the Schwinger-Keldysh formalism.
The Hamiltonian for the BHM, allowing for a time dependent hopping
term, is

\begin{eqnarray}
\hat{H}_{\text{BHM}}\left(t\right) & = & \hat{H}_{J}\left(t\right)+\hat{H}_{0},\label{eq:BHM Hamiltonian - 1}
\end{eqnarray}

\noindent where

\begin{eqnarray}
\hat{H}_{J}\left(t\right) & = & -\sum_{\left\langle \vec{r}_{1},\vec{r}_{2}\right\rangle }J_{\vec{r}_{1}\vec{r}_{2}}\left(t\right)\left(\hat{a}_{\vec{r}_{1}}^{\dagger}\hat{a}_{\vec{r}_{2}}+\hat{a}_{\vec{r}_{2}}\hat{a}_{\vec{r}_{1}}^{\dagger}\right),\label{eq:H_J defined - 1}
\end{eqnarray}

\begin{equation}
\hat{H}_{0}=\hat{H}_{U}-\mu\hat{N}=\frac{U}{2}\sum_{\vec{r}}\hat{n}_{\vec{r}}\left(\hat{n}_{\vec{r}}-1\right)-\mu\sum_{\vec{r}}\hat{n}_{\vec{r}},\label{eq:H_0 defined - 1}
\end{equation}

\noindent with $\hat{a}_{\vec{r}}^{\dagger}$ and $\hat{a}_{\vec{r}}$
annihilation and creation operators for bosons on lattice site $\vec{r}$
respectively, $\hat{n}_{\vec{r}}\equiv\hat{a}_{\vec{r}}^{\dagger}\hat{a}_{\vec{r}}$
the number operator, $U$ the interaction strength, and $\mu$ the
chemical potential. The notation $\left\langle \vec{r}_{1},\vec{r}_{2}\right\rangle $
indicates a sum over nearest neighbours only. We allow $J_{\vec{r}_{1}\vec{r}_{2}}\left(t\right)$,
the hopping amplitude between sites $\vec{r}_{1}$ and $\vec{r}_{2}$,
to be time dependent.

\subsection{Contour-time formalism\label{sub:Contour formalism - 1}}

\begin{figure}
\begin{centering}
\includegraphics[scale=0.6]{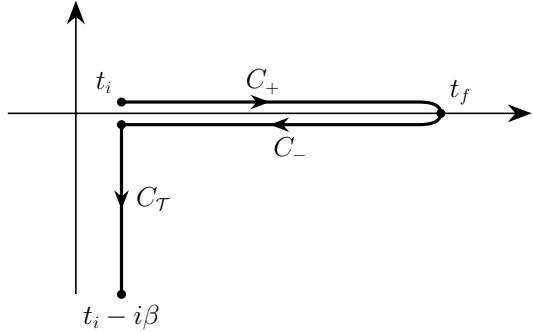}
\par\end{centering}

\caption{Contour for a system initially prepared at time $t_{i}$ in a thermal
state with inverse temperature $\beta$. $t_{f}$ is the maximum real-time
considered in the problem, which may be set to $t_{f}\to\infty$ without
loss of generality.\label{fig:fig1}}
\end{figure}

We use the contour-time formalism \citep{Schwinger,Keldysh,Rammer,Semenoff,Landsman,Chou},
which treats time as a complex variable lying along a contour. For
systems initially prepared in thermal states, which we consider here,
one can work with a contour $C$ of the form illustrated in Fig.~\ref{fig:fig1}.
One obtains the imaginary-time Matsubara formalism, which is restricted
to equilibrium problems, by setting $t_{f}=t_{i}$. If one does not
work in the Matsubara formalism, $t_{f}$ can be set to $\infty$
without loss of generality \citep{RammerText}. Furthermore, if one were
to set instead $t_{i}\to-\infty$, then one can obtain the real-time Schwinger-Keldysh closed-time path,
which is suitable for both equilibrium and out of equilibrium problems,
as the imaginary part of the contour would not contribute anything
to the dynamics of the system. By setting $t_{i}\to-\infty$, one
is effectively discarding transient effects. Since we are interested
in studying transient phenomena, we do not set $t_{i}\to-\infty$
and instead work with the general contour illustrated in Fig.~\ref{fig:fig1}.
A number of authors have applied contour-time approaches to the BHM
\citep{Robertson,Kennett,Gras,Grass2,Grassthesis,Rey1,Rey2,Temme,Calzetta,PolkovnikovCTP,DellAnna}
-- our work differs from previous approaches in that we apply a 2PI
approach within the contour formalism that is appropriate for strong
coupling as well as weak coupling \citep{Rey1,Temme}. 

\subsection{Green's functions and the 1PI generating functionals\label{sub:Green's fncs and the 1PI Gen Func - 1}}
To characterize spatio-temporal correlations in the BHM we calculate
contour-ordered Green's functions (COGFs). We generalize the work
in Ref.~\citep{Kennett} to include Green's functions with unequal
numbers of annihilation and creation operators to allow for the study
of broken symmetry phases. We frequently use the compact notation
$\hat{a}_{\vec{r}}^{a}$ for the bosonic fields, defined by

\begin{equation}
\hat{a}_{\vec{r}}^{1}\equiv\hat{a}_{\vec{r}},\quad\hat{a}_{\vec{r}}^{2}\equiv\hat{a}_{\vec{r}}^{\dagger}.\label{eq:introducing nambu indices - 1}
\end{equation}

\noindent We define the $n$-point COGF as \citep{Chou}

\begin{eqnarray}
G_{\vec{r}_{1}\ldots\vec{r}_{n}}^{a_{1}\ldots a_{n}}\left(\tau_{1},\ldots,\tau_{n}\right) & \equiv & \left(-i\right)^{n-1}\text{Tr}\left\{ \hat{\rho}_{i}T_{C}\left[\hat{a}_{\vec{r}_{1}}^{a_{1}}\left(\tau_{1}\right)\ldots\hat{a}_{\vec{r}_{n}}^{a_{n}}\left(\tau_{n}\right)\right]\right\} \nonumber \\
 & \equiv & \left(-i\right)^{n-1}\left\langle T_{C}\left[\hat{a}_{\vec{r}_{1}}^{a_{1}}\left(\tau_{1}\right)\ldots\hat{a}_{\vec{r}_{n}}^{a_{n}}\left(\tau_{n}\right)\right]\right\rangle _{\hat{\rho}_{i}},\label{eq:COGFs defined - 1}
\end{eqnarray}

\noindent where $\hat{\rho}_{i}$ is the state operator for a thermal
state representing the initial state of our system

\begin{eqnarray}
\hat{\rho}_{i} & = & \frac{e^{-\beta\hat{H}_{\text{BHM}}\left(t_{i}\right)}}{\text{Tr}\left\{ e^{-\beta\hat{H}_{\text{BHM}}\left(t_{i}\right)}\right\} },\label{eq:State operator for initial thermal state - 1}
\end{eqnarray}

\noindent and $\hat{a}_{\vec{r}}^{a}\left(\tau\right)$ are the bosonic
fields in the Heisenberg picture with respect to $\hat{H}_{\text{BHM}}\left(\tau\right)$
{[}Eq.~(\ref{eq:BHM Hamiltonian - 1}){]}

\noindent 
\begin{eqnarray}
\hat{a}_{\vec{r}}^{a}\left(\tau\right) & = & U^{\dagger}\left(\tau,\tau_{i}\right)\hat{a}_{\vec{r}}^{a}U\left(\tau,\tau_{i}\right),\label{eq:Heisenberg fields - 1}\\
U\left(\tau,\tau^{\prime}\right) & = & T_{C}\left[e^{-i\int_{C\left(\tau,\tau^{\prime}\right)}d\tau^{\prime\prime}\hat{H}_{\text{BHM}}\left(\tau^{\prime\prime}\right)}\right].\label{eq:evolution operator - 1}
\end{eqnarray}
Here we have introduced explicitly the complex contour time argument
$\tau$, the sub-contour $C\left(\tau,\tau^{\prime}\right)$ which
goes from $\tau$ to $\tau^{\prime}$ along the contour $C$, and
the contour time ordering operator $T_{C}$, which orders strings
of operators according to their position on the contour, with operators
at earlier contour times placed to the right. Note that the presence
of $T_{C}$ in Eq.~(\ref{eq:COGFs defined - 1}) leads to symmetry
under permutations $\left\{ p_{1},\ldots,p_{n}\right\} $ of the sequence
$\left\{ 1,\ldots,n\right\} $:

\begin{eqnarray}
G_{\vec{r}_{1}\ldots\vec{r}_{n}}^{a_{1}\ldots a_{n}}\left(\tau_{1},\ldots,\tau_{n}\right) & = & G_{\vec{r}_{p_{1}}\ldots\vec{r}_{p_{n}}}^{a_{p_{1}}\ldots a_{p_{n}}}\left(\tau_{p_{1}},\ldots,\tau_{p_{n}}\right).\label{eq:COGF permutation symmetry - 1}
\end{eqnarray}

At times it will be useful to express the contour time $\tau$ in
terms of a contour label $\alpha$ (commonly called a Keldysh index)
indicating a contour time located on $C_\alpha$
and a positive real parameter $s$ such that

\begin{equation}
\tau=\left(\alpha,s\right)=\begin{cases}
t_{i}+s+i0^{+}, & \text{if }\alpha=+,\\
t_{i}+s+i0^{-}, & \text{if }\alpha=-,\\
t_{i}-is+i0^{-}, & \text{if }\alpha=\mathcal{T},
\end{cases}\label{eq:tau in terms of alpha and s - 1}
\end{equation}

\noindent e.g. we can rewrite the bosonic fields $\hat{a}_{\vec{r}}^{a}\left(\tau\right)$
as

\begin{eqnarray}
\hat{a}_{\vec{r},\alpha}^{a}\left(s\right) & \equiv & \hat{a}_{\vec{r}}^{a}\left(\tau\right),\label{eq:Rewrite bosonic fields - 1}
\end{eqnarray}

\noindent and the COGFs in Eq.~(\ref{eq:COGFs defined - 1}) as

\begin{eqnarray}
G_{\vec{r}_{1}\ldots\vec{r}_{n},\alpha_{1}\ldots\alpha_{n}}^{a_{1}\ldots a_{n}}\left(s_{1},\ldots,s_{n}\right) & \equiv & G_{\vec{r}_{1}\ldots\vec{r}_{n}}^{a_{1}\ldots a_{n}}\left(\tau_{1},\ldots,\tau_{n}\right)\nonumber \\
 & = & \left(-i\right)^{n-1}\left\langle T_{C}\left[\hat{a}_{\vec{r}_{1},\alpha_{1}}^{a_{1}}\left(s_{1}\right)\ldots\hat{a}_{\vec{r}_{n},\alpha_{n}}^{a_{n}}\left(s_{n}\right)\right]\right\rangle _{\hat{\rho}_{i}}.\label{eq:Rewrite COGFs - 1}
\end{eqnarray}

\noindent In order for the Heisenberg fields $\hat{a}_{\vec{r}}^{a}\left(\tau\right)$
to be well-defined, we need to analytically continue the BHM Hamiltonian
{[}Eq.~(\ref{eq:BHM Hamiltonian - 1}){]}. For the contour considered
in this paper, $\hat{H}_{\text{BHM}}\left(\tau\right)$ is analytically
continued as follows

\begin{equation}
\hat{H}_{\text{BHM}}\left(\tau\right)=\hat{H}_{\text{BHM},\alpha}\left(s\right)\equiv\begin{cases}
\hat{H}_{\text{BHM}}\left(s\right), & \text{if }\alpha=+,\\
\hat{H}_{\text{BHM}}\left(s\right), & \text{if }\alpha=-,\\
\hat{H}_{\text{BHM}}\left(t_{i}\right), & \text{if }\alpha=\mathcal{T}.
\end{cases}\label{eq:Analytic continuation of H_BHM - 1}
\end{equation}
The COGFs above can be derived from a generating functional $\mathcal{Z}\left[f\right]$
defined as

\begin{eqnarray}
\mathcal{Z}\left[f\right] & \equiv & \text{Tr}\left\{ \hat{\rho}_{i}T_{C}\left[e^{i\int_{C}d\tau\sum_{\vec{r}}f_{\vec{r}}^{\overline{a}}\left(\tau\right)\hat{a}_{\vec{r}}^{a}\left(\tau\right)}\right]\right\} \nonumber \\
 & = & \text{Tr}\left\{ \hat{\rho}_{i}T_{C}\left[e^{i\left(\int_{C_{+}}+\int_{C_{-}}+\int_{C_{\mathcal{T}}}\right)d\tau\sum_{\vec{r}}f_{\vec{r}}^{\overline{a}}\left(\tau\right)\hat{a}_{\vec{r}}^{a}\left(\tau\right)}\right]\right\} \nonumber \\
 & = & \text{Tr}\left\{ \hat{\rho}_{i}T_{C}\left[e^{i\left(\int_{0}^{\infty}ds\sum_{\vec{r}}f_{\vec{r},+}^{\overline{a}}\left(s\right)\hat{a}_{\vec{r},+}^{a}\left(s\right)+\int_{0}^{\infty}\left(-ds\right)\sum_{\vec{r}}f_{\vec{r},-}^{\overline{a}}\left(s\right)\hat{a}_{\vec{r},-}^{a}\left(s\right)+\int_{0}^{\beta}\left(-ids\right)\sum_{\vec{r}}f_{\vec{r},\mathcal{T}}^{\overline{a}}\left(s\right)\hat{a}_{\vec{r},\mathcal{T}}^{a}\left(s\right)\right)}\right]\right\} \nonumber \\
 & = & \text{Tr}\left\{ \hat{\rho}_{i}T_{C}\left[e^{i\int_{0}^{s_{\alpha\alpha^{\prime}}^{f}}ds\sum_{\vec{r}}\tau_{\alpha\alpha^{\prime}}^{3}f_{\vec{r},\alpha}^{\overline{a}}\left(s\right)\hat{a}_{\vec{r},\alpha^{\prime}}^{a}\left(s\right)}\right]\right\} ,\label{eq:Z generating functional defined - 1}
\end{eqnarray}

\noindent where

\begin{eqnarray}
\hat{\tau}^{3} & = & \left(\begin{array}{ccc}
1 & 0 & 0\\
0 & -1 & 0\\
0 & 0 & -i
\end{array}\right)\label{eq:tau_3 defined - 1}
\end{eqnarray}

\noindent in the $\left(+,-,\mathcal{T}\right)$ basis,

\begin{eqnarray}
s_{\alpha\alpha^{\prime}}^{f} & = & \begin{cases}
\infty, & \text{if }\alpha=\alpha^{\prime}=+ \, {\rm or}\, -,\\
\beta, & \text{if }\alpha=\alpha^{\prime}=\mathcal{T},\\
0, & \text{otherwise},
\end{cases}\label{eq:Contour integration limits - 1}
\end{eqnarray}

\noindent the $f$s are source currents, the overscored index in $f_{\vec{r},\alpha}^{\overline{a}}\left(s\right)$
is defined by

\begin{eqnarray}
f_{\vec{r},\alpha}^{\overline{a}}\left(s\right) & = & \sigma_{1}^{aa^{\prime}}f_{\vec{r},\alpha}^{a^{\prime}}\left(s\right),\label{eq:Overscored index defined - 1}
\end{eqnarray}

\noindent and $\sigma_{i}$ is the $i^{\text{th}}$ Pauli matrix,
i.e. $\overline{1}=2$ and $\overline{2}=1$. We use the Einstein
summation convention for both the Keldysh and Nambu indices, i.e.
matching indices implies a summation over all possible values of those
indices. It is clear from the definition above that the generating
functional is normalized such that $\mathcal{Z}\left[f=0\right]=1$.

To derive the COGFs in Eq.~(\ref{eq:Rewrite COGFs - 1}) from $\mathcal{Z}\left[f\right]$,
we take appropriate functional derivatives with respect to the sources
and set the sources to zero afterwards

\begin{eqnarray}
G_{\vec{r}_{1}\ldots\vec{r}_{n},\alpha_{1}\ldots\alpha_{n}}^{a_{1}\ldots a_{n}}\left(s_{1},\ldots,s_{n}\right) & = & i\left(-1\right)^{n}\left(\left[\tau^{3}\right]_{\alpha_{1}\alpha_{1}^{\prime}}^{\dagger}\ldots\left[\tau^{3}\right]_{\alpha_{n}\alpha_{n}^{\prime}}^{\dagger}\right)\nonumber \\
 &  & \quad\times\left.\frac{1}{\mathcal{Z}\left[f=0\right]}\frac{\delta^{n}\mathcal{Z}\left[f\right]}{\delta f_{\vec{r}_{1},\alpha_{1}^{\prime}}^{\overline{a_{1}}}\left(s_{1}\right)\ldots\delta f_{\vec{r}_{n},\alpha_{n}^{\prime}}^{\overline{a_{n}}}\left(s_{n}\right)}\right|_{f=0}.\label{eq:COGFs from Z - 1}
\end{eqnarray}

\subsection{Path integral form of $\mathcal{Z}\left[f\right]$\label{sub:Path integral form of Z - 1}}

We cast the generating functional $\mathcal{Z}\left[f\right]$ in
the path integral form \citep{Semenoff}, which for the case of the
BHM is \citep{Kennett}

\begin{eqnarray}
\mathcal{Z}\left[f\right] & = & \int\left[\mathcal{D}a^{a}\right]e^{iS_{\text{BHM}}\left[a\right]+iS_{f}\left[a\right]},\label{eq:Path integral form of Z - 1}
\end{eqnarray}

\noindent where $S_{\text{BHM}}$ is the action for the BHM, and $\int\left[\mathcal{D}a^{a}\right]$
is the coherent-state measure. We absorb overall constants into the
measure as they will cancel out in the calculation of the COGFs due
to the factor of $1/\mathcal{Z}\left[f=0\right]$ in Eq.~(\ref{eq:COGFs from Z - 1}).
Note that in the path-integral formalism $a_{\vec{r},\alpha}^{1}=a_{\vec{r},\alpha}$
and $a_{\vec{r},\alpha}^{2}=a_{\vec{r},\alpha}^{*}$. In this formalism,
we can rewrite averages of the form $\left\langle T_{C}\left[\ldots\right]\right\rangle _{\hat{\rho}_{i}}$
as follows

\begin{eqnarray}
\left\langle T_{C}\left[\hat{a}_{\vec{r}_{1},\alpha_{1}}^{a_{1}}\left(s_{1}\right)\ldots\hat{a}_{\vec{r}_{n},\alpha_{n}}^{a_{n}}\left(s_{n}\right)\right]\right\rangle _{\hat{\rho}_{i}} & \equiv & \left\langle a_{\vec{r}_{1},\alpha_{1}}^{a_{1}}\left(s_{1}\right)\ldots a_{\vec{r}_{n},\alpha_{n}}^{a_{n}}\left(s_{n}\right)\right\rangle _{S_{\text{BHM}}},\label{eq:Rewriting averages - 1}
\end{eqnarray}

\noindent where contour ordering is now implicit in the path integral
representation \citep{Negele}. In addition to the generating functional,
we make extensive use of the generator of connected COGFs (CCOGFs)
defined by

\begin{eqnarray}
W\left[f\right] & \equiv & -i\ln\mathcal{Z}\left[f\right].\label{eq:W generating functional defined - 1}
\end{eqnarray}

The $n$-point CCOGF $G_{\vec{r}_{1}\ldots\vec{r}_{n},\alpha_{1}\ldots\alpha_{n}}^{a_{1}\ldots a_{n},c}\left(s_{1},\ldots,s_{n}\right)$
can be obtained from $W\left[f\right]$ by calculating

\begin{eqnarray}
G_{\vec{r}_{1}\ldots\vec{r}_{n},\alpha_{1}\ldots\alpha_{n}}^{a_{1}\ldots a_{n},c}\left(s_{1},\ldots,s_{n}\right) & = & \left.\left(-1\right)^{n-1}\left(\left[\tau^{3}\right]_{\alpha_{1}\alpha_{1}^{\prime}}^{\dagger}\ldots\left[\tau^{3}\right]_{\alpha_{n}\alpha_{n}^{\prime}}^{\dagger}\right)\frac{\delta^{n}W\left[f\right]}{\delta f_{\vec{r}_{1},\alpha_{1}^{\prime}}^{\overline{a_{1}}}\left(s_{1}\right)\ldots\delta f_{\vec{r}_{n},\alpha_{n}^{\prime}}^{\overline{a_{n}}}\left(s_{n}\right)}\right|_{f=0}\nonumber \\
 & \equiv & \left(-i\right)^{n-1}\left\langle a_{\vec{r}_{1},\alpha_{1}}^{a_{1}}\left(s_{1}\right)\ldots a_{\vec{r}_{n},\alpha_{n}}^{a_{n}}\left(s_{n}\right)\right\rangle _{S_{\text{BHM}}}^{c},\label{eq:CCOGFs from W - 1}
\end{eqnarray}

\noindent where $\left\langle \ldots\right\rangle ^{c}$ indicates
that only connected diagrams are kept. Note that the CCOGFs satisfy
the same symmetry property as the COGFs

\begin{eqnarray}
G_{\vec{r}_{1}\ldots\vec{r}_{n},\alpha_{1}\ldots\alpha_{n}}^{a_{1}\ldots a_{n},c}\left(s_{1},\ldots,s_{n}\right) & = & G_{\vec{r}_{p_{1}}\ldots\vec{r}_{p_{n}},\alpha_{p_{1}}\ldots\alpha_{p_{n}}}^{a_{p_{1}}\ldots a_{p_{n}},c}\left(s_{p_{1}},\ldots,s_{p_{n}}\right).\label{eq:CCOGF permutation symmetry - 1}
\end{eqnarray}

\subsection{Keldysh rotation\label{sub:Keldysh rotation - 1}}

For the $n$-point CCOGF defined in Eq.~(\ref{eq:CCOGFs from W - 1})
there are $3^{n}$ Keldysh components. However, as a consequence of
causality, we can eliminate $\sum_{m=0}^{n-1}\left(\begin{smallmatrix}n\\
m
\end{smallmatrix}\right)$ of these components by performing the following transformation on
the bosonic fields \citep{Keldysh}

\begin{equation}
\left(\begin{array}{c}
a_{+}\left(t\right)\\
a_{-}\left(t\right)\\
a_{\mathcal{T}}\left(t\right)
\end{array}\right)\longrightarrow\left(\begin{array}{c}
\tilde{a}_{q}\left(t\right)\\
\tilde{a}_{c}\left(t\right)\\
\tilde{a}_{\mathcal{T}}\left(t\right)
\end{array}\right)=\hat{L}\left(\begin{array}{c}
a_{+}\left(t\right)\\
a_{-}\left(t\right)\\
a_{\mathcal{T}}\left(t\right)
\end{array}\right),\label{eq:Keldysh rotation - 1}
\end{equation}

\noindent with

\begin{eqnarray}
\hat{L} & = & \frac{1}{\sqrt{2}}\left(\begin{array}{ccc}
1 & -1 & 0\\
1 & 1 & 0\\
0 & 0 & \sqrt{2}
\end{array}\right),\label{eq:L matrix - 1}
\end{eqnarray}

\noindent where $\tilde{a}_{q}$ and $\tilde{a}_{c}$ are the quantum
and classical components of the field respectively \citep{Grassthesis,Weert,Cugliandolo,KCY},
and $\tilde{a}_{\mathcal{T}}=a_{\mathcal{T}}$. After the above basis
transformation $\left(+,-,\mathcal{T}\right)\to\left(q,c,\mathcal{T}\right)$,
the matrix $\tau^{3}$ becomes

\begin{eqnarray}
\hat{\tau}^{1} & = & \left(\begin{array}{ccc}
0 & 1 & 0\\
1 & 0 & 0\\
0 & 0 & -i
\end{array}\right),\label{eq:tau_1 defined - 1}
\end{eqnarray}

\noindent the limits of integration become

\begin{eqnarray}
s_{\alpha\alpha^{\prime}}^{f} & = & \begin{cases}
\infty, & \text{if }\left\{ \alpha,\alpha^{\prime}\right\} \in P\left(\left\{ q,c\right\} \right),\\
\beta, & \text{if }\alpha=\alpha^{\prime}=\mathcal{T},\\
0, & \text{otherwise},
\end{cases}\label{eq:Contour integration limits - 2}
\end{eqnarray}

\noindent and $P\left(\left\{ x_{m}\right\} _{m=1}^{n}\right)$ is
the set of all permutations of the sequence $\left\{ x_{m}\right\} _{m=1}^{n}$.

After performing the above Keldysh transformation, any COGFs $\tilde{G}_{\vec{r}_{1}\ldots\vec{r}_{n},\alpha_{1}\ldots\alpha_{n}}^{a_{1}\ldots a_{n}}\left(s_{1},\ldots,s_{n}\right)$
with at least one quantum $\alpha$-index and no classical $\alpha$-indices
will vanish. To see this, consider the following COGF

\begin{eqnarray}
 &  & \tilde{G}_{\vec{r}_{1}\ldots\vec{r}_{n},\underbrace{\mathcal{T}\ldots\mathcal{T}}_{m\text{ terms}}\underbrace{q\ldots q}_{n-m\text{ terms}}}^{a_{1}\ldots a_{n}}\left(s_{1},\ldots,s_{n}\right)\nonumber \\
 &  & \quad=\left(-i\right)^{n-1}\left\langle T_{C}\left[\hat{\tilde{a}}_{\vec{r}_{1},\mathcal{T}}^{a_{1}}\left(s_{1}\right)\ldots\hat{\tilde{a}}_{\vec{r}_{m},\mathcal{T}}^{a_{m}}\left(s_{m}\right)\hat{\tilde{a}}_{\vec{r}_{m+1},q}^{a_{m+1}}\left(s_{m+1}\right)\ldots\hat{\tilde{a}}_{\vec{r}_{n},q}^{a_{n}}\left(s_{n}\right)\right]\right\rangle _{\hat{\rho}_{i}}\nonumber \\
 &  & \quad=\frac{\left(-i\right)^{n-1}}{2^{(n-m)/2}}\left\langle T_{C}\left[\hat{a}_{\vec{r}_{1},\mathcal{T}}^{a_{1}}\left(s_{1}\right)\ldots\hat{a}_{\vec{r}_{m},\mathcal{T}}^{a_{m}}\left(s_{m}\right)\right.\right.\nonumber \\
 &  & \left.\left.\phantom{\quad=\frac{\left(-i\right)^{n-1}}{2^{(n-m)/2}}}\quad\left\{ \hat{a}_{\vec{r}_{m+1},+}^{a_{m+1}}\left(s_{m+1}\right)-\hat{a}_{\vec{r}_{m+1},-}^{a_{m+1}}\left(s_{m+1}\right)\right\} \ldots\left\{ \hat{a}_{\vec{r}_{n},+}^{a_{n}}\left(s_{n}\right)-\hat{a}_{\vec{r}_{n},-}^{a_{n}}\left(s_{n}\right)\right\} \right]\right\rangle _{\hat{\rho}_{i}}\nonumber \\
 &  & \quad=\frac{\left(-i\right)^{n-1}}{2^{(n-m)/2}}\left\langle T_{C}\left[\hat{a}_{\vec{r}_{1},\mathcal{T}}^{a_{1}}\left(s_{1}\right)\ldots\hat{a}_{\vec{r}_{m},\mathcal{T}}^{a_{m}}\left(s_{m}\right)\right]\right.\nonumber \\
 &  & \left.\phantom{\quad=\frac{\left(-i\right)^{n-1}}{2^{(n-m)/2}}}\quad\times T_{C}\left[\left\{ \hat{a}_{\vec{r}_{m+1},+}^{a_{m+1}}\left(s_{m+1}\right)-\hat{a}_{\vec{r}_{m+1},-}^{a_{m+1}}\left(s_{m+1}\right)\right\} \ldots\left\{ \hat{a}_{\vec{r}_{n},+}^{a_{n}}\left(s_{n}\right)-\hat{a}_{\vec{r}_{n},-}^{a_{n}}\left(s_{n}\right)\right\} \right]\right\rangle _{\hat{\rho}_{i}}.\label{eq:Vanishing COGF proof - 1}
\end{eqnarray}

\noindent Following the argument given in Ref.~\citep{Grassthesis},
multiplying out the products in the second $T_{C}\left[\ldots\right]$
yields $2^{n-m}$ path-ordered terms. The key point to note is that
within any one of these path-ordered products the position of the
field with the largest $s$ does not depend on its Keldysh index.
This implies that for each path-ordered product there is another path-ordered
product is which is identical except with opposite sign. Therefore
every term cancels out. It immediately follows that the associated
CCOGFs vanish as well:

\begin{eqnarray}
\tilde{G}_{\vec{r}_{1}\ldots\vec{r}_{n},\underbrace{\mathcal{T}\ldots\mathcal{T}}_{m\text{ terms}}\underbrace{q\ldots q}_{n-m\text{ terms}}}^{a_{1}\ldots a_{n},c}\left(s_{1},\ldots,s_{n}\right) & = & 0,\quad0\le m<n.\label{eq:Vanishing CCOGF statement - 1}
\end{eqnarray}

\noindent Moreover, any permutation of the Keldysh indices in Eq.
(\ref{eq:Vanishing CCOGF statement - 1}) will also yield a vanishing
CCOGF. Since there are $\left(\begin{smallmatrix}n\\
m
\end{smallmatrix}\right)$ distinct permutations for fixed $n$ and $m$, there are $\sum_{m=0}^{n-1}\left(\begin{smallmatrix}n\\
m
\end{smallmatrix}\right)$ components that will vanish in total. This completes the proof. Note
that if we were working with a closed-time path, where there is no
imaginary appendix to the contour, we recover the special
case where only $\left(\begin{smallmatrix}n\\
0
\end{smallmatrix}\right) = 1$ Keldysh component vanishes, namely $\tilde{G}_{\vec{r}_{1}\ldots\vec{r}_{n},q\ldots q}^{a_{1}\ldots a_{n},c}\left(s_{1},\ldots,s_{n}\right)$
\citep{Keldysh,Grassthesis}.

After performing the Keldysh transformation, the BHM action takes
the form \citep{Kennett} (dropping tildes)

\begin{eqnarray}
S_{\text{BHM}} & = & \frac{1}{2}\int_{0}^{s_{\alpha_{1}\alpha_{2}}^{f}}ds\sum_{\vec{r}}\left[a_{\vec{r},\alpha_{1}}^{a_{1}}\left(s\right)\left(\left[\tau^{0}\right]_{\alpha_{1}\alpha_{3}}^{\dagger}\tau_{\alpha_{3}\alpha_{2}}^{1}\sigma_{2}^{a_{1}a_{2}}\partial_{s}\right)a_{\vec{r},\alpha_{2}}^{a_{2}}\left(s\right)\right]+S_{J}+S_{U},\label{eq:S_BHM - 1}
\end{eqnarray}

\noindent where

\begin{eqnarray}
S_{J} & = & \frac{1}{2}\int_{0}^{s_{\alpha_{1}\alpha_{2}}^{f}}ds\sum_{\left\langle \vec{r}_{1}\vec{r}_{2}\right\rangle }a_{\vec{r}_{1},\alpha_{1}}^{a_{1}}\left(s\right)\left(2J_{\vec{r}_{1}\vec{r}_{2}}\tau_{\alpha_{1}\alpha_{2}}^{1}\sigma_{1}^{a_{1}a_{2}}\right)a_{\vec{r}_{2},\alpha_{2}}^{a_{2}}\left(s\right),\label{eq:S_J - 1}
\end{eqnarray}

\begin{eqnarray}
S_{U} & = & \frac{1}{4!}\int_{0}^{s_{\alpha_{1}\alpha_{2}\alpha_{3}\alpha_{4}}^{f}}ds\sum_{\vec{r}}\left(-U\zeta_{\alpha_{1}\alpha_{2}\alpha_{3}\alpha_{4}}^{a_{1}a_{2}a_{3}a_{4}}\right)a_{\vec{r},\alpha_{1}}^{a_{1}}\left(s\right)a_{\vec{r},\alpha_{2}}^{a_{2}}\left(s\right)a_{\vec{r},\alpha_{3}}^{a_{3}}\left(s\right)a_{\vec{r},\alpha_{4}}^{a_{4}}\left(s\right),\label{eq:S_U - 1}
\end{eqnarray}

\begin{eqnarray}
\hat{\tau}^{0} & = & \left(\begin{array}{ccc}
1 & 0 & 0\\
0 & 1 & 0\\
0 & 0 & -i
\end{array}\right),\label{eq:tau_0 defined - 1}
\end{eqnarray}

\begin{eqnarray}
\zeta_{\alpha_{1}\alpha_{2}\alpha_{3}\alpha_{4}}^{a_{1}a_{2}a_{3}a_{4}} & = & 2\tau_{\alpha_{1}\alpha_{2}\alpha_{3}\alpha_{4}}\sigma^{a_{1}a_{2}a_{3}a_{4}},\label{eq:zeta tensor defined - 1}
\end{eqnarray}

\begin{eqnarray}
\tau_{\alpha_{1}\alpha_{2}\alpha_{3}\alpha_{4}} & = & \begin{cases}
\frac{1}{2}, & \text{if }\left\{ \alpha_{m}\right\} _{m=1}^{4}\in P\left(\left\{ q,c,c,c\right\} \right)\bigcup P\left(\left\{ c,q,q,q\right\} \right), \\
-i, & \text{if }\left\{ \alpha_{m}\right\} _{m=1}^{4}=\left\{ \mathcal{T},\mathcal{T},\mathcal{T},\mathcal{T}\right\}, \\
0, & \text{otherwise},
\end{cases}\label{eq:generalized tau tensor defined - 1}
\end{eqnarray}

\begin{eqnarray}
\sigma^{a_{1}a_{2}a_{3}a_{4}} & = & \begin{cases}
1, & \text{if }\left\{ a_{m}\right\} _{m=1}^{4}\in P\left(\left\{ 1,1,2,2\right\} \right), \\
0, & \text{otherwise},
\end{cases}\label{eq:generalized sigma tensor defined - 1}
\end{eqnarray}

\begin{eqnarray}
s_{\alpha_{1}\alpha_{2}\alpha_{3}\alpha_{4}}^{f} & = & \begin{cases}
\infty, & \text{if }\left\{ \alpha_{m}\right\} _{m=1}^{4}\in P\left(\left\{ q,c,c,c\right\} \right)\bigcup P\left(\left\{ c,q,q,q\right\} \right), \\
\beta, & \text{if }\left\{ \alpha_{m}\right\} _{m=1}^{4}=\left\{ \mathcal{T},\mathcal{T},\mathcal{T},\mathcal{T}\right\}, \\
0, & \text{otherwise}.
\end{cases}\label{eq:generalized integration limits - 1}
\end{eqnarray}

\noindent In the $\left(q,c,\mathcal{T}\right)$ basis, the source
term becomes

\begin{eqnarray}
S_{f} & = & \int_{0}^{s_{\alpha_{1}\alpha_{2}}^{f}}ds\sum_{\vec{r}}\tau_{\alpha_{1}\alpha_{2}}^{1}f_{\vec{r},\alpha_{1}}^{\overline{a}}\left(s\right)a_{\vec{r},\alpha_{2}}^{a}\left(s\right),\label{eq:S_f in qcT basis - 1}
\end{eqnarray}

\noindent and the CCOGFs are

\begin{eqnarray}
G_{\vec{r}_{1}\ldots\vec{r}_{n},\alpha_{1}\ldots\alpha_{n}}^{a_{1}\ldots a_{n},c}\left(s_{1},\ldots,s_{n}\right) & = & \left.\left(-1\right)^{n-1}\left(\left[\tau^{1}\right]_{\alpha_{1}\alpha_{1}^{\prime}}^{\dagger}\ldots\left[\tau^{1}\right]_{\alpha_{n}\alpha_{n}^{\prime}}^{\dagger}\right)\frac{\delta^{n}W\left[f\right]}{\delta f_{\vec{r}_{1},\alpha_{1}^{\prime}}^{\overline{a_{1}}}\left(s_{1}\right)\ldots\delta f_{\vec{r}_{n},\alpha_{n}^{\prime}}^{\overline{a_{n}}}\left(s_{n}\right)}\right|_{f=0}.\label{eq:CCOGFs in the qcT basis - 1}
\end{eqnarray}

\subsection{Effective theory for the Bose-Hubbard model\label{sub:Effective theory for the BHM - 1}}

In order to study quench dynamics in the BHM, we make use of an effective
theory that can describe both the weak and strong coupling limits
of the model in the same formalism. Such an approach was developed
in imaginary time by Sengupta and Dupuis \citep{SenguptaDupuis} by
using two Hubbard-Stratonovich transformations and generalized to
real-time in Ref.~\citep{Kennett}. A similar real-time theory was
also obtained based on a Ginzburg-Landau approach using the Schwinger-Keldysh
technique \citep{Gras,Grass2,Grassthesis}. A brief discussion of
the derivation of the effective theory along with minor corrections
to several expressions presented in Ref.~\citep{Kennett} is given
in \ref{app:deriving effective theory - 1}. The effective theory
obtained in Ref.~\citep{Kennett} for the $z$ fields (which are obtained
after two Hubbard Stratonovich transformations and have the same correlations
as the original $a$ fields \citep{SenguptaDupuis}) is

\begin{eqnarray}
S\left[z\right] & = & \frac{1}{2}\int_{0}^{s_{\alpha\alpha^{\prime}}^{f}}\left(\tau_{\alpha\alpha^{\prime}}^{1}ds\right)\sum_{\left\langle \vec{r}_{1}\vec{r}_{2}\right\rangle }z_{\vec{r}_{1},\alpha}^{\overline{a}}\left(s\right)\left[2J_{\vec{r}_{1}\vec{r}_{2}}\left(s\right)\right]z_{\vec{r}_{2},\alpha^{\prime}}^{a}\left(s\right)\nonumber \\
 &  & +\frac{1}{2}\sum_{\vec{r}}\prod_{m=1}^{2}\left[\int_{0}^{s_{\alpha_{m}\alpha_{m}^{\prime}}^{f}}\left(\tau_{\alpha_{m}\alpha_{m}^{\prime}}^{1}ds_{m}\right)z_{\vec{r},\alpha_{m}}^{a_{m}}\left(s_{m}\right)\right]\left[\left(\mathcal{G}^{c}\right)^{-1}\right]_{\alpha_{1}^{\prime}\alpha_{2}^{\prime}}^{\overline{a_{1}}\overline{a_{2}}}\left(s_{1},s_{2}\right)\nonumber \\
 &  & +\frac{1}{4!}\sum_{\vec{r}}\prod_{m=1}^{4}\left[\int_{0}^{s_{\alpha_{m}\alpha_{m}^{\prime}}^{f}}\left(\tau_{\alpha_{m}\alpha_{m}^{\prime}}^{1}ds_{m}\right)z_{\vec{r},\alpha_{m}}^{a_{m}}\left(s_{m}\right)\right]u_{\alpha_{1}^{\prime}\alpha_{2}^{\prime}\alpha_{3}^{\prime}\alpha_{4}^{\prime}}^{\overline{a_{1}}\overline{a_{2}}\overline{a_{3}}\overline{a_{4}}}\left(s_{1},s_{2},s_{3},s_{4}\right),\label{eq:Effective theory of BHM - 1}
\end{eqnarray}

\noindent where $\left(\mathcal{G}^{c}\right)^{-1}$ is the inverse of the two-point
CCOGF in the atomic limit (i.e. $J=0$), $u^{\left(4\right)}$ is 

\begin{eqnarray}
u_{\alpha_{1}\alpha_{2}\alpha_{3}\alpha_{4}}^{a_{1}a_{2}a_{3}a_{4}}\left(s_{1},s_{2},s_{3},s_{4}\right) & = & -\prod_{m=1}^{4}\left[\int_{0}^{s_{\alpha_{m}^{\prime}\alpha_{m}^{\prime\prime}}^{f}}\left(\tau_{\alpha_{m}^{\prime}\alpha_{m}^{\prime\prime}}^{1}ds_{m}^{\prime}\right)\left[\left(\mathcal{G}^{c}\right)^{-1}\right]_{\alpha_{m}\alpha_{m}^{\prime}}^{a_{m}a_{m}^{\prime}}\left(s_{m},s_{m}^{\prime}\right)\right]\nonumber \\
 &  & \left.\phantom{-\prod_{m=1}^{4}}\quad\times\mathcal{G}_{\alpha_{1}^{\prime\prime}\alpha_{2}^{\prime\prime}\alpha_{3}^{\prime\prime}\alpha_{4}^{\prime\prime}}^{\overline{a_{1}^{\prime}}\overline{a_{2}^{\prime}}\overline{a_{3}^{\prime}}\overline{a_{4}^{\prime}},c}\left(s_{1}^{\prime},s_{2}^{\prime},s_{3}^{\prime},s_{4}^{\prime}\right)\right.,\label{eq:u-vertex defined - 1}
\end{eqnarray}

\noindent and the inverse of an arbitrary two-point function $X$ satisfies

\begin{eqnarray}
 &  & \int_{0}^{s_{\alpha_{3}\alpha_{3}^{\prime}}^{f}}ds_{3}\sum_{\vec{r}_{3}}\left[X^{-1}\right]_{\vec{r}_{1}\vec{r}_{3},\alpha_{1}\alpha_{3}}^{a_{1}a_{3}}\left(s_{1},s_{3}\right)\left(\tau_{\alpha_{3}\alpha_{3}^{\prime}}^{1}\tau_{\alpha_{2}\alpha_{2}^{\prime}}^{1}X_{\vec{r}_{3}\vec{r}_{2},\alpha_{3}^{\prime}\alpha_{2}^{\prime}}^{\overline{a_{3}}\overline{a_{2}}}\left(s_{3},s_{2}\right)\right)\nonumber \\
 &  & \quad\equiv\delta_{\vec{r}_{1}\vec{r}_{2}}\delta_{\alpha_{1}\alpha_{2}}\delta^{a_{1}a_{2}}\delta\left(s_{1}-s_{2}\right).\label{eq:inverse of a two-point fn defined - 1}
\end{eqnarray}

\noindent Both $\left(\mathcal{G}^{c}\right)^{-1}$ and $u^{\left(4\right)}$ are independent
of site index $\vec{r}$, hence we write them without site labels.
However, throughout this paper we occasionally include the site labels
when it serves to provide more clarity to the reader. One would have
to include the site labels if for instance one considers the BHM with
a harmonic potential as is realised experimentally. 

Equation~(\ref{eq:Effective theory of BHM - 1}) is the key result from
Ref.~\citep{Kennett} that we use to develop the 2PI formalism in
Section \ref{sub:2PI Formalism and the effective action - 1}. However, before
applying the 2PI formalism to this action, we need to include an additional
correction term:

\begin{eqnarray}
S_{\text{correction}}\left[z\right] & = & \frac{1}{2}\sum_{\vec{r}}\prod_{m=1}^{2}\left[\int_{0}^{s_{\alpha_{m}\alpha_{m}^{\prime}}^{f}}\left(\tau_{\alpha_{m}\alpha_{m}^{\prime}}^{1}ds_{m}\right)z_{\vec{r},\alpha_{m}}^{a_{m}}\left(s_{m}\right)\right]\tilde{u}_{\alpha_{1}^{\prime}\alpha_{2}^{\prime}}^{\overline{a_{1}}\overline{a_{2}}}\left(s_{1},s_{2}\right),\label{eq:correction term to effective theory - 1}
\end{eqnarray}

\noindent where $\tilde{u}^{\left(2\right)}$ contains an infinite set of 
diagrams, although here we truncate it keeping only the lowest order term:

\begin{eqnarray}
\tilde{u}_{\alpha_{1}\alpha_{2}}^{a_{1}a_{2}}\left(s_{1},s_{2}\right) & = & -\frac{1}{2!}\prod_{m=3}^{4}\left[\int_{0}^{s_{\alpha_{m}^{\prime}\alpha_{m}^{\prime\prime}}^{f}}\left(\tau_{\alpha_{m}^{\prime}\alpha_{m}^{\prime\prime}}^{1}ds_{m}\right)\right]u_{\alpha_{1}\alpha_{2}\alpha_{3}\alpha_{4}}^{a_{1}a_{2}a_{3}a_{4}}\left(s_{1},s_{2},s_{3},s_{4}\right)\left\{ i\mathcal{G}_{\tau_{3}\tau_{4}}^{\overline{a_{3}}\overline{a_{4}},c}\left(s_{3},s_{4}\right)\right\} .\label{eq:approximation of 2-point u tilde - 1}
\end{eqnarray}

\noindent This correction term ensures that our equations of motion
are accurate to first order in $\mathcal{G}^{\left(4\right),c}$ (see
\ref{app:deriving effective theory - 1} for further discussion).
Moreover, it ensures that the equations of motion for the two-point
CCOGF we derive in Section \ref{sec:eqns of motion - 1} are exact
in the atomic ($J=0$) limit, which is essential when considering
quenches beginning in the atomic limit. This action also gives the
exact two-point CCOGF in the noninteracting ($U=0$) limit \citep{SenguptaDupuis}.
These features make this theory particularly appealing for the study
of quench dynamics, since it gives the hope that one can accurately
describe the behaviour of the system in both the superfluid and Mott-insulating
regimes \citep{KennettRev}. 

Using the symmetry relation in Eq.~(\ref{eq:CCOGF permutation symmetry - 1}),
we also note that $\left(G^{c}\right)^{-1}$, $\tilde{u}^{\left(2\right)}$ and $u^{\left(4\right)}$
satisfy the following symmetry relations (correcting Ref.~\citep{Kennett})

\begin{eqnarray}
\left[\left(G^{c}\right)^{-1}\right]_{\vec{r}_{1}\vec{r}_{2},\alpha_{1}\alpha_{2}}^{a_{1}a_{2}}\left(s_{1},s_{2}\right) & = & \left[\left(G^{c}\right)^{-1}\right]_{\vec{r}_{p_{1}}\vec{r}_{p_{2}},\alpha_{p_{1}}\alpha_{p_{2}}}^{a_{p_{1}}a_{p_{2}}}\left(s_{p_{1}},s_{p_{2}}\right),\label{eq:G^(-1) permutation symmetry - 1}\\
\tilde{u}_{\alpha_{1}\alpha_{2}}^{a_{1}a_{2}}\left(s_{1},s_{2}\right) & = & \tilde{u}_{\alpha_{p_{1}}\alpha_{p_{2}}}^{a_{p_{1}}a_{p_{2}}}\left(s_{p_{1}},s_{p_{2}}\right),\label{eq:2-point u tilde permutation symmetry - 1}\\
u_{\alpha_{1}\alpha_{2}\alpha_{3}\alpha_{4}}^{a_{1}a_{2}a_{3}a_{4}}\left(s_{1},s_{2},s_{3},s_{4}\right) & = & u_{\alpha_{p_{1}}\alpha_{p_{2}}\alpha_{p_{3}}\alpha_{p_{4}}}^{a_{p_{1}}a_{p_{2}}a_{p_{3}}a_{p_{4}}}\left(s_{p_{1}},s_{p_{2}},s_{p_{3}},s_{p_{4}}\right).\label{eq:u-vertex permutation symmetry - 1}
\end{eqnarray}

\noindent Similar symmetry relations for four-point functions were
noted in Refs. \citep{KennettRev,Grass2,Grassthesis}.

\subsection{2PI Formalism and the effective action\label{sub:2PI Formalism and the effective action - 1}}

In order to obtain the full two-point CCOGF (the ``full propagator''
from now on), which encodes non-local spatial and temporal correlations,
we adopt a 2PI approach. Unlike 1PI approaches
\citep{Kennett,Gras,Grass2,Grassthesis}, the 2PI formalism describes
the evolution of the mean field (i.e. superfluid order parameter for
the BHM) and the full propagator on equal footing \citep{Rey1}. Several
authors \citep{Rey1,Rey2,Temme} have applied the 2PI formalism to
the BHM to derive equations of motion for the mean field and the full
propagator for weak interactions.

Here, we develop a real-time 2PI approach based on the strong-coupling
theory of Sengupta and Dupuis \citep{SenguptaDupuis,Kennett} to capture
behaviour of correlations across a quantum quench. We adopt a compact
notation where we write an arbitrary function $X$ as 

\begin{equation}
X_{\vec{r}_{1}\ldots\vec{r}_{n},\tau_{1}\ldots\tau_{n}}^{a_{1}\ldots a_{n}}\equiv X_{\vec{r}_{1}\ldots\vec{r}_{n}}^{a_{1}\ldots a_{n}}\left(\tau_{1}\ldots\tau_{n}\right)=X_{\vec{r}_{1}\ldots\vec{r}_{n},\alpha_{1}\ldots\alpha_{n}}^{a_{1}\ldots a_{n}}\left(s_{1}\ldots s_{n}\right).\label{eq:compact notation for 2PI calculations - 1}
\end{equation}

\noindent We extend the Einstein summation convention to the $\tau$
subindices such that for two arbitrary functions $X$ and $Y$ we
have

\begin{eqnarray}
\sum_{\vec{r}}X_{\vec{r},\tau}^{a}Y_{\vec{r},\tau}^{\overline{a}} & = & \sum_{\vec{r}}\int_{0}^{s_{\alpha\alpha^{\prime}}^{f}}\left(\tau_{\alpha\alpha^{\prime}}^{1}ds\right)X_{\vec{r},\alpha}^{a}\left(s\right)Y_{\vec{r},\alpha^{\prime}}^{\overline{a}}\left(s\right).\label{eq:Einstein summation convention for tau - 1}
\end{eqnarray}

We can rewrite Eq.~(\ref{eq:Effective theory of BHM - 1}) (with the
correction term {[}Eq.~(\ref{eq:correction term to effective theory - 1}){]}
included) in the condensed notation as

\begin{eqnarray}
S\left[z\right] & = & \frac{1}{2!}\sum_{\vec{r}_{1}\vec{r}_{2}}\left[g_{0}^{-1}\right]_{\vec{r}_{1}\vec{r}_{2},\tau_{1}\tau_{2}}^{a_{1}a_{2}}z_{\vec{r}_{1},\tau_{1}}^{\overline{a_{1}}}z_{\vec{r}_{2},\tau_{2}}^{\overline{a_{2}}}+\frac{1}{4!}u_{\tau_{1}\tau_{2}\tau_{3}\tau_{4}}^{a_{1}a_{2}a_{3}a_{4}}\sum_{\vec{r}}z_{\vec{r},\tau_{1}}^{\overline{a_{1}}}z_{\vec{r},\tau_{2}}^{\overline{a_{2}}}z_{\vec{r},\tau_{3}}^{\overline{a_{3}}}z_{\vec{r},\tau_{4}}^{\overline{a_{4}}},\label{eq:Effective theory in condensed notation - 1}
\end{eqnarray}

\noindent where we have introduced the generalized inverse bare propagator
$g_{0}^{-1}$

\begin{eqnarray}
\left[g_{0}^{-1}\right]_{\vec{r}_{1}\vec{r}_{2},\tau_{1}\tau_{2}}^{a_{1}a_{2}} & = & \left[\left(\mathcal{G}^{c}\right)^{-1}\right]_{\vec{r}_{1}\vec{r}_{2},\tau_{1}\tau_{2}}^{a_{1}a_{2}}+2J_{\vec{r}_{1}\vec{r}_{2},\tau_{1}\tau_{2}}^{a_{1}a_{2}}-\frac{1}{2!}\delta_{\vec{r}_{1}\vec{r}_{2}}u_{\tau_{1}\tau_{2}\tau_{3}\tau_{4}}^{a_{1}a_{2}a_{3}a_{4}}\left(i\mathcal{G}_{\vec{r}_{1}\vec{r}_{1},\tau_{3}\tau_{4}}^{\overline{a_{3}}\overline{a_{4}},c}\right),\label{eq:generalized bare propagator - 1}
\end{eqnarray}

\noindent with

\begin{eqnarray}
\left[\left(\mathcal{G}^{c}\right)^{-1}\right]_{\vec{r}_{1}\vec{r}_{2},\tau_{1}\tau_{2}}^{a_{1}a_{2}} & = & \delta_{\vec{r}_{1}\vec{r}_{2}}\left[\left(\mathcal{G}^{c}\right)^{-1}\right]_{\alpha_{1}\alpha_{2}}^{a_{1}a_{2}}\left(s_{1},s_{2}\right),\label{eq:2-pt G^(-1) in atomic limit - 1}
\end{eqnarray}

\begin{eqnarray}
J_{\vec{r}_{1}\vec{r}_{2},\tau_{1}\tau_{2}}^{a_{1}a_{2}} & = & J_{\vec{r}_{1}\vec{r}_{2}}\left(s_{1}\right)\left[\tau^{1}\right]_{\alpha_{1}\alpha_{2}}^{\dagger}\sigma_{1}^{a_{1}a_{2}}\delta\left(s_{1}-s_{2}\right).\label{eq:J tensor defined - 1}
\end{eqnarray}

In the 2PI formalism \citep{Cornwall,RammerText}, physical quantities
are expressed in terms of the mean field $\phi$ and the full propagator
$G^{c}$

\begin{eqnarray}
\phi_{\vec{r}_{1},\tau_{1}}^{a_{1}} & \equiv & \left\langle z_{\vec{r}_{1},\tau_{1}}^{a_{1}}\right\rangle ,\label{eq:phi defined - 1}\\
iG_{\vec{r}_{1}\vec{r}_{2},\tau_{1}\tau_{2}}^{a_{1}a_{2},c} & = & \left\langle z_{\vec{r}_{1},\tau_{1}}^{a_{1}}z_{\vec{r}_{2},\tau_{2}}^{a_{2}}\right\rangle -\left\langle z_{\vec{r}_{1},\tau_{1}}^{a_{1}}\right\rangle \left\langle z_{\vec{r}_{2},\tau_{2}}^{a_{2}}\right\rangle .\label{eq:2-point CCOGF reintroduced - 1}
\end{eqnarray}

\noindent Note that $G^{c}$ is symmetric: $G_{\vec{r}_{1}\vec{r}_{2},\tau_{1}\tau_{2}}^{a_{1}a_{2},c}=G_{\vec{r}_{2}\vec{r}_{1},\tau_{2}\tau_{1}}^{a_{2}a_{1},c}$.
The equations of motion for $\phi$ and $G^{c}$ are obtained by requiring
the 2PI effective action $\Gamma\left[\phi,G^{c}\right]$ be stationary
with respect to variations of $\phi$ and $G^{c}$. This is similar to
the 1PI case where the equations of motion for $\phi$ are obtained
by requiring the 1PI effective action $\Gamma\left[\phi\right]$ to
be stationary with respect to variations of $\phi$. The full propagator
from the 2PI effective action allows one to take into account broken
symmetry states \citep{Cornwall,RammerText}, which is necessary to
describe quenches in the superfluid regime. 

To obtain the effective action we define the 2PI generating functional
for Green's functions $\mathcal{Z}\left[f,K\right]$

\begin{equation}
\mathcal{Z}\left[f,K\right]=e^{iW\left[f,K\right]}=\int\left[\mathcal{D}z^{a}\right]e^{iS\left[z\right]+i\sum_{\vec{r}_{1}}f_{\vec{r}_{1},\tau_{1}}^{a_{1}}z_{\vec{r}_{1},\tau_{1}}^{\overline{a_{1}}}+\frac{i}{2}\sum_{\vec{r}_{1}\vec{r}_{2}}K_{\vec{r}_{1}\vec{r}_{2},\tau_{1}\tau_{2}}^{a_{1}a_{2}}z_{\vec{r}_{1},\tau_{1}}^{\overline{a_{1}}}z_{\vec{r}_{2},\tau_{2}}^{\overline{a_{2}}}},\label{eq:2PI Z generating functional - 1}
\end{equation}

\noindent where in addition to the single-particle source current
$f$, we have included a (symmetric) two-particle source current $K$.
Note that $\phi$ and $G^{c}$ are obtained by calculating the following
functional derivatives of $W\left[f,K\right]$:

\begin{equation}
\phi_{\vec{r}_{1},\tau_{1}}^{a_{1}}=\frac{\delta W\left[f,K\right]}{\delta f_{\vec{r}_{1},\tau_{1}}^{\overline{a_{1}}}},\qquad\frac{1}{2}\left(\phi_{\vec{r}_{1},\tau_{1}}^{a_{1}}\phi_{\vec{r}_{2},\tau_{2}}^{a_{2}}+iG_{\vec{r}_{1}\vec{r}_{2},\tau_{1}\tau_{2}}^{a_{1}a_{2},c}\right)=\frac{\delta W\left[f,K\right]}{\delta K_{\vec{r}_{1}\vec{r}_{2},\tau_{1}\tau_{2}}^{\overline{a_{1}}\overline{a_{2}}}}.\label{eq:phi-G-W identities - 1}
\end{equation}

\noindent These equations implicitly give $f$ and $K$ as functions
of $\phi$ and $G^{c}$: $f=f\left[\phi,G^{c}\right]$ and $K=K\left[\phi,G^{c}\right]$.
The 2PI effective action $\Gamma\left[\phi,G^{c}\right]$ is formally
defined as the double Legendre transform of $W\left[f,K\right]$

\begin{eqnarray}
\Gamma\left[\phi,G\right] & = & W\left[f,K\right]-\sum_{\vec{r}_{1}}f_{\vec{r}_{1},\tau_{1}}^{a_{1}}\phi_{\vec{r}_{1},\tau_{1}}^{\overline{a_{1}}}-\frac{1}{2}\sum_{\vec{r}_{1}\vec{r}_{2}}K_{\vec{r}_{1}\vec{r}_{2},\tau_{1}\tau_{2}}^{a_{1}a_{2}}\left(\phi_{\vec{r}_{1},\tau_{1}}^{\overline{a_{1}}}\phi_{\vec{r}_{2},\tau_{2}}^{\overline{a_{2}}}+iG_{\vec{r}_{1}\vec{r}_{2},\tau_{1}\tau_{2}}^{\overline{a_{1}}\overline{a_{2}},c}\right),\label{eq:Gamma generating functional defined - 1}
\end{eqnarray}

\noindent where $f$ and $K$ should be understood as being expressed
in terms of $\phi$ and $G^{c}$. The following identities can be derived
\citep{Cornwall,RammerText} from Eq.~(\ref{eq:Gamma generating functional defined - 1}) 

\begin{eqnarray}
\frac{\delta\Gamma\left[\phi,G^{c}\right]}{\delta\phi_{\vec{r}_{1},\tau_{1}}^{\overline{a_{1}}}} & = & -f_{\vec{r}_{1},\tau_{1}}^{a_{1}}-\sum_{\vec{r}_{1}\vec{r}_{2}}K_{\vec{r}_{1}\vec{r}_{2},\tau_{1}\tau_{2}}^{a_{1}a_{2}}\phi_{\vec{r}_{2},\tau_{2}}^{\overline{a_{2}}},\label{eq:phi derivative of Gamma - 1}
\end{eqnarray}

\begin{eqnarray}
\frac{\delta\Gamma\left[\phi,G^{c}\right]}{\delta G_{\vec{r}_{1}\vec{r}_{2},\tau_{1}\tau_{2}}^{\overline{a_{1}}\overline{a_{2}},c}} & = & -\frac{i}{2}K_{\vec{r}_{1}\vec{r}_{2},\tau_{1}\tau_{2}}^{a_{1}a_{2}}.\label{eq:G derivative of Gamma - 1}
\end{eqnarray}

\noindent Defining

\begin{eqnarray}
\left[D^{-1}\right]_{\vec{r}_{1}\vec{r}_{2},\tau_{1}\tau_{2}}^{a_{1}a_{2}} & = & \frac{\delta^{2}S\left[\phi\right]}{\delta\phi_{\vec{r}_{1},\tau_{1}}^{\overline{a_{1}}}\delta\phi_{\vec{r}_{2},\tau_{2}}^{\overline{a_{2}}}}\nonumber \\
 & = & \left[g_{0}^{-1}\right]_{\vec{r}_{1}\vec{r}_{2},\tau_{1}\tau_{2}}^{a_{1}a_{2}}+\frac{1}{2!}\delta_{\vec{r}_{1}\vec{r}_{2}}u_{\tau_{1}\tau_{2}\tau_{3}\tau_{4}}^{a_{1}a_{2}a_{3}a_{4}}\phi_{\vec{r}_{1},\tau_{3}}^{\overline{a_{3}}}\phi_{\vec{r}_{1},\tau_{4}}^{\overline{a_{4}}},\label{eq:D^(-1) defined - 1}
\end{eqnarray}

\noindent the effective action can be shown to take the form \citep{Cornwall,RammerText}

\begin{eqnarray}
\Gamma\left[\phi,G^{c}\right] & = & S\left[\phi\right]+\frac{i}{2}\text{Tr}\left\{ \ln \left[\left(G^{c}\right)^{-1}\right]\right\} +\frac{i}{2}\sum_{\vec{r}_{1}\vec{r}_{2}}\left[D^{-1}\right]_{\vec{r}_{1}\vec{r}_{2},\tau_{1}\tau_{2}}^{a_{1}a_{2}}G_{\vec{r}_{2}\vec{r}_{1},\tau_{2}\tau_{1}}^{\overline{a_{2}}\overline{a_{1}},c}+\Gamma_{2}\left[\phi,G^{c}\right]+\text{const},\label{eq:Gamma rewritten - 1}
\end{eqnarray}

\noindent where $\Gamma_{2}\left[\phi,G^{c}\right]$ is the sum of all
2PI connected vacuum diagrams in the theory with vertices determined
by the action

\begin{eqnarray}
S_{\text{int}}\left[\varphi;\phi\right] & = & u_{\tau_{1}\tau_{2}\tau_{3}\tau_{4}}^{a_{1}a_{2}a_{3}a_{4}}\sum_{\vec{r}}\left\{ \frac{1}{3!}\varphi_{\vec{r},\tau_{1}}^{\overline{a_{1}}}\varphi_{\vec{r},\tau_{2}}^{\overline{a_{2}}}\varphi_{\vec{r},\tau_{3}}^{\overline{a_{3}}}\phi_{\vec{r},\tau_{4}}^{\overline{a_{4}}}+\frac{1}{4!}\varphi_{\vec{r},\tau_{1}}^{\overline{a_{1}}}\varphi_{\vec{r},\tau_{2}}^{\overline{a_{2}}}\varphi_{\vec{r},\tau_{3}}^{\overline{a_{3}}}\varphi_{\vec{r},\tau_{4}}^{\overline{a_{4}}}\right\} ,\label{eq:2PI S_int - 1}
\end{eqnarray}

\noindent and the propagator lines determined by $G^{c}$, i.e. 

\begin{eqnarray}
\Gamma_{2}\left[\phi,G^{c}\right] & = & -i\ln\left\{ \left(\det\left\{ iG^{c}\right\} \right)^{-1/2}\int\mathcal{D}\left[\varphi\right]e^{\frac{i}{2!}\sum_{\vec{r}_{1}\vec{r}_{2}}\left[\left(G^{c}\right)^{-1}\right]_{\vec{r}_{1}\vec{r}_{2},\tau_{1}\tau_{2}}^{a_{1}a_{2}}\varphi_{\vec{r}_{1},\tau_{1}}^{\overline{a_{1}}}\varphi_{\vec{r}_{2},\tau_{2}}^{\overline{a_{2}}}}e^{iS_{\text{int}}\left[\varphi;\phi\right]}\right\} ^{\text{2PI}}.\label{eq:Gamma_2 expression - 1}
\end{eqnarray}

\noindent One can use Eq.~(\ref{eq:Gamma_2 expression - 1}) along
with Wick's theorem to generate all the diagrams in $\Gamma_{2}\left[\phi,G^{c}\right]$.

\begin{figure}
\begin{centering}
\includegraphics[scale=0.8]{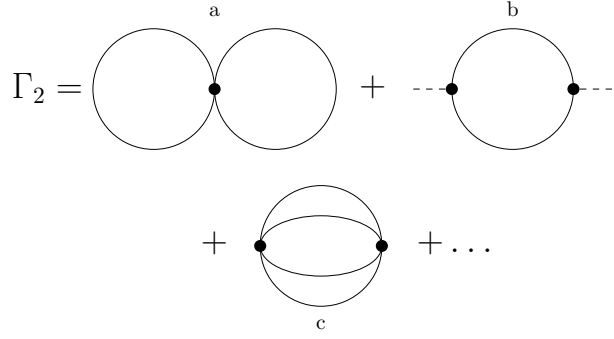}
\par\end{centering}

\caption{Diagrammatic expansion of $\Gamma_{2}$ up to second-order in the
four-point vertex $u^{\left(4\right)}$ (as shown as a solid dot),
showing (a) the \emph{double-bubble} diagram, (b) the \emph{setting
sun} diagram, and (c) the \emph{basketball} diagram.\label{fig:fig2}}
\end{figure}

\begin{figure}
\begin{centering}
\includegraphics[scale=0.8]{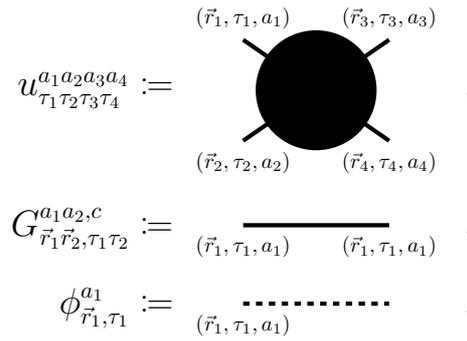}
\par\end{centering}

\caption{Diagrammatic representation of $u_{\tau_{1}\tau_{2}\tau_{3}\tau_{4}}^{a_{1}a_{2}a_{3}a_{4}}$,
$G_{\vec{r}_{1}\vec{r}_{2},\tau_{1}\tau_{2}}^{a_{1}a_{2},c}$, and $\phi_{\vec{r}_{1},\tau_{1}}^{a_{1}}$.\label{fig:fig3}}
\end{figure}

The diagrammatic expansion of $\Gamma_{2}\left[\phi,G^{c}\right]$ is
shown in Fig. \ref{fig:fig2} up to second-order in the four-point
vertex $u^{\left(4\right)}$. The solid dots represent the interaction
vertices $u^{\left(4\right)}$, the solid lines represent $G^{c}$, and
the dashed lines represent $\phi$ (as illustrated in Fig.~\ref{fig:fig3}).
In this paper, we only consider the first diagram in Fig.~\ref{fig:fig2},
i.e. the double-bubble (D.B.) diagram, which was also considered (along
with the remaining two diagrams) in Refs.~\citep{Rey1,Temme} where
the BHM was studied at weak coupling. However, there is an important
distinction between the calculations here and those in Refs.~\citep{Rey1,Temme}, 
which is that the interaction
vertices in Refs.~\citep{Rey1,Temme} are local in both space and
time, whereas the interaction vertices we consider are local in space
but \emph{nonlocal} in time -- this leads to additional features in
the equations of motion. The contribution from the D.B. diagram is

\begin{eqnarray}
\Gamma_{2}^{\left(\text{D.B.}\right)} & = & \frac{1}{8}u_{\tau_{1}\tau_{2}\tau_{3}\tau_{4}}^{a_{1}a_{2}a_{3}a_{4}}\sum_{\vec{r}}\left(iG_{\vec{r}\vec{r},\tau_{1}\tau_{2}}^{\overline{a_{1}}\overline{a_{2}},c}\right)\left(iG_{\vec{r}\vec{r},\tau_{3}\tau_{4}}^{\overline{a_{3}}\overline{a_{4}},c}\right).\label{eq:Gamma_2^(D.B.) - 1}
\end{eqnarray}

\section{Equations of motion\label{sec:eqns of motion - 1}}
To calculate the equations of motion, first we use Eqs.~(\ref{eq:phi derivative of Gamma - 1})
and (\ref{eq:G derivative of Gamma - 1}) and set the sources to zero,
giving

\begin{equation}
\frac{\delta S}{\delta\phi_{\vec{r}_{1},\tau_{1}}^{\overline{a_{1}}}}+\frac{i}{2}\left[\sum_{\vec{r}_{2}\vec{r}_{3}}\frac{\delta\left[D^{-1}\right]_{\vec{r}_{2}\vec{r}_{3},\tau_{2}\tau_{3}}^{a_{2}a_{3}}}{\delta\phi_{\vec{r}_{1},\tau_{1}}^{\overline{a_{1}}}}G_{\vec{r}_{3}\vec{r}_{2},\tau_{3}\tau_{2}}^{\overline{a_{3}}\overline{a_{2}},c}\right]+\frac{\delta\Gamma_{2}}{\delta\phi_{\vec{r}_{1},\tau_{1}}^{\overline{a_{1}}}}=0,\label{eq:phi eqn of motion - 1}
\end{equation}

\noindent and

\begin{eqnarray}
i\left[\left(G^{c}\right)^{-1}\right]_{\vec{r}_{1}\vec{r}_{2},\tau_{1}\tau_{2}}^{a_{1}a_{2}} & = & i\left[D^{-1}\right]_{\vec{r}_{1}\vec{r}_{2},\tau_{1}\tau_{2}}^{a_{1}a_{2}}-i\left[\Sigma^{\left(2\text{PI}\right)}\right]_{\vec{r}_{1}\vec{r}_{2},\tau_{1}\tau_{2}}^{a_{1}a_{2}},\label{eq:G eqn of motion - 1}
\end{eqnarray}

\noindent where the second equation is Dyson's equation with 

\begin{eqnarray}
\left[\Sigma^{\left(2\text{PI}\right)}\right]_{\vec{r}_{1}\vec{r}_{2},\tau_{1}\tau_{2}}^{a_{1}a_{2}} & \equiv & 2i\frac{\delta\Gamma_{2}}{\delta G_{\vec{r}_{1}\vec{r}_{2},\tau_{1}\tau_{2}}^{\overline{a_{1}}\overline{a_{2}},c}},\label{eq:self energy - 1}
\end{eqnarray}

\noindent the 2PI self energy. 

Given the form of the bare propagator in our strong-coupling theory,
the equations of motion Eq.~(\ref{eq:phi eqn of motion - 1}) and
(\ref{eq:G eqn of motion - 1}) in their above formulations are not
suitable for dynamical calculations. We begin by reformulating Eq.~(\ref{eq:phi eqn of motion - 1}). First, we explicitly calculate
the first term in Eq.~(\ref{eq:phi eqn of motion - 1})

\begin{eqnarray}
\frac{\delta S}{\delta\phi_{\vec{r}_{1},\tau_{1}}^{\overline{a_{1}}}} & = & \sum_{\vec{r}_{2}}\left[\left(\mathcal{G}^{c}\right)^{-1}\right]_{\vec{r}_{1}\vec{r}_{2},\tau_{1}\tau_{2}}^{a_{1}a_{2}}\phi_{\vec{r}_{2},\tau_{2}}^{\overline{a_{2}}}+\sum_{\vec{r}_{2}}2J_{\vec{r}_{1}\vec{r}_{2},\tau_{1}\tau_{2}}^{a_{1}a_{2}}\phi_{\vec{r}_{2},\tau_{2}}^{\overline{a_{2}}}\nonumber \\
 &  & -\frac{1}{2!}u_{\tau_{1}\tau_{2}\tau_{3}\tau_{4}}^{a_{1}a_{2}a_{3}a_{4}}\phi_{\vec{r}_{1},\tau_{2}}^{\overline{a_{2}}}\left(i\mathcal{G}_{\vec{r}_{1}\vec{r}_{1},\tau_{3}\tau_{4}}^{\overline{a_{3}}\overline{a_{4}},c}\right)+\frac{1}{3!}u_{\tau_{1}\tau_{2}\tau_{3}\tau_{4}}^{a_{1}a_{2}a_{3}a_{4}}\phi_{\vec{r}_{1},\tau_{2}}^{\overline{a_{2}}}\phi_{\vec{r}_{1},\tau_{3}}^{\overline{a_{3}}}\phi_{\vec{r}_{1},\tau_{4}}^{\overline{a_{4}}}.\label{eq:dS/dphi - 1}
\end{eqnarray}

\noindent The second term in Eq.~(\ref{eq:phi eqn of motion - 1})
can be written as

\begin{eqnarray}
\frac{i}{2}\left[\sum_{\vec{r}_{2}\vec{r}_{3}}\frac{\delta\left[D^{-1}\right]_{\vec{r}_{2}\vec{r}_{3},\tau_{2}\tau_{3}}^{a_{2}a_{3}}}{\delta\phi_{\vec{r}_{1},\tau_{1}}^{\overline{a_{1}}}}G_{\vec{r}_{3}\vec{r}_{2},\tau_{3}\tau_{2}}^{\overline{a_{3}}\overline{a_{2}},c}\right] & = & \frac{1}{2!}u_{\tau_{1}\tau_{2}\tau_{3}\tau_{4}}^{a_{1}a_{2}a_{3}a_{4}}\phi_{\vec{r}_{1},\tau_{2}}^{\overline{a_{2}}}\left(iG_{\vec{r}_{1}\vec{r}_{1},\tau_{3}\tau_{4}}^{\overline{a_{3}}\overline{a_{4}},c}\right).\label{eq:second term in phi eqn of motion - 1}
\end{eqnarray}

\noindent We act on both sides of Eq.~(\ref{eq:phi eqn of motion - 1})
with $\mathcal{G}^c$ from the left and rearrange terms to get

\begin{eqnarray}
\phi_{\vec{r}_{1},\tau_{1}}^{a_{1}} & = & \mathcal{G}_{\vec{r}_{1}\vec{r}_{2},\tau_{1}\tau_{2}}^{a_{1}a_{2},c}\Omega_{\vec{r}_{1},\tau_{2}}^{\overline{a_{2}}},\label{eq:phi eqn of motion - 2}
\end{eqnarray}

\noindent where we have introduced the quantity

\begin{eqnarray}
\Omega_{\vec{r}_{1},\tau_{1}}^{a_{1}} & = & -\sum_{\vec{r}_{2}}2J_{\vec{r}_{1}\vec{r}_{2},\tau_{1}\tau_{2}}^{a_{1}a_{2}}\phi_{\vec{r}_{2},\tau_{2}}^{\overline{a_{2}}}-\frac{1}{3!}u_{\tau_{1}\tau_{2}\tau_{3}\tau_{4}}^{a_{1}a_{2}a_{3}a_{4}}\phi_{\vec{r}_{1},\tau_{2}}^{\overline{a_{2}}}\phi_{\vec{r}_{1},\tau_{3}}^{\overline{a_{3}}}\phi_{\vec{r}_{1},\tau_{4}}^{\overline{a_{4}}}\nonumber \\
 &  & -\frac{1}{2!}u_{\tau_{1}\tau_{2}\tau_{3}\tau_{4}}^{a_{1}a_{2}a_{3}a_{4}}\phi_{\vec{r}_{1},\tau_{2}}^{\overline{a_{2}}}\left(iG_{\vec{r}_{1}\vec{r}_{1},\tau_{3}\tau_{4}}^{\overline{a_{3}}\overline{a_{4}},c}-i\mathcal{G}_{\vec{r}_{1}\vec{r}_{1},\tau_{3}\tau_{4}}^{\overline{a_{3}}\overline{a_{4}},c}\right)-\frac{\delta\Gamma_{2}}{\delta\phi_{\vec{r}_{1},\tau_{1}}^{\overline{a_{1}}}}.\label{eq:sigma^(phi) defined - 1}
\end{eqnarray}

\noindent Eq.~(\ref{eq:phi eqn of motion - 2}) is a much more suitable
form for dynamical calculations.

Next we reformulate Eq.~(\ref{eq:G eqn of motion - 1}) into a more
appropriate form. First, we separate $\left[D^{-1}\right]_{\vec{r}_{1}\vec{r}_{2},\tau_{1}\tau_{2}}^{a_{1}a_{2}}$
as follows

\begin{eqnarray}
\left[D^{-1}\right]_{\vec{r}_{1}\vec{r}_{2},\tau_{1}\tau_{2}}^{a_{1}a_{2}} & = & \left[\left(\mathcal{G}^{c}\right)^{-1}\right]_{\vec{r}_{1}\vec{r}_{2},\tau_{1}\tau_{2}}^{a_{1}a_{2}}-\left[\Sigma^{\left(1\right)}\right]_{\vec{r}_{1}\vec{r}_{2},\tau_{1}\tau_{2}}^{a_{1}a_{2}},\label{eq:separate D^(-1) - 1}
\end{eqnarray}

\noindent where 

\begin{eqnarray}
\left[\Sigma^{\left(1\right)}\right]_{\vec{r}_{1}\vec{r}_{2},\tau_{1}\tau_{2}}^{a_{1}a_{2}} & = & -2J_{\vec{r}_{1}\vec{r}_{2},\tau_{1}\tau_{2}}^{a_{1}a_{2}}+\frac{1}{2!}\delta_{\vec{r}_{1}\vec{r}_{2}}u_{\tau_{1}\tau_{2}\tau_{3}\tau_{4}}^{a_{1}a_{2}a_{3}a_{4}}\left(i\mathcal{G}_{\vec{r}_{1}\vec{r}_{1},\tau_{3}\tau_{4}}^{\overline{a_{3}}\overline{a_{4}},c}\right)-\frac{1}{2!}\delta_{\vec{r}_{1}\vec{r}_{2}}u_{\tau_{1}\tau_{2}\tau_{3}\tau_{4}}^{a_{1}a_{2}a_{3}a_{4}}\phi_{\vec{r}_{1},\tau_{3}}^{\overline{a_{3}}}\phi_{\vec{r}_{1},\tau_{4}}^{\overline{a_{4}}},\label{eq:1-loop self energy - 1}
\end{eqnarray}

\noindent is the $1$-loop contribution to the total self
energy. If we define the full self energy as 

\begin{eqnarray}
\Sigma_{\vec{r}_{1}\vec{r}_{2},\tau_{1}\tau_{2}}^{a_{1}a_{2}} & \equiv & \left[\Sigma^{\left(1\right)}\right]_{\vec{r}_{1}\vec{r}_{2},\tau_{1}\tau_{2}}^{a_{1}a_{2}}+\left[\Sigma^{\left(2\text{PI}\right)}\right]_{\vec{r}_{1}\vec{r}_{2},\tau_{1}\tau_{2}}^{a_{1}a_{2}},\label{eq:full self energy - 1}
\end{eqnarray}

\noindent then Eq.~(\ref{eq:G eqn of motion - 1}) becomes

\begin{eqnarray}
i\left[\left(G^{c}\right)^{-1}\right]_{\vec{r}_{1}\vec{r}_{2},\tau_{1}\tau_{2}}^{a_{1}a_{2}} & = & i\left[\left(\mathcal{G}^{c}\right)^{-1}\right]_{\vec{r}_{1}\vec{r}_{2},\tau_{1}\tau_{2}}^{a_{1}a_{2}}-i\Sigma_{\vec{r}_{1}\vec{r}_{2},\tau_{1}\tau_{2}}^{a_{1}a_{2}}.\label{eq:G eqn of motion - 2}
\end{eqnarray}

\noindent After rearranging a few terms, one obtains

\begin{eqnarray}
G_{\vec{r}_{1}\vec{r}_{2}\tau_{1}\tau_{2}x_{2}}^{a_{1}a_{2},c} & = & \mathcal{G}_{\vec{r}_{1}\vec{r}_{2},\tau_{1}\tau_{2}}^{a_{1}a_{2},c}+\sum_{\vec{r}_{3}\vec{r}_{4}}\mathcal{G}_{\vec{r}_{1}\vec{r}_{3},\tau_{1}\tau_{3}}^{a_{1}a_{3},c}\Sigma_{\vec{r}_{3}\vec{r}_{4},\tau_{3}\tau_{4}}^{\overline{a_{3}}\overline{a_{4}}}G_{\vec{r}_{4}\vec{r}_{2},\tau_{4}\tau_{2}}^{a_{4}a_{2},c},\label{eq:G eqn of motion - 3}
\end{eqnarray}

\noindent which is a more suitable form for dynamical calculations.
That being said, the form shown here is still not particularly amenable
to solution. We now discuss simplifications that allow us to obtain
equations of motion that are easier to solve.

\subsection{Low-frequency approximation\label{sub:low freq approx - 1}}

Equations~(\ref{eq:phi eqn of motion - 2}) and (\ref{eq:G eqn of motion - 3}),
whilst having a compact form in our notation, contain as many as four
time-integrals, making it computationally expensive to solve the equations
numerically. This suggests that some level of approximation is required
in order to obtain more physical insight from the equations above.
Following Refs.~\citep{Kennett}, we focus on the low frequency components of the equations of
motion.  In a quench protocol this would correspond to considering changes that 
are slow enough that the equations of motion are dominated by low frequency terms. 
The approximation also applies to equilibrium calculations where there is no quench at all.

The low-frequency approximation we consider involves taking the static-limit
of the four-point vertex $u^{\left(4\right)}$. If we only consider
values of the chemical potential away from the degeneracy points between
adjacent Mott lobes, i.e. $\mu\not\approx Ur$, with $r$ an integer,
then the static limit of $u^{\left(4\right)}$ can be expressed as
\citep{SenguptaDupuis,Kennett,Grassthesis}

\begin{eqnarray}
u_{\tau_{1}\tau_{2}\tau_{3}\tau_{4}}^{a_{1}a_{2}a_{3}a_{4}} & \approx & -u_{1}\delta\left(s_{1}-s_{2}\right)\delta\left(s_{1}-s_{3}\right)\delta\left(s_{1}-s_{4}\right)\zeta_{\alpha_{1}\alpha_{2}\alpha_{3}\alpha_{4}}^{a_{1}a_{2}a_{3}a_{4}}\nonumber \\
 &  & +iu_{2}^{2}\left[\delta\left(s_{1}-s_{2}\right)\delta\left(s_{3}-s_{4}\right)\eta_{\alpha_{1}\alpha_{2}\alpha_{3}\alpha_{4}}^{a_{1}a_{2}a_{3}a_{4}}+\left\{ 2\leftrightarrow3\right\} +\left\{ 2\leftrightarrow4\right\} \right],\label{eq:static lim of u^(4) - 1}
\end{eqnarray}

\noindent where $u_{1}$ and $u_{2}^{2}$ are defined in \ref{app:low frequency approximation - 1},
$\zeta_{\alpha_{1}\alpha_{2}\alpha_{3}\alpha_{4}}^{a_{1}a_{2}a_{3}a_{4}}$
is defined in Eq.~(\ref{eq:zeta tensor defined - 1}) and

\begin{eqnarray}
\eta_{\alpha_{1}\alpha_{2}\alpha_{3}\alpha_{4}}^{a_{1}a_{2}a_{3}a_{4}} & \equiv & \sigma_{1}^{a_{1}a_{2}}\sigma_{1}^{a_{3}a_{4}}\begin{cases}
\tau_{\alpha_{1}\alpha_{2}}^{1}\tau_{\alpha_{3}\alpha_{4}}^{1} & \text{if }\alpha_{m}=q\text{ or }c\text{ for }m=1,\ldots4\\
0 & \text{otherwise}
\end{cases}.\label{eq:eta tensor defined - 1}
\end{eqnarray}

\begin{figure}
\begin{centering}
\includegraphics[scale=0.4]{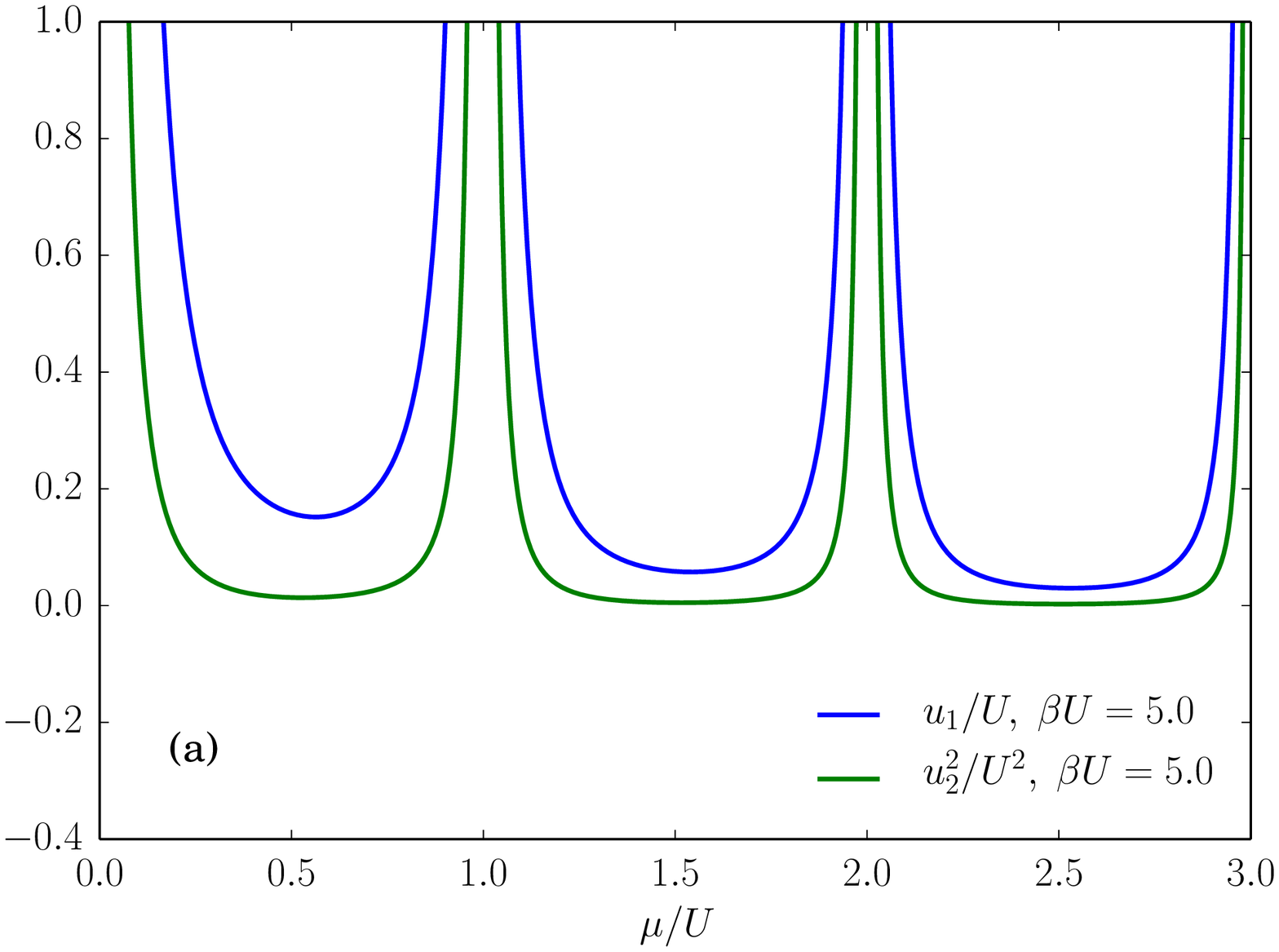}\includegraphics[scale=0.4]{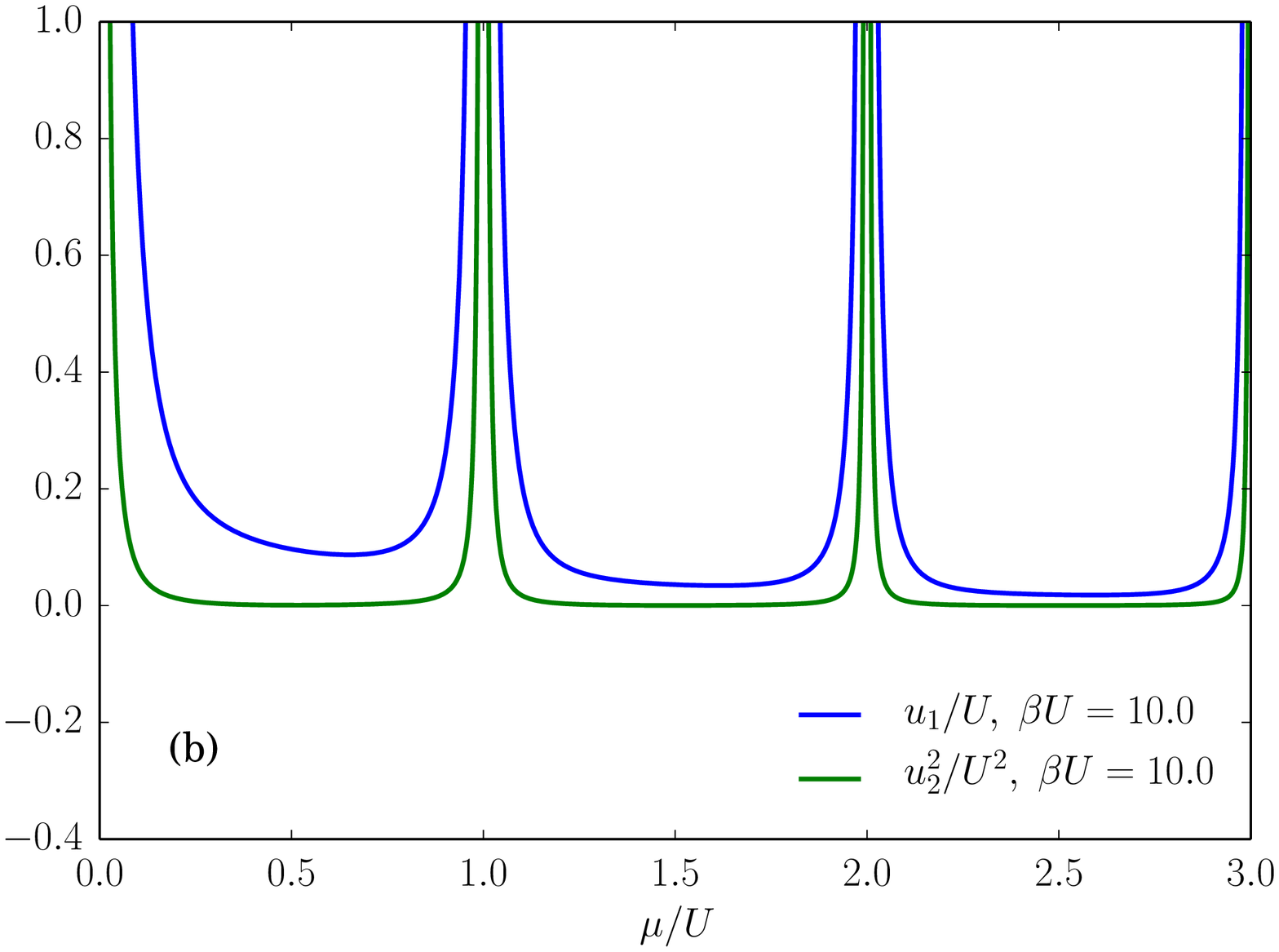}
\par\end{centering}

\caption{(Color online) (a) Plot of $u_{1}$ and $u_{2}^{2}$ as a function of
$\mu/U$ for inverse temperature $\beta U=5.0$; and (b) for $\beta U=10.0$.\label{fig:fig4}}
\end{figure}

\noindent Numerical evaluation of $u_{1}$ and $u_{2}^{2}$ for a homogeneous
system, shown in Fig.~\ref{fig:fig4} demonstrates that unless $\mu/U$
is close to an integer, the $u_{1}$ terms will dominate the $u_{2}^{2}$
terms. Moreover, for low temperatures, $u_{2}^{2}$ becomes negligible and
goes to zero as $\beta\to\infty$. Hence, to simplify the equations
of motion, we further assume that the temperature is sufficiently
low such that $u_{2}^{2}$ can be safely ignored. The end result is that
the equations of motion contain single time-integrals only.

\subsection{Keldysh structure of $\phi$, $G^{c}$, $\Omega$, $\Sigma$\label{sub:keldysh structure - 1}}

Before presenting numerical results, it is worth discussing the explicit
Keldysh structure of the mean field $\phi$, full propagator $G^{c}$,
and their respective interaction terms $\Sigma$ and $\Omega$. Starting
with the mean field $\phi$, we have

\begin{eqnarray}
\left[\phi\right] & = & \left(\begin{array}{c}
0\\
\sqrt{2}\phi_{\vec{r}_{1}}^{a_{1}}\left(s_{1}\right)\\
\phi_{\vec{r}_{1}}^{a_{1}}\left(s^{\prime}=0\right)
\end{array}\right),\label{eq:keldysh struct of phi - 1}
\end{eqnarray}

\noindent where $\phi_{\vec{r}_{1}}^{a_{1}}\left(s_{1}\right)$ is
the superfluid order parameter

\begin{eqnarray}
\phi_{\vec{r}_{1}}^{a_{1}}\left(s_{1}\right) & = & \left\langle \hat{a}_{\vec{r}_{1}}^{a_{1}}\left(t_{i}+s_{1}\right)\right\rangle _{\hat{\rho}_{i}}.\label{eq:superfluid order parameter defined - 1}
\end{eqnarray}

\noindent Note that $\phi_{\vec{r}_{1}}^{2}\left(s_{1}\right)=\left[\phi_{\vec{r}_{1}}\left(s_{1}\right)\right]^{*}$.
Then, following Ref.~\citep{Stefanucci}, we can express $G^{c}$ as follows

\begin{eqnarray}
\left[G^{c}\right] & = & \left(\begin{array}{ccc}
0 & G_{\vec{r}_{1}\vec{r}_{2}}^{a_{1}a_{2},\left(A\right)}\left(s_{1},s_{2}\right) & 0\\
G_{\vec{r}_{1}\vec{r}_{2}}^{a_{1}a_{2},\left(R\right)}\left(s_{1},s_{2}\right) & G_{\vec{r}_{1}\vec{r}_{2}}^{a_{1}a_{2},\left(K\right)}\left(s_{1},s_{2}\right) & \sqrt{2}G_{\vec{r}_{1}\vec{r}_{2}}^{a_{1}a_{2},\left(\rceil\right)}\left(s_{1},s_{2}\right)\\
0 & \sqrt{2}G_{\vec{r}_{1}\vec{r}_{2}}^{a_{1}a_{2},\left(\lceil\right)}\left(s_{1},s_{2}\right) & iG_{\vec{r}_{1}\vec{r}_{2}}^{a_{1}a_{2},\left(M\right)}\left(s_{1},s_{2}\right)
\end{array}\right),\label{eq:keldysh struct of G - 1}
\end{eqnarray}

\noindent with

\begin{eqnarray}
G_{\vec{r}_{1}\vec{r}_{2}}^{a_{1}a_{2},\left(R\right)}\left(s_{1},s_{2}\right) & \equiv & -i\Theta\left(s_{1}-s_{2}\right)\left\langle \hat{a}_{\vec{r}_{1}}^{a_{1}}\left(t_{i}+s_{1}\right)\hat{a}_{\vec{r}_{2}}^{a_{2}}\left(t_{i}+s_{2}\right)-\hat{a}_{\vec{r}_{2}}^{a_{2}}\left(t_{i}+s_{2}\right)\hat{a}_{\vec{r}_{1}}^{a_{1}}\left(t_{i}+s_{1}\right)\right\rangle _{\hat{\rho}_{i}}^{c},\label{eq:G^(R) defined - 1}\\
G_{\vec{r}_{1}\vec{r}_{2}}^{a_{1}a_{2},\left(A\right)}\left(s_{1},s_{2}\right) & \equiv & i\Theta\left(s_{2}-s_{1}\right)\left\langle \hat{a}_{\vec{r}_{1}}^{a_{1}}\left(t_{i}+s_{1}\right)\hat{a}_{\vec{r}_{2}}^{a_{2}}\left(t_{i}+s_{2}\right)-\hat{a}_{\vec{r}_{2}}^{a_{2}}\left(t_{i}+s_{2}\right)\hat{a}_{\vec{r}_{1}}^{a_{1}}\left(t_{i}+s_{1}\right)\right\rangle _{\hat{\rho}_{i}}^{c},\label{eq:G^(A) defined - 1}\\
G_{\vec{r}_{1}\vec{r}_{2}}^{a_{1}a_{2},\left(K\right)}\left(s_{1},s_{2}\right) & \equiv & -i\left\langle \hat{a}_{\vec{r}_{1}}^{a_{1}}\left(t_{i}+s_{1}\right)\hat{a}_{\vec{r}_{2}}^{a_{2}}\left(t_{i}+s_{2}\right)+\hat{a}_{\vec{r}_{2}}^{a_{2}}\left(t_{i}+s_{2}\right)\hat{a}_{\vec{r}_{1}}^{a_{1}}\left(t_{i}+s_{1}\right)\right\rangle _{\hat{\rho}_{i}}^{c},\label{eq:G^(K) defined - 1}\\
G_{\vec{r}_{1}\vec{r}_{2}}^{a_{1}a_{2},\left(\lceil\right)}\left(s_{1},s_{2}\right) & \equiv & -i\left\langle \hat{a}_{\vec{r}_{1}}^{a_{1}}\left(t_{i}-is_{1}\right)\hat{a}_{\vec{r}_{2}}^{a_{2}}\left(t_{i}+s_{2}\right)\right\rangle _{\hat{\rho}_{i}}^{c},\label{eq:G^(left) defined - 1}\\
G_{\vec{r}_{1}\vec{r}_{2}}^{a_{1}a_{2},\left(\rceil\right)}\left(s_{1},s_{2}\right) & \equiv & -i\left\langle \hat{a}_{\vec{r}_{2}}^{a_{2}}\left(t_{i}-is_{2}\right)\hat{a}_{\vec{r}_{1}}^{a_{1}}\left(t_{i}+s_{1}\right)\right\rangle _{\hat{\rho}_{i}}^{c},\label{eq:G^(right) defined - 1}\\
G_{\vec{r}_{1}\vec{r}_{2}}^{a_{1}a_{2},\left(M\right)}\left(s_{1},s_{2}\right) & \equiv & -\left(\Theta\left(s_{1}-s_{2}\right)\left\langle \hat{a}_{\vec{r}_{1}}^{a_{1}}\left(t_{i}-is_{1}\right)\hat{a}_{\vec{r}_{2}}^{a_{2}}\left(t_{i}-is_{2}\right)\right\rangle _{\hat{\rho}_{i}}^{c}\right.\nonumber \\
 &  & \phantom{-}\left.\quad+\Theta\left(s_{2}-s_{1}\right)\left\langle \hat{a}_{\vec{r}_{2}}^{a_{2}}\left(t_{i}-is_{2}\right)\hat{a}_{\vec{r}_{1}}^{a_{1}}\left(t_{i}-is_{1}\right)\right\rangle _{\hat{\rho}_{i}}^{c}\right),\label{eq:G^(M) defined - 1}
\end{eqnarray}

\noindent where $G^{\left(R\right)}$ and $G^{\left(A\right)}$ are
the retarded and advanced Green's functions respectively, $G^{\left(K\right)}$
is the Keldysh or Kinetic Green's function, $G^{\left(\lceil\right)}$
and $G^{\left(\rceil\right)}$ are the left and right Green's functions respectively,
and $G^{\left(M\right)}$ is the Matsubara Green's function. 

Next we have $\Omega$, which takes on the following Keldysh structure

\begin{eqnarray}
\left[\Omega\right] & = & \left(\begin{array}{c}
0\\
\sqrt{2}\Omega_{\vec{r}_{1}}^{a_{1}}\left(s_{1}\right)\\
\Omega_{\vec{r}_{1}}^{a_{1}}\left(s^{\prime}=0\right)
\end{array}\right),\label{eq:keldysh struct of Ohm - 1}
\end{eqnarray}

\noindent where to first order in $u_{1}$ we have

\begin{eqnarray}
\Omega_{\vec{r}_{1}}^{a_{1}}\left(s_{1}\right) & \approx & -\sum_{\vec{r}_{2}}2J_{\vec{r}_{1}\vec{r}_{2}}\left(t_{i}+s_{1}\right)\phi_{\vec{r}_{2}}^{a_{1}}\left(s_{1}\right)+u_{1}\left|\phi_{\vec{r}_{1}}\left(s_{1}\right)\right|^{2}\phi_{\vec{r}_{1}}^{a_{1}}\left(s_{1}\right)\nonumber \\
 &  & \phantom{-\sum_{\vec{r}_{2}}}+\frac{u_{1}}{2}\sigma^{a_{1}a_{2}a_{3}a_{4}}\phi_{\vec{r}_{1}}^{\overline{a_{2}}}\left(s_{1}\right)\left\{ iG_{\vec{r}_{1}\vec{r}_{1}}^{\overline{a_{3}}\overline{a_{4}},\left(K\right)}\left(s_{1},s_{1}\right)-i\mathcal{G}^{\overline{a_{3}}\overline{a_{4}},\left(K\right)}\left(s^{\prime}=0\right)\right\} .\label{eq:Ohm_c - 1}
\end{eqnarray}

\noindent The self energy $\Sigma$ is similar in structure to $G$
where we have

\begin{eqnarray}
\left[\Sigma\right] & = & \left(\begin{array}{ccc}
0 & \Sigma_{\vec{r}_{1}\vec{r}_{2}}^{a_{1}a_{2},\left(A\right)}\left(s_{1},s_{2}\right) & 0\\
\Sigma_{\vec{r}_{1}\vec{r}_{2}}^{a_{1}a_{2},\left(R\right)}\left(s_{1},s_{2}\right) & \Sigma_{\vec{r}_{1}\vec{r}_{2}}^{a_{1}a_{2},\left(K\right)}\left(s_{1},s_{2}\right) & \sqrt{2}\Sigma_{\vec{r}_{1}\vec{r}_{2}}^{a_{1}a_{2},\left(\rceil\right)}\left(s_{1},s_{2}\right)\\
0 & \sqrt{2}\Sigma_{\vec{r}_{1}\vec{r}_{2}}^{a_{1}a_{2},\left(\lceil\right)}\left(s_{1},s_{2}\right) & i\Sigma_{\vec{r}_{1}\vec{r}_{2}}^{a_{1}a_{2},\left(M\right)}\left(s_{1},s_{2}\right)
\end{array}\right),\label{eq:keldysh struct of self energy - 1}
\end{eqnarray}

\noindent where $\Sigma^{\left(R\right)}$ and $\Sigma^{\left(A\right)}$
have the same properties of causality as $G^{\left(R\right)}$ and
$G^{\left(A\right)}$ respectively. To first order in $u_{1}$, we
have

\begin{eqnarray}
 &  & \Sigma_{\vec{r}_{1}\vec{r}_{2}}^{a_{1}a_{2},\left(R,A\right)}\left(s_{1},s_{2}\right)\nonumber \\
 &  & \quad\approx\delta\left(s_{1}-s_{2}\right)\left(-2\sigma_{1}^{a_{1}a_{2}}J_{\vec{r}_{1}\vec{r}_{2}}\left(t_{i}+s_{1}\right)+u_{1}\delta_{\vec{r}_{1}\vec{r}_{2}}\sigma^{a_{1}a_{2}a_{3}a_{4}}\phi_{\vec{r}_{1}}^{\overline{a_{3}}}\left(s_{1}\right)\phi_{\vec{r}_{1}}^{\overline{a_{4}}}\left(s_{1}\right)\right.\nonumber \\
 &  & \left.\quad\phantom{\delta\left(s_{1}-s_{2}\right)}\quad+\frac{u_{1}}{2}\delta_{\vec{r}_{1}\vec{r}_{2}}\sigma^{a_{1}a_{2}a_{3}a_{4}}\left\{ iG_{\vec{r}_{1}\vec{r}_{1}}^{\overline{a_{3}}\overline{a_{4}},\left(K\right)}\left(s_{1},s_{1}\right)-i\mathcal{G}^{\overline{a_{3}}\overline{a_{4}},\left(K\right)}\left(s^{\prime}=0\right)\right\} \right),\label{eq:Sigma^(R,A) - 1}\\
 &  & \Sigma_{\vec{r}_{1}\vec{r}_{2}}^{a_{1}a_{2},\left(M\right)}\left(s_{1},s_{2}\right)\nonumber \\
 &  & \quad\approx\delta\left(s_{1}-s_{2}\right)\left(-2\sigma_{1}^{a_{1}a_{2}}J_{\vec{r}_{1}\vec{r}_{2}}\left(t_{i}\right)+u_{1}\delta_{\vec{r}_{1}\vec{r}_{2}}\sigma^{a_{1}a_{2}a_{3}a_{4}}\phi_{\vec{r}_{1}}^{\overline{a_{3}}}\left(s^{\prime}=0\right)\phi_{\vec{r}_{1}}^{\overline{a_{4}}}\left(s^{\prime}=0\right)\right.\\
 &  & \left.\phantom{\quad\approx\delta\left(s_{1}-s_{2}\right)}\quad+\frac{u_{1}}{2}\delta_{\vec{r}_{1}\vec{r}_{2}}\sigma^{a_{1}a_{2}a_{3}a_{4}}\times\left\{ iG_{\vec{r}_{1}\vec{r}_{1}}^{\overline{a_{3}}\overline{a_{4}},\left(K\right)}\left(s^{\prime}=0,s^{\prime}=0\right)-i\mathcal{G}^{\overline{a_{3}}\overline{a_{4}},\left(K\right)}\left(s^{\prime}=0\right)\right\} \right),\label{eq:Sigma^(M) - 1}
\end{eqnarray}

\noindent and

\begin{eqnarray}
\Sigma_{\vec{r}_{1}\vec{r}_{2}}^{a_{1}a_{2},\left(K,\lceil,\rceil\right)}\left(s_{1},s_{2}\right) & \approx & 0.\label{eq:Sigmas that vanish - 1}
\end{eqnarray}

Lastly, we rewrite the equations of motion Eqs.~(\ref{eq:phi eqn of motion - 2})
and (\ref{eq:G eqn of motion - 3}) explicitly in terms of the various
Keldysh components (i.e. $R,A,K,\lceil,\rceil,M$)

\begin{eqnarray}
\phi_{\vec{r}_{1}}^{a_{1}}\left(s_{1}\right) & = & \sum_{\vec{r}_{2}}\int_{0}^{\infty}ds_{2}\,\mathcal{G}_{\vec{r}_{1}\vec{r}_{2}}^{a_{1}a_{2},\left(R\right)}\left(s_{1},s_{2}\right)\Omega_{\vec{r}_{2}}^{\overline{a_{2}}}\left(s_{2}\right)-i\sum_{\vec{r}_{2}}\left\{ \int_{0}^{\beta}ds_{2} \, 
\mathcal{G}_{\vec{r}_{1}\vec{r}_{2}}^{a_{1}a_{2},\left(\rceil\right)}\left(s_{1},s_{2}\right)\right\} \Omega_{\vec{r}_{2}}^{\overline{a_{2}}}\left(s^{\prime}=0\right),\label{eq:phi eqn of motion - 3}
\end{eqnarray}

\begin{eqnarray}
G_{\vec{r}_{1}\vec{r}_{2}}^{a_{1}a_{2},\left(R\right)}\left(s_{1},s_{2}\right) & = & \mathcal{G}_{\vec{r}_{1}\vec{r}_{2}}^{a_{1}a_{2},\left(R\right)}\left(s_{1},s_{2}\right)\nonumber \\
 &  & +\sum_{\vec{r}_{3}\vec{r}_{4}}\int_{0}^{\infty}\int_{0}^{\infty}ds_{3}ds_{4}\, 
	\mathcal{G}_{\vec{r}_{1}\vec{r}_{3}}^{a_{1}a_{3},\left(R\right)}\left(s_{1},s_{3}\right)\Sigma_{\vec{r}_{3}\vec{r}_{4}}^{\overline{a_{3}}\overline{a_{4}},\left(R\right)}\left(s_{3},s_{4}\right)G_{\vec{r}_{4}\vec{r}_{2}}^{a_{4}a_{2},\left(R\right)}\left(s_{4},s_{2}\right),\label{eq:G^(R) eqn of motion - 1}
\end{eqnarray}

\begin{eqnarray}
G_{\vec{r}_{1}\vec{r}_{2}}^{a_{1}a_{2},\left(A\right)}\left(s_{1},s_{2}\right) & = & \mathcal{G}_{\vec{r}_{1}\vec{r}_{2}}^{a_{1}a_{2},\left(A\right)}\left(s_{1},s_{2}\right)\nonumber \\
 &  & +\sum_{\vec{r}_{3}\vec{r}_{4}}\int_{0}^{\infty}\int_{0}^{\infty}ds_{3}ds_{4}\, 
	\mathcal{G}_{\vec{r}_{1}\vec{r}_{3}}^{a_{1}a_{3},\left(A\right)}\left(s_{1},s_{3}\right)\Sigma_{\vec{r}_{3}\vec{r}_{4}}^{\overline{a_{3}}\overline{a_{4}},\left(A\right)}\left(s_{3},s_{4}\right)G_{\vec{r}_{4}\vec{r}_{2}}^{a_{4}a_{2},\left(A\right)}\left(s_{4},s_{2}\right),\label{eq:G^(A) eqn of motion - 1}
\end{eqnarray}

\begin{eqnarray}
G_{\vec{r}_{1}\vec{r}_{2}}^{a_{1}a_{2},\left(K\right)}\left(s_{1},s_{2}\right) & = & \mathcal{G}_{\vec{r}_{1}\vec{r}_{2}}^{a_{1}a_{2},\left(K\right)}\left(s_{1},s_{2}\right)\nonumber \\
 &  & +\sum_{\vec{r}_{3}\vec{r}_{4}}\int_{0}^{\infty}\int_{0}^{\infty}ds_{3}ds_{4}\, 
	\mathcal{G}_{\vec{r}_{1}\vec{r}_{3}}^{a_{1}a_{3},\left(R\right)}\left(s_{1},s_{3}\right)\Sigma_{\vec{r}_{3}\vec{r}_{4}}^{\overline{a_{3}}\overline{a_{4}},\left(R\right)}\left(s_{3},s_{4}\right)G_{\vec{r}_{4}\vec{r}_{2}}^{a_{4}a_{2},\left(K\right)}\left(s_{4},s_{2}\right)\nonumber \\
 &  & +\sum_{\vec{r}_{3}\vec{r}_{4}}\int_{0}^{\infty}\int_{0}^{\infty}ds_{3}ds_{4}\,
	\mathcal{G}_{\vec{r}_{1}\vec{r}_{3}}^{a_{1}a_{3},\left(K\right)}\left(s_{1},s_{3}\right)\Sigma_{\vec{r}_{3}\vec{r}_{4}}^{\overline{a_{3}}\overline{a_{4}},\left(A\right)}\left(s_{3},s_{4}\right)G_{\vec{r}_{4}\vec{r}_{2}}^{a_{4}a_{2},\left(A\right)}\left(s_{4},s_{2}\right)\nonumber \\
 &  & -2i\sum_{\vec{r}_{3}\vec{r}_{4}}\int_{0}^{\beta}\int_{0}^{\beta}ds_{3}ds_{4}\, 
	\mathcal{G}_{\vec{r}_{1}\vec{r}_{3}}^{a_{1}a_{3},\left(\rceil\right)}\left(s_{1},s_{3}\right)\Sigma_{\vec{r}_{3}\vec{r}_{4}}^{\overline{a_{3}}\overline{a_{4}},\left(M\right)}\left(s_{3},s_{4}\right)G_{\vec{r}_{4}\vec{r}_{2}}^{a_{4}a_{2},\left(\lceil\right)}\left(s_{4},s_{2}\right),\label{eq:G^(K) eqn of motion - 1}
\end{eqnarray}

\begin{eqnarray}
G_{\vec{r}_{1}\vec{r}_{2}}^{a_{1}a_{2},\left(\lceil\right)}\left(s_{1},s_{2}\right) & = & \mathcal{G}_{\vec{r}_{1}\vec{r}_{2}}^{a_{1}a_{2},\left(\lceil\right)}\left(s_{1},s_{2}\right)\nonumber \\
 &  & +\sum_{\vec{r}_{3}\vec{r}_{4}}\int_{0}^{\beta}\int_{0}^{\beta}ds_{3}ds_{4}\, 
	\mathcal{G}_{\vec{r}_{1}\vec{r}_{3}}^{a_{1}a_{3},\left(M\right)}\left(s_{1},s_{3}\right)\Sigma_{\vec{r}_{3}\vec{r}_{4}}^{\overline{a_{3}}\overline{a_{4}},\left(M\right)}\left(s_{3},s_{4}\right)G_{\vec{r}_{4}\vec{r}_{2}}^{a_{4}a_{2},\left(\lceil\right)}\left(s_{4},s_{2}\right)\nonumber \\
 &  & +\sum_{\vec{r}_{3}\vec{r}_{4}}\int_{0}^{\infty}\int_{0}^{\infty}ds_{3}ds_{4}\, 
	\mathcal{G}_{\vec{r}_{1}\vec{r}_{3}}^{a_{1}a_{3},\left(\lceil\right)}\left(s_{1},s_{3}\right)\Sigma_{\vec{r}_{3}\vec{r}_{4}}^{\overline{a_{3}}\overline{a_{4}},\left(A\right)}\left(s_{3},s_{4}\right)G_{\vec{r}_{4}\vec{r}_{2}}^{a_{4}a_{2},\left(A\right)}\left(s_{4},s_{2}\right),\label{eq:G^(left) eqn of motion - 1}
\end{eqnarray}

\begin{eqnarray}
G_{\vec{r}_{1}\vec{r}_{2}}^{a_{1}a_{2},\left(M\right)}\left(s_{1},s_{2}\right) & = & \mathcal{G}_{\vec{r}_{1}\vec{r}_{2}}^{a_{1}a_{2},\left(M\right)}\left(s_{1},s_{2}\right)\nonumber \\
 &  & +\sum_{\vec{r}_{3}\vec{r}_{4}}\int_{0}^{\beta}\int_{0}^{\beta}ds_{3}ds_{4}\, 
	\mathcal{G}_{\vec{r}_{1}\vec{r}_{3}}^{a_{1}a_{3},\left(M\right)}\left(s_{1},s_{3}\right)\Sigma_{\vec{r}_{3}\vec{r}_{4}}^{\overline{a_{3}}\overline{a_{4}},\left(M\right)}\left(s_{3},s_{4}\right)G_{\vec{r}_{4}\vec{r}_{2}}^{a_{4}a_{2},\left(M\right)}\left(s_{4},s_{2}\right),\label{eq:G^(M) eqn of motion - 1}
\end{eqnarray}

\noindent where the various Keldysh components of $\mathcal{G}^{c}$ can
be found in \ref{app:keldysh components of G0 - 1}. Equations~(\ref{eq:phi eqn of motion - 3})--(\ref{eq:G^(M) eqn of motion - 1}),
along with Eqs.~(\ref{eq:Sigma^(R,A) - 1})--(\ref{eq:Sigmas that vanish - 1})
and Eq.~(\ref{eq:Ohm_c - 1}) together form one of the main results
of this paper. These can be readily used to study out of equilibrium
dynamics for strongly interacting systems. By considering only terms
up to first order in $u_{1}$, our approximation can be thought of
in some sense as a Hartree-Fock-Bogoliubov (HFB) approximation in
the strong-coupling regime. In future works we will study these equations
of motion for various nonequilibrium scenarios. In the remainder of
this paper however, we study the equilibrium solutions to the equations
of motion above, going beyond the work in Ref.~\citep{SenguptaDupuis}
in which only the equilibrium solutions at the
one-loop level in the imaginary-time formalism were studied.

\section{Equilibrium solution\label{sec:equilibrium solution - 1}}

In studying the equilibrium solution to the equations of motion derived
in the previous section we consider a homogeneous system at zero temperature.
As a result, it is easier to work in $\vec{k}$-space rather
than real space. In equilibrium, the mean field equation of
motion Eq.~(\ref{eq:phi eqn of motion - 3}) reduces to \citep{Stefanucci}

\begin{eqnarray}
\phi & = & \mathcal{G}^{12,\left(R\right)}\left(\omega^{\prime}=0\right)\Omega^{2}\left(s^{\prime}=0\right),\label{eq:phi equilibrium eqn of motion - 1}
\end{eqnarray}

\noindent where we used the fact that the superfluid order parameter
is constant in time, $\phi^{1}\left(s_{1}\right)=\phi$. Expressions
for $\mathcal{G}^{12,\left(R\right)}\left(\omega\right)$ and $\mathcal{G}^{12,\left(R\right)}\left(\omega^{\prime}=0\right)$
are given by Eqs.~(\ref{eq:G0^(R)(w) - 1}) and (\ref{eq:static limit of G0^(R) - 1})
respectively. We also have that in equilibrium all the various real-time
Green's functions may be expressed in terms of the spectral function
$G^{\left(\rho\right)}$ 

\begin{eqnarray}
G_{\vec{k}}^{12,\left(\rho\right)}\left(\omega\right) & = & -2\ \text{Im}\left[G_{\vec{k}}^{12,\left(R\right)}\left(\omega\right)\right].\label{eq:G^(rho) from G(R) - 1}
\end{eqnarray}

\noindent One can calculate $G^{\left(K\right)}$ from $G^{\left(\rho\right)}$
via the fluctuation dissipation theorem (FDT) \citep{RammerText,Stefanucci},
which at zero temperature is

\begin{eqnarray}
G_{\vec{k}}^{12,\left(K\right)}\left(\omega\right) & = & -iG_{\vec{k}}^{12,\left(\rho\right)}\left(\omega\right)\text{sgn}\left(\omega\right),\label{eq:FDT - 1}
\end{eqnarray}

\noindent hence one need only focus on the $G^{\left(R\right)}$ equation
of motion directly. In equilibrium, it is easier to work in frequency
space, hence we may rewrite the $G^{\left(R\right)}$ equation of
motion as \citep{Stefanucci}

\begin{eqnarray}
G_{\vec{k}}^{a_{1}a_{2},\left(R\right)}\left(\omega\right) & = & \mathcal{G}^{a_{1}a_{2},\left(R\right)}\left(\omega\right)+\sum_{a_{3}a_{4}}\mathcal{G}^{a_{1}a_{3},\left(R\right)}\left(\omega\right)\Sigma_{\vec{k}}^{\overline{a_{3}}\overline{a_{4}},\left(R\right)}G_{\vec{k}}^{a_{4}a_{2},\left(R\right)}\left(\omega\right),\label{eq:G^(R) equilibrium eqn of motion - 1}
\end{eqnarray}

\noindent where

\begin{equation}
\Sigma_{\vec{k}}^{12,\left(R\right)}=\Sigma_{\vec{k}}^{21,\left(R\right)}=\epsilon_{\vec{k}}+2u_{1}\left\{ \left|\phi\right|^{2}+\left(n-n_{0}\right)\right\} ,\label{eq:Sigma_12^(R) - 1}
\end{equation}

\begin{eqnarray}
\Sigma_{\vec{k}}^{11,\left(R\right)} & = & \frac{1}{2}u_{1}\left\{ 2\left(\phi^{1}\right)^{2}+iG_{\vec{r}^{\prime}=\mathbf{0}}^{11\left(K\right)}\left(s^{\prime}=0\right)\right\} ,\label{eq:Sigma_11^(R) - 1}
\end{eqnarray}

\begin{eqnarray}
\Sigma_{\vec{k}}^{22,\left(R\right)} & = & \frac{1}{2}u_{1}\left\{ 2\left(\phi^{2}\right)^{2}+iG_{\vec{r}^{\prime}=\mathbf{0}}^{22\left(K\right)}\left(s^{\prime}=0\right)\right\} ,\label{eq:Sigma_22^(R) - 1}
\end{eqnarray}

\begin{eqnarray}
\epsilon_{\vec{k}} & = & -2J\sum_{i=0}^{d}\cos\left(k_{i}a\right),\label{eq:epsilon_k - 1}
\end{eqnarray}

\noindent and $n$ and $n_{0}$ are the average particle densities
for $J\neq0$ and $J=0$ respectively. Note that 

\noindent 
\begin{eqnarray}
n_{0} & = & \left\lceil \mu/U\right\rceil .\label{eq:n0 - 1}
\end{eqnarray}
With a bit of algebra, one can show that

\begin{eqnarray}
G_{\vec{k}}^{12,\left(R\right)}\left(\omega\right) & = & \frac{\left[\left\{\mathcal{G}_{\vec{k}}^{21,\left(R\right)}\left(\omega\right)\right\}^{-1}-\Sigma_{\vec{k}}^{21,\left(R\right)}\right]}{\left[\left\{ \mathcal{G}^{21,\left(R\right)}\left(\omega\right)\right\} ^{-1}-\Sigma_{\vec{k}}^{21,\left(R\right)}\right]\left[\left\{ \mathcal{G}^{12,\left(R\right)}\left(\omega\right)\right\} ^{-1}-\Sigma_{\vec{k}}^{12,\left(R\right)}\right]-\left|\Sigma_{\vec{k}}^{22,\left(R\right)}\right|^{2}},\label{eq:G_12^(R) equilibrium eqn of motion - 1}\\
G_{\vec{k}}^{22,\left(R\right)}\left(\omega\right) & = & \frac{\Sigma_{\vec{k}}^{22,\left(R\right)}}{\left[\left\{ \mathcal{G}^{21,\left(R,\right)}\left(\omega\right)\right\} ^{-1}-\Sigma_{\vec{k}}^{21,\left(R\right)}\right]\left[\left\{ \mathcal{G}^{12,\left(R\right)}\left(\omega\right)\right\} ^{-1}-\Sigma_{\vec{k}}^{12,\left(R\right)}\right]-\left|\Sigma_{\vec{k}}^{22,\left(R\right)}\right|^{2}}.\label{eq:G_22^(R) equilibrium eqn of motion - 1}
\end{eqnarray}

\noindent From here, the next step is to simplify $G_{\vec{k}}^{12,\left(R\right)}\left(\omega\right)$
by starting from Eq.~(\ref{eq:G_12^(R) equilibrium eqn of motion - 1})
and then applying Eq.~(\ref{eq:G^(rho) from G(R) - 1}) to obtain an
expression for $G_{\vec{k}}^{12,\left(\rho\right)}\left(\omega\right)$.
One can then express $G_{\vec{k}}^{12,\left(\rho\right)}\left(\omega\right)$
in the Lehmann representation

\begin{eqnarray}
G_{\vec{k}}^{12,\left(\rho\right)}\left(\omega\right) & = & 2\pi\sum_{s}\left\{ z_{\vec{k}}^{\left(s,+\right)}\delta\left(\omega-\Delta E_{\vec{k}}^{\left(s,+\right)}\right)-z_{\vec{k}}^{\left(s,-\right)}\delta\left(\omega+\Delta E_{\vec{k}}^{\left(s,-\right)}\right)\right\} ,\label{eq:general lehmann rep - 1}
\end{eqnarray}

\noindent where $s$ is the branch number, $\Delta E_{\vec{k}}^{\left(s,+\right)}$
and $\Delta E_{\vec{k}}^{\left(s,-\right)}$ are the particle and
hole excitation energies respectively, and $z_{\vec{k}}^{\left(s,\pm\right)}$
are the corresponding spectral weights. Once written in this form,
we can simply read off the expressions for the desired quantities. We do
this in the following by considering the Mott insulator and superfluid
cases separately.

\subsection{Mott insulator phase\label{sub:MI phase - 1}}

In the Mott insulator phase, $\phi=\left|\Sigma_{\vec{k}}^{22,\left(R,A\right)}\right|^{2}=0$
and Eq.~(\ref{eq:G_12^(R) equilibrium eqn of motion - 1}) reduces
to 

\begin{eqnarray}
G_{\vec{k}}^{12,\left(R\right)}\left(\omega\right) & = & \frac{1}{\left[\left\{ \mathcal{G}^{12,\left(R\right)}\left(\omega\right)\right\} ^{-1}-\Sigma_{\vec{k}}^{12,\left(R\right)}\left(\omega\right)\right]}.\label{eq:MI G_12^(R) equilibrium eqn of motion - 1}
\end{eqnarray}

\noindent One can rewrite Eq.~(\ref{eq:MI G_12^(R) equilibrium eqn of motion - 1})
as

\begin{eqnarray}
G_{\vec{k}}^{12,\left(R\right)}\left(\omega\right) & = & z_{\text{MI},\vec{k}}^{\left(+\right)}\frac{1}{\left(\omega-\Delta E_{\text{MI},\vec{k}}^{\left(+\right)}\right)+i0^{+}}-z_{\text{MI},\vec{k}}^{\left(-\right)}\frac{1}{\left(\omega+\Delta E_{\text{MI},\vec{k}}^{\left(-\right)}\right)+i0^{+}},\label{eq:MI G_12^(R) equilibrium eqn of motion - 2}
\end{eqnarray}

\noindent where

\begin{eqnarray}
\Delta E_{\text{MI},\vec{k}}^{\left(\pm\right)} & = & \frac{\mp B_{\vec{k}}+\sqrt{\left(B_{\vec{k}}\right)^{2}-4C_{\vec{k}}}}{2},\label{eq:E_MI - 1}\\
B_{\vec{k}} & = & -\left\{ \Delta\mathcal{E}^{\left(+\right)}-\Delta\mathcal{E}^{\left(-\right)}\right\} -\Sigma_{\vec{k}}^{12,\left(R\right)},\label{eq:B_k - 1}\\
C_{\vec{k}} & = & -\left(U+\mu\right)\left\{ \Sigma_{\vec{k}}^{12,\left(R\right)}-\left\{ \mathcal{G}^{12,\left(R\right)}\left(\omega^{\prime}=0\right)\right\} ^{-1}\right\} ,\label{eq:C_k - 1}
\end{eqnarray}

\begin{eqnarray}
z_{\text{MI},\vec{k}}^{\left(\pm\right)} & = & \frac{\left(U+\mu\right)\pm\Delta E_{\text{MI},\vec{k}}^{\left(\pm\right)}}{\Delta E_{\text{MI},\vec{k}}^{\left(+\right)}+\Delta E_{\text{MI},\vec{k}}^{\left(-\right)}},\label{eq:z_MI - 1}
\end{eqnarray}

\noindent and $\Delta\mathcal{E}^{\left(\pm\right)}$ are the excitation
energies in the atomic limit (i.e $J=0$)

\begin{eqnarray}
\Delta\mathcal{E}^{\left(+\right)} & \equiv & \mathcal{E}_{n_{0}+1}-\mathcal{E}_{n_{0}},\label{eq:E_atomic^(+) - 1}\\
\Delta\mathcal{E}^{\left(-\right)} & \equiv & \mathcal{E}_{n_{0}-1}-\mathcal{E}_{n_{0}},\label{eq:E_atomic^(-) - 1}\\
\mathcal{E}_{n} & \equiv & \frac{U}{2}n\left(n-1\right)-n\mu.\label{eq:E_atomic - 1}
\end{eqnarray}

\noindent Using Eq.~(\ref{eq:G^(rho) from G(R) - 1}) along with the
Sokhotski-Plemelj theorem 

\begin{eqnarray}
\frac{1}{x+i0^{\pm}} & = & \mp i\pi\delta\left(x\right)+\mathcal{P}\left(\frac{1}{x}\right),\label{eq:Sokhotski-Plemelj theorem - 1}
\end{eqnarray}

\noindent we obtain for the spectral function

\begin{eqnarray}
G_{\vec{k}}^{12,\left(\rho\right)}\left(\omega\right) & = & 2\pi\left\{ z_{\text{MI},\vec{k}}^{\left(+\right)}\delta\left(\omega-\Delta E_{\text{MI},\vec{k}}^{\left(+\right)}\right)-z_{\text{MI},\vec{k}}^{\left(-\right)}\delta\left(\omega+\Delta E_{\text{MI},\vec{k}}^{\left(-\right)}\right)\right\} .\label{eq:MI G^(rho) equilibrium - 1}
\end{eqnarray}

\noindent By comparing Eq.~(\ref{eq:MI G^(rho) equilibrium - 1}) to Eq.~(\ref{eq:general lehmann rep - 1}), 
it is clear that $\Delta E_{\text{MI},\vec{k}}^{\left(\pm\right)}$
and $z_{\text{MI},\vec{k}}^{\left(\pm\right)}$ are the excitation
energies and spectral weights respectively.

\subsubsection{Calculating $n_{\vec{k}}$ and $n$ \label{sub:MI calculating n_k and n - 1}}

At the HFB level, one needs to calculate $\Delta E_{\text{MI},\vec{k}}^{\left(\pm\right)}$
and $z_{\text{MI},\vec{k}}^{\left(\pm\right)}$ in a self-consistent
way since there is no closed-form expression for the self energy $\Sigma_{\vec{k}}^{12,\left(R\right)}$.
This becomes evident when one notes that $\Sigma_{\vec{k}}^{12,\left(R\right)}$
depends on $n$, which in turn depends on $n_{\vec{k}}$ through

\begin{eqnarray}
n & = & \int_{1^{\text{st}}\text{B.Z.}}\frac{d\vec{k}}{\left(2\pi\right)^{d}}n_{\vec{k}},\label{eq:n from n_k - 1}
\end{eqnarray}

\noindent which in turn depends on $G_{\vec{k}}^{12,\left(K\right)}\left(s^{\prime}=0\right)$
through

\begin{eqnarray}
n_{\vec{k}} & = & \frac{1}{2}\left\{ iG_{\vec{k}}^{12,\left(K\right)}\left(s^{\prime}=0\right)-1\right\} ,\label{eq:MI n_k depends on G^(K) - 1}
\end{eqnarray}

\noindent in the Mott insulator phase. Using Eq.~(\ref{eq:FDT - 1})
we obtain for $G_{\vec{k}}^{12,\left(K\right)}\left(\omega\right)$

\begin{eqnarray}
G_{\vec{k}}^{12,\left(K\right)}\left(\omega\right) & = & -2\pi i\left\{ z_{\text{MI},\vec{k}}^{\left(+\right)}\delta\left(\omega-\Delta E_{\text{MI},\vec{k}}^{\left(+\right)}\right)+z_{\text{MI},\vec{k}}^{\left(-\right)}\delta\left(\omega+\Delta E_{\text{MI},\vec{k}}^{\left(-\right)}\right)\right\} ,\label{eq:MI G_12^(K) equilibrium eqn of motion - 1}
\end{eqnarray}

\noindent and therefore

\begin{eqnarray}
n_{\vec{k}} & = & \frac{1}{2}\left\{ i\int_{-\infty}^{\infty}\frac{d\omega}{2\pi}G_{\vec{k}}^{a_{1}a_{2},\left(K\right)}\left(\omega\right)-1\right\} \nonumber \\
 & = & \frac{1}{2}\left(z_{\text{MI},\vec{k}}^{\left(+\right)}+z_{\text{MI},\vec{k}}^{\left(-\right)}-1\right).\label{eq:MI n_k - 1}
\end{eqnarray}

\begin{figure}
\begin{centering}
\includegraphics[scale=0.4]{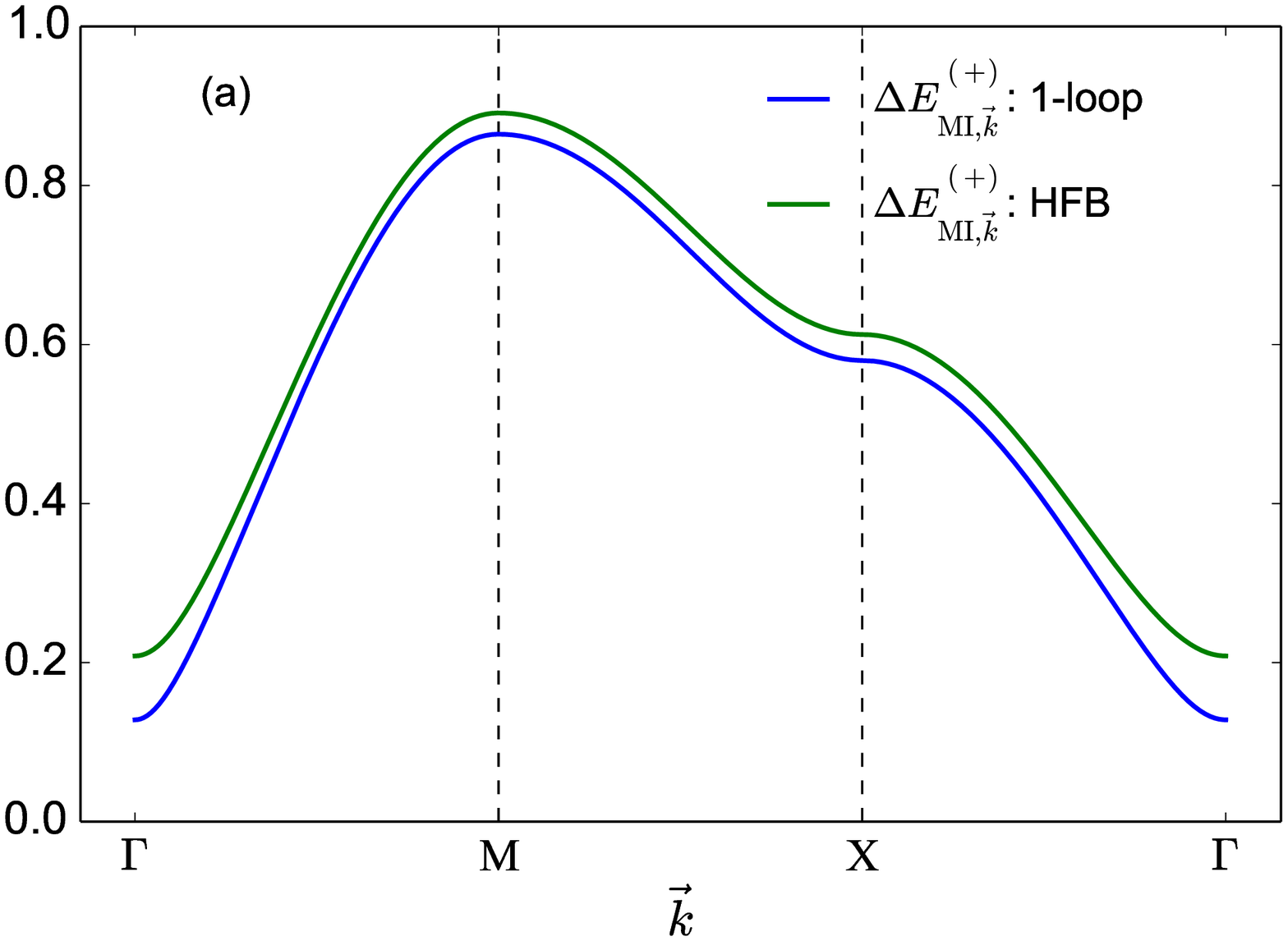}\includegraphics[scale=0.4]{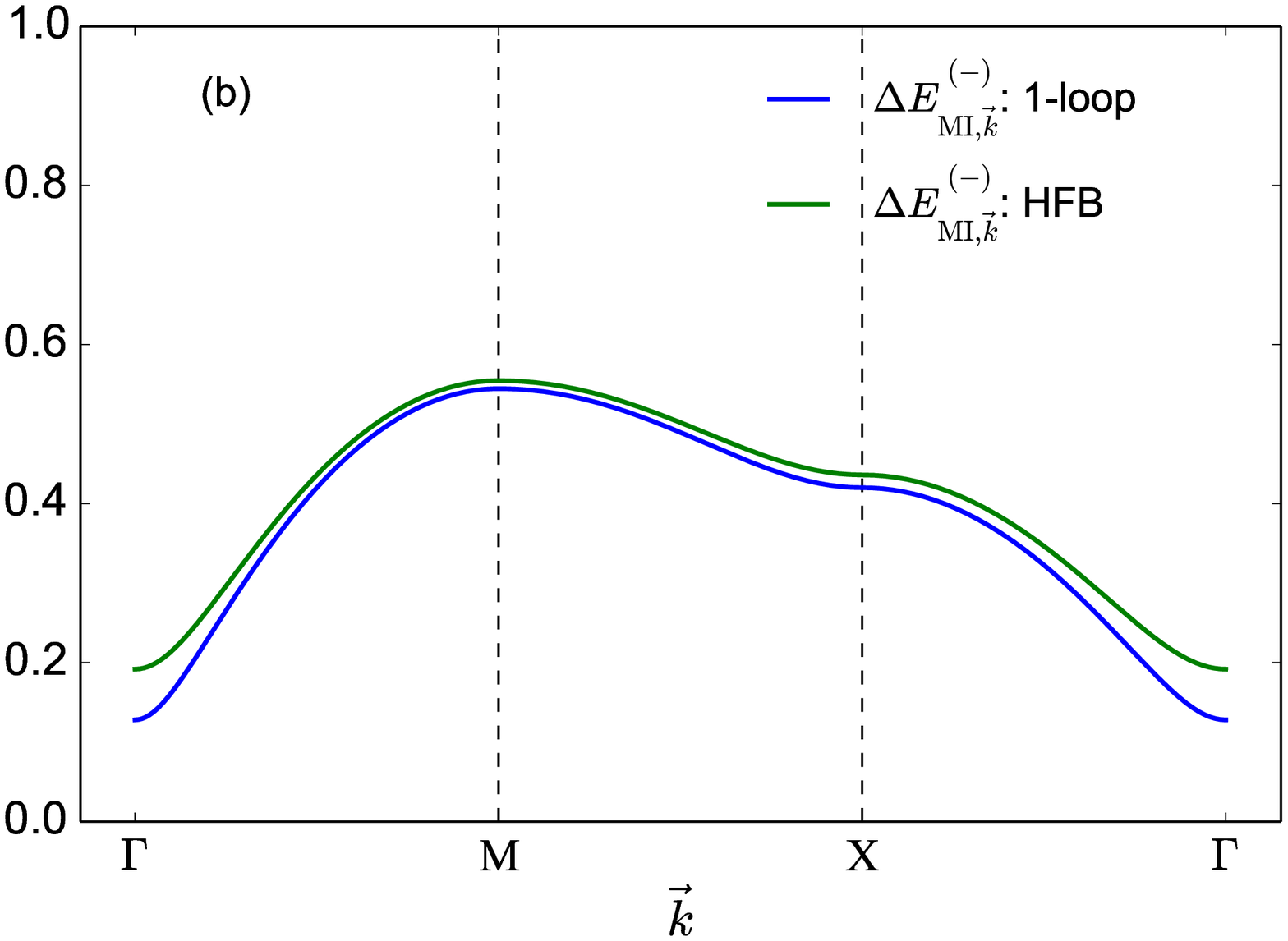}
\par\end{centering}

\begin{centering}
\includegraphics[scale=0.4]{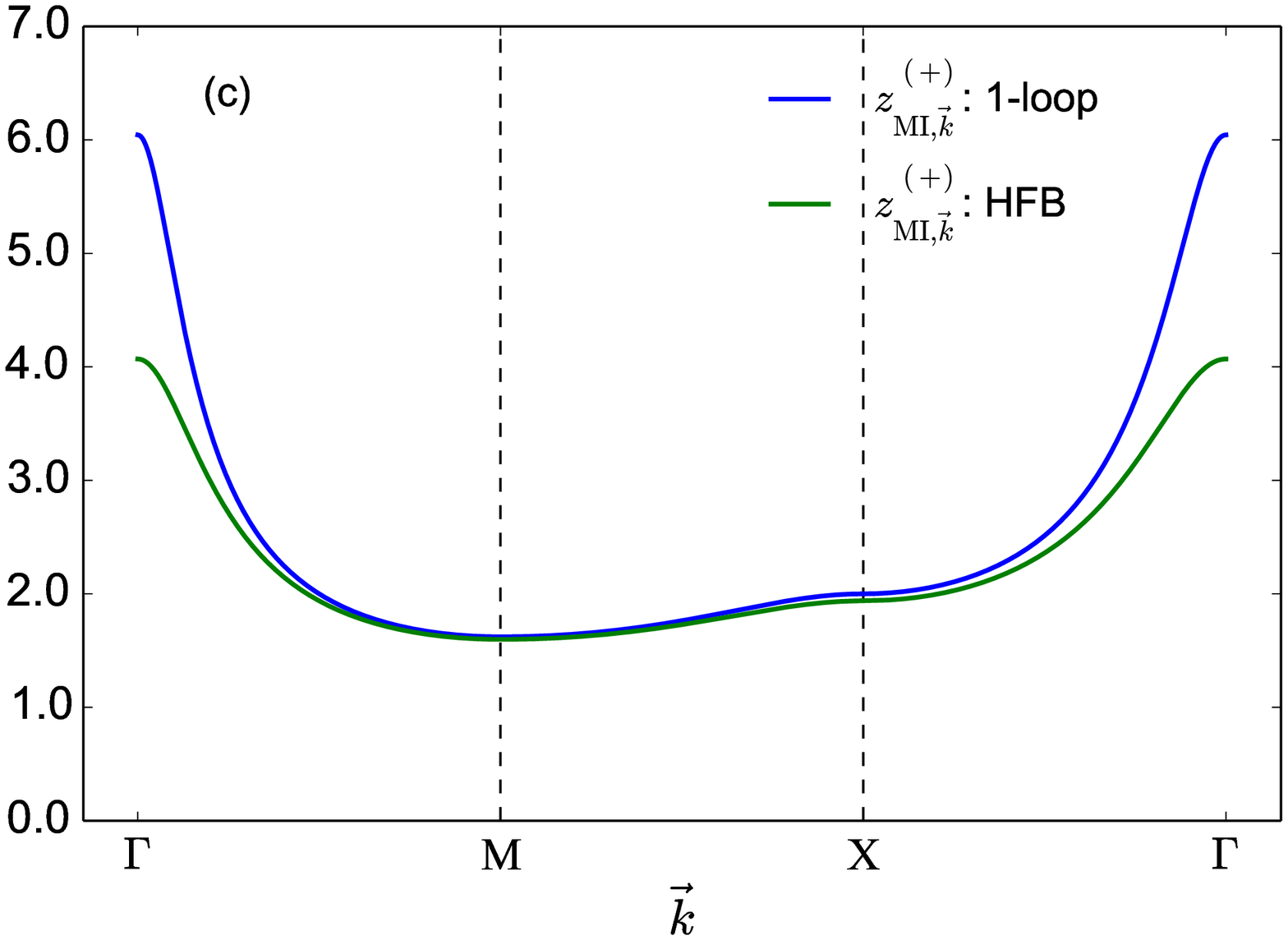}\includegraphics[scale=0.4]{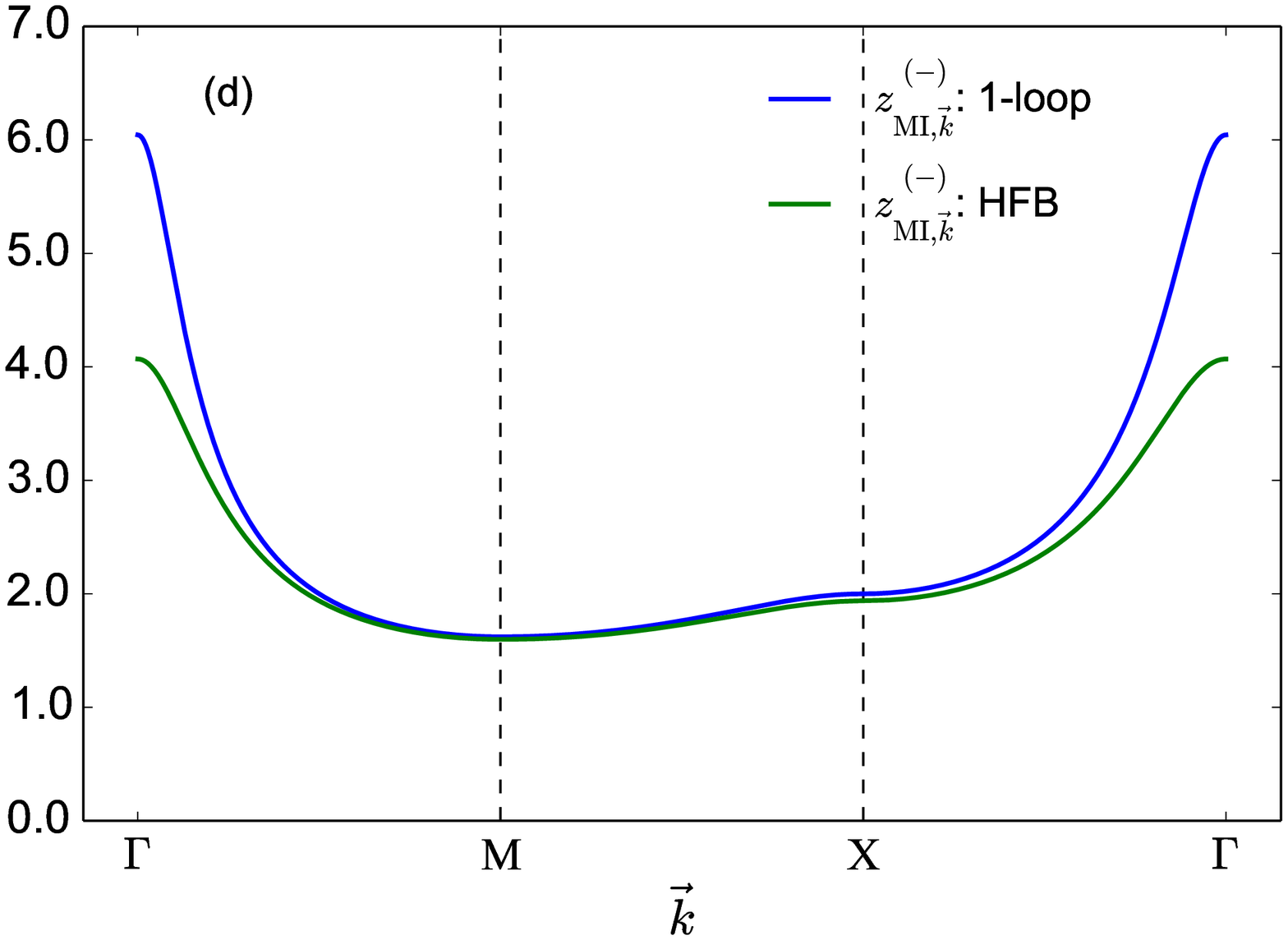}
\par\end{centering}

\begin{centering}
\includegraphics[scale=0.4]{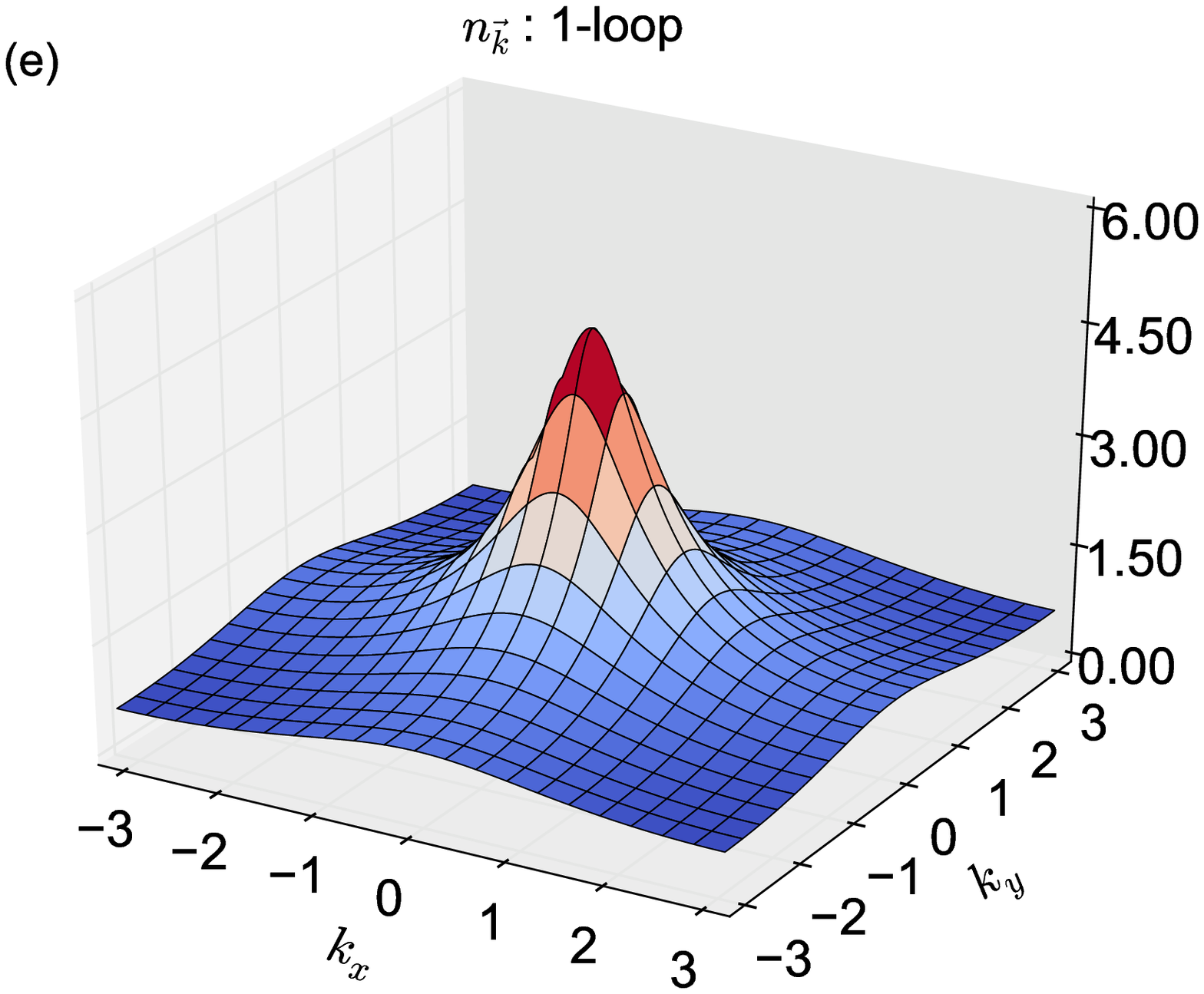}\includegraphics[scale=0.4]{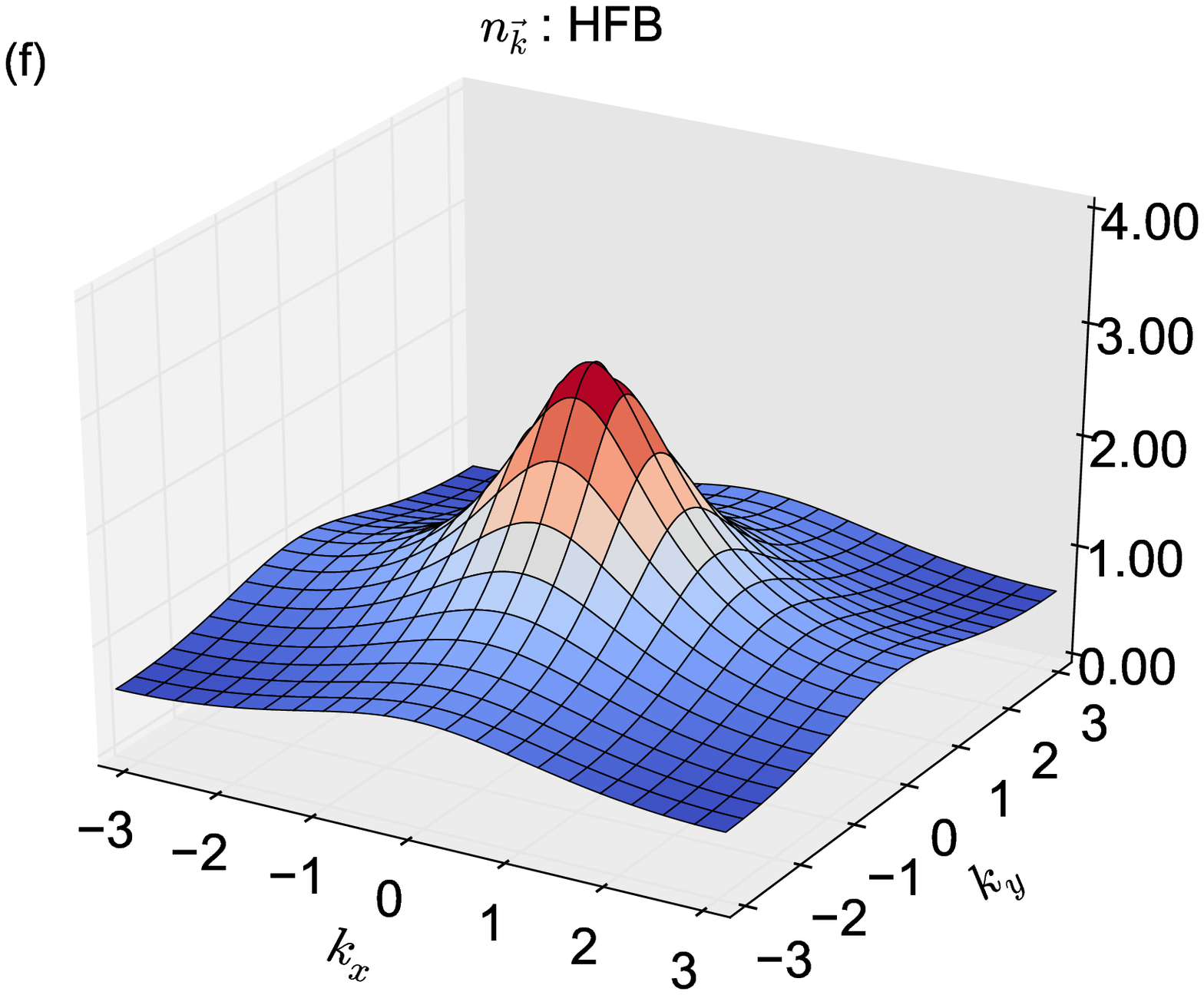}
\par\end{centering}

\caption{(Color online) Comparisons between the 1-loop and the HFB equilibrium
solution in the Mott-insulating phase. The parameters used were $d=2$,
$N_{s}=1000^{2}$, $\mu/U=0.42$, $J/U=0.04$, $\beta U=\infty$.
(a) The particle excitation energy $\Delta E_{\text{MI},\vec{k}}^{\left(+\right)}$,
(b) the hole excitation energy $\Delta E_{\text{MI},\vec{k}}^{\left(-\right)}$,
(c) the particle spectral weight $z_{\text{MI},\vec{k}}^{\left(+\right)}$,
(d) the hole spectral weight $z_{\text{MI},\vec{k}}^{\left(-\right)}$,
(e) the quasi-momentum distribution $n_{\vec{k}}$ in the 1-loop approximation,
(f) $n_{\vec{k}}$ in the HFB approximation. Note that $\Gamma=\left(0,0\right)$,
$M=\left(\pi,\pi\right)$, and $X=\left(\pi,0\right)$.\label{fig:fig5}}
\end{figure}

\noindent Hence the self-consistent solution can be formulated as
follows:
\begin{enumerate}
\item Make an initial guess for $n$.
\item Use $n$ to calculate $\Sigma_{\vec{k}}^{12,\left(R\right)}$ via
Eq.~(\ref{eq:Sigma_12^(R) - 1}).
\item Use $\Sigma_{\vec{k}}^{12,\left(R\right)}$ to calculate $\Delta E_{\text{MI},\vec{k}}^{\left(\pm\right)}$
via Eqs.~(\ref{eq:E_MI - 1})--(\ref{eq:C_k - 1}).
\item Use $\Delta E_{\text{MI},\vec{k}}^{\left(\pm\right)}$ to calculate
$z_{\text{MI},\vec{k}}^{\left(\pm\right)}$ via Eq.~(\ref{eq:z_MI - 1}).
\item Use $z_{\text{MI},\vec{k}}^{\left(\pm\right)}$ to calculate $n_{k}$
via Eq.~(\ref{eq:MI n_k - 1}).
\item Use $n_{\vec{k}}$ to recalculate $n$ via Eq.~(\ref{eq:n from n_k - 1}).
\item Repeat steps 2 to 6 until self-consistency is reached.
\end{enumerate}
In Fig.~\ref{fig:fig5}, we compare the 1-loop and HFB equilibrium
solutions in the Mott-insulating phase by calculating the excitation
energies $\Delta E_{\text{MI},\vec{k}}^{\left(\pm\right)}$, the spectral
weights $z_{\text{MI},\vec{k}}^{\left(\pm\right)}$, and the quasi-momentum
distribution $n_{\vec{k}}$ for a square lattice system with $\mu/U=0.42$,
$J/U=0.04$, and $\beta U=\infty$. The 1-loop solution, which was
studied in Ref.~\citep{SenguptaDupuis}, amounts to approximating
the self-energy by $\Sigma_{\vec{k}}^{12,\left(R\right)}=\epsilon_{\vec{k}}$
in the Mott-insulating phase. From Fig. \ref{fig:fig5} we see that
there is little qualitative change in the excitation energies between the two
approximations. The same can be said for the spectral weights for
values of $\vec{k}$ well away from zero, however there are appreciable
differences in the long-wavelength limit. These differences can be more clearly 
visualised in the quasi-momentum distribution where
we see that the $\vec{k}=0$ peak is sharper in the 1-loop approximation 
than the HFB approximation.

\begin{figure}
\begin{centering}
\includegraphics[scale=0.4]{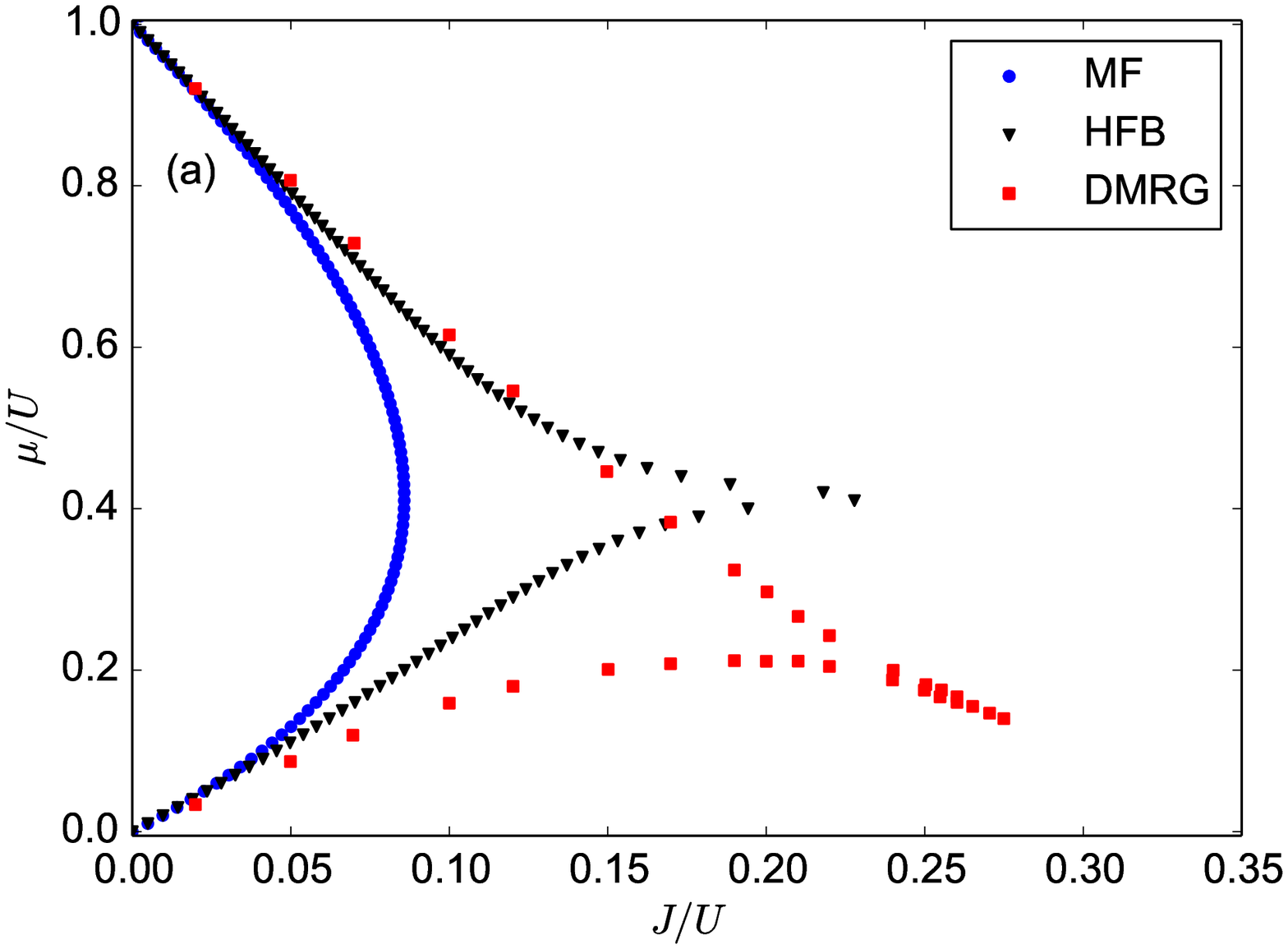}\includegraphics[scale=0.4]{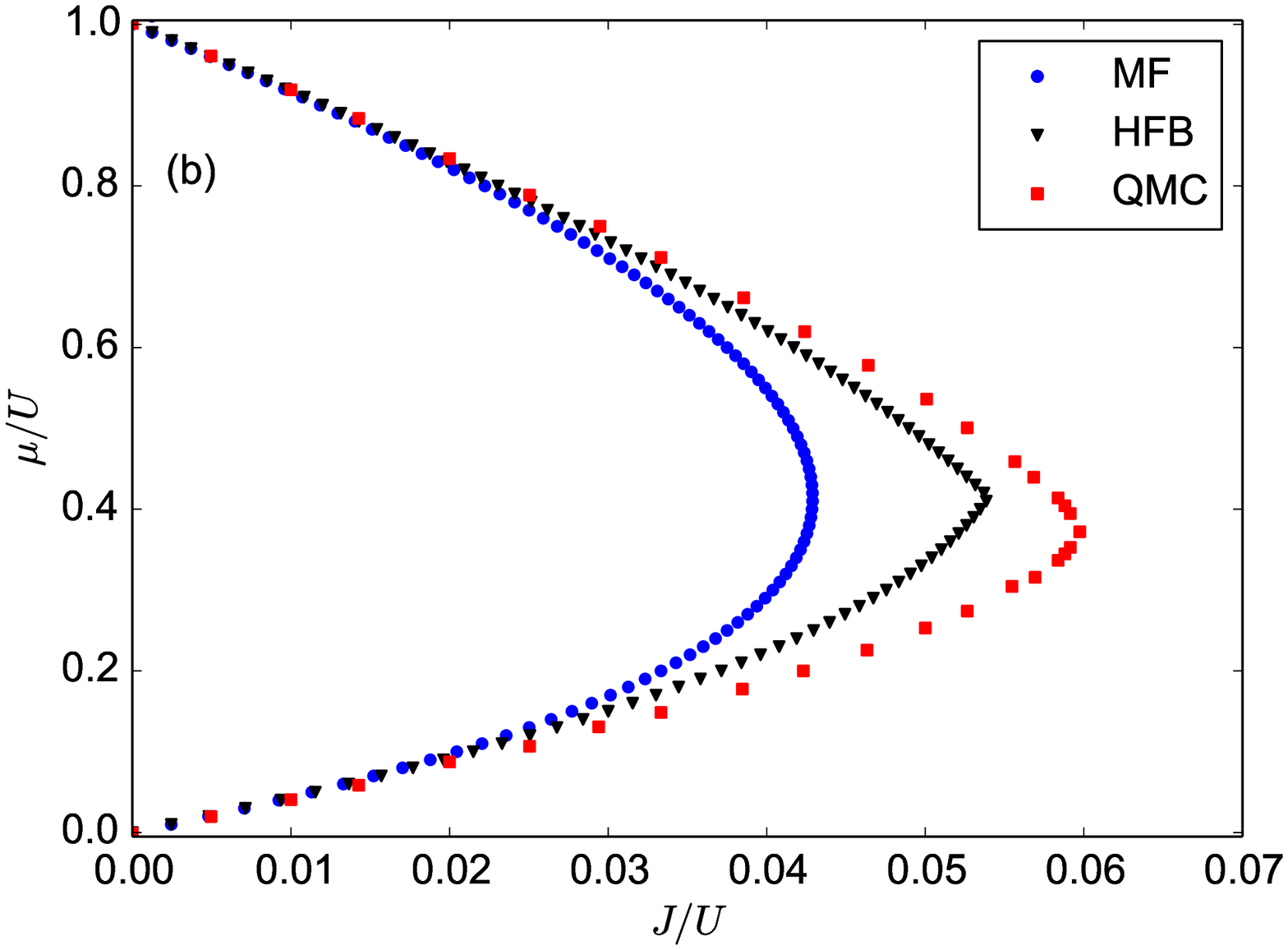}
\par\end{centering}

\begin{centering}
\includegraphics[scale=0.4]{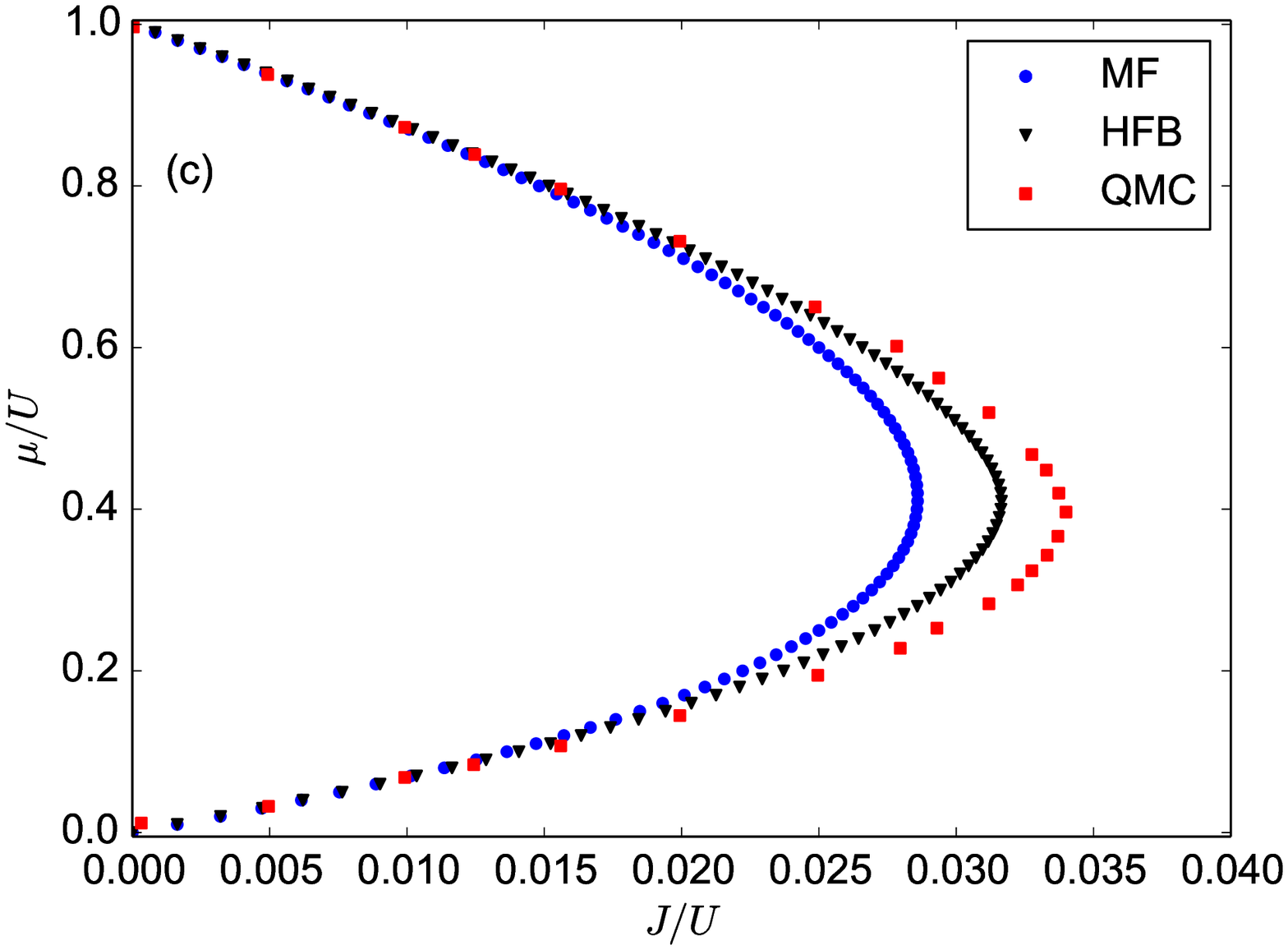}
\par\end{centering}

\caption{(Color online) Comparisons between the MF and the HFB approximations
of the phase boundary along with the exact solution for $\beta U=\infty$:
(a) $d=1$, (b) $d=2$, (c) $d=3$. The exact data was taken from
Fig. 3 in Ref.~\citep{Kuhner} for $d=1$, Fig. 1 in Ref.~\citep{Capogrosso1}
for $d=2$, and Fig. 3 in Ref.~\citep{Capogrosso2} for $d=3$.\label{fig:fig6}}
\end{figure}

One way to account for the differences in the spectral weights is
to consider how well each solution scheme approximates the phase boundary between 
Mott insulating and superfluid phases.
In Fig.~\ref{fig:fig6} we compare the mean-field (MF) and HFB approximations
of the phase boundary along with the exact calculation. Figure~\ref{fig:fig6}
clearly shows that there is significant quantitative improvement in
the phase boundary calculation when going from the MF level to the
HFB level. Moreover, in 1 dimension, where the MF approximation is
expected to be poor, we have a clear qualitative improvement in the
phase boundary calculation. The closer we are to the phase boundary
(in the Mott-insulator phase), the sharper the $\vec{k}=0$ peak is
in $n_{\vec{k}}$. Since the MF approximation always underestimates
the location of the phase boundary more than the HFB approximation,
the 1-loop approximation -- which uses the MF approximation of $\phi$
-- will wrongly predict a sharper peak as compared to that in the
HFB case. Equivalently, the 1-loop approximation will always overestimate
the values of the spectral weights in the neighbourhood of $\vec{k}=0$.

Another way to assess the accuracy of the two approximation schemes
in the Mott-insulating phase is to look at the average particle density
$n$ {[}Eq.~(\ref{eq:n from n_k - 1}){]}. In the Mott-insulating
phase, $n=\left\lceil \mu/U\right\rceil $. For the same parameter
values mentioned above, we have 

\begin{eqnarray}
n & \approx & 1.22,\qquad\text{(1-loop)},\label{eq:MI n 1-loop approx - 1}\\
n & \approx & 1.08,\qquad\text{(HFB)},\label{eq:MI n HFB approx - 1}\\
n & = & 1.00,\qquad\text{(exact)},\label{eq:MI n exact - 1}
\end{eqnarray}

\noindent where we see that the HFB approximation yields a significant
improvement as compared to the 1-loop approximation.

\subsection{Superfluid phase\label{sub:SF phase - 1}}

In the superfluid phase, $\phi$ and $\Sigma_{\vec{k}}^{22,\left(R,A\right)}$
are non-zero, hence we must use the full form of Eq.~(\ref{eq:G_12^(R) equilibrium eqn of motion - 1}).
We begin by calculating $\phi$ from Eqs.~(\ref{eq:phi equilibrium eqn of motion - 1})
and (\ref{eq:Ohm_c - 1}). Without loss of generality, we can assume
that $\phi$ is real which further implies that the quantities $iG_{\vec{r}=\mathbf{0}}^{11,\left(K\right)}\left(s^{\prime}=0\right)$
and $iG_{\vec{r}=\mathbf{0}}^{22,\left(K\right)}\left(s^{\prime}=0\right)$
are real. Based on these assumptions we obtain

\begin{eqnarray}
\phi & = & \sqrt{\frac{\left\{ \mathcal{G}^{12,\left(R\right)}\left(\omega^{\prime}=0\right)\right\} ^{-1}+2dJ}{u_{1}}-2\left(n-n_{0}\right)-\frac{1}{2}\left\{ iG_{\vec{r}=\mathbf{0}}^{22,\left(K\right)}\left(s^{\prime}=0\right)\right\} }.\label{eq:phi equilibrium eqn of motion - 2}
\end{eqnarray}

\noindent As is clear from Eq.~(\ref{eq:phi equilibrium eqn of motion - 2})
the mean field $\phi$ needs to be solved self-consistently along
with the full propagator $G$.  We now calculate $G^{\left(R\right)}$. 
Starting from Eq.~(\ref{eq:G_12^(R) equilibrium eqn of motion - 1}),
one can show that

\begin{eqnarray}
G_{\vec{k}}^{12,\left(R\right)}\left(\omega\right) & = & \frac{\left(\omega^{+}+\Delta E_{\text{MI},\vec{k}}^{\left(+\right)}\right)\left(\omega^{+}-\Delta E_{\text{MI},\vec{k}}^{\left(-\right)}\right)\left(\omega^{+}+\left\{ U+\mu\right\} \right)}{\left(\omega^{+}-\Delta E_{\text{SF},\vec{k}}^{\left(1\right)}\right)\left(\omega^{+}+\Delta E_{\text{SF},\vec{k}}^{\left(1\right)}\right)\left(\omega^{+}-\Delta E_{\text{SF},\vec{k}}^{\left(2\right)}\right)\left(\omega^{+}+\Delta E_{\text{SF},\vec{k}}^{\left(2\right)}\right)},\label{eq:SF G^{R}_12 result - 1}
\end{eqnarray}

\noindent where

\begin{eqnarray}
\Delta E_{\text{SF},\vec{k}}^{\left(s\right)} & = & \sqrt{\frac{-\tilde{B}_{\vec{k}}-\left(-1\right)^{s}\sqrt{\left(\tilde{B}_{\vec{k}}\right)^{2}-4\tilde{C}_{\vec{k}}}}{2}},\label{eq:E_SF - 1}
\end{eqnarray}

\begin{eqnarray}
\tilde{B}_{\vec{k}} & = & \left|\Sigma_{\vec{k}}^{22,\left(R\right)}\right|^{2}-\left(\Delta E_{\text{MI},\vec{k}}^{\left(+\right)}\right)^{2}-\left(\Delta E_{\text{MI},\vec{k}}^{\left(-\right)}\right)^{2},\label{eq:B_k tilde defined - 1}\\
\tilde{C}_{\vec{k}} & = & \left(\Delta E_{\text{MI},\vec{k}}^{\left(+\right)}\Delta E_{\text{MI},\vec{k}}^{\left(-\right)}\right)^{2}-\left(U+\mu\right)^{2}\left|\Sigma_{\vec{k}}^{22,\left(R\right)}\right|^{2},\label{eq:C_k tilde defined - 1}
\end{eqnarray}

\noindent In a moment we will show that the $\Delta E_{\text{SF},\vec{k}}^{\left(s\right)}$
are the excitation energies in the SF phase. Before doing so, it is
worth commenting on our approximation for the self energy in the superfluid
phase. In \ref{app:gapless spectrum in the HFBP approx - 1} we show
that in the full HFB approximation the excitation spectrum is not
gapless, violating Goldstone's Theorem, whereas if we ignore contributions from the anomalous
Keldysh Green's function $iG_{\vec{r}=0}^{22,\left(K\right)}\left(s^{\prime}=0\right)$
there is a gapless spectrum. The latter scheme is called the HFB-Popov
(HFBP) approximation \citep{PopovText}. Thus in the HFBP approximation
we have

\begin{eqnarray}
\Sigma_{\vec{k}}^{22,\left(R\right)} & = & u_{1}\left(\phi\right)^{2},\label{eq:HFB-Popov Sigma_22^(RA) - 1}
\end{eqnarray}

\begin{eqnarray}
\phi & = & \sqrt{\frac{\left\{ \mathcal{G}^{12,\left(R\right)}\left(\omega^{\prime}=0\right)\right\} ^{-1}+2dJ}{u_{1}}-2\left(n-n_{0}\right)}.\label{eq:HFB-Popov phi equation of motion - 1}
\end{eqnarray}

\noindent The HFBP approximation is most accurate for values of the
chemical potential away from integer values which is evident from
the fact that $G_{\vec{k}}^{22,\left(R\right)}\left(\omega\right)$ (and hence
$G_{\vec{r}=\mathbf{0}}^{22,\left(K\right)}\left(s^{\prime}=0\right)$) is proportional
to $\Sigma_{\vec{k}}^{22,\left(R\right)}$, which in turn is proportional
to $u_{1}$, which is small for values of the chemical potential away
from integer values. Therefore $iG_{\vec{r}=\mathbf{0}}^{22,\left(K\right)}\left(s^{\prime}=0\right)$
ought to be smaller than the average particle density $n$ by a factor
of $u_{1}$.

For the remainder of this section, we apply the HFBP approximation.
Since the energy spectrum is gapless in this approximation, i.e. $\Delta E_{\text{SF},\vec{k}\to0}^{\left(2\right)}\to0$,
care must be taken in calculating the spectral function from the retarded
Green's function. Hence we will break the calculations up into two
cases: the general case $\vec{k}\neq0$ and the special case $\vec{k}=0$.
We start with the general case.

\subsubsection{$\vec{k}\protect\neq0$\label{sub:SF k neq 0 - 1}}

When $\vec{k}\neq0$, we can derive the spectral function from the
retarded Green's function as we did above in Sec.~\ref{sub:MI phase - 1} using the Sokhotski-Plemelj
formula as we did in the MI case {[}Eq.~(\ref{eq:Sokhotski-Plemelj theorem - 1}){]}

\begin{eqnarray}
G_{\vec{k}}^{12,\left(\rho\right)}\left(\omega\right) & = & 2\pi\left\{ z_{\text{SF},\vec{k}}^{\left(1,+\right)}\delta\left(\omega-\Delta E_{\text{SF},\vec{k}}^{\left(1\right)}\right)-z_{\text{SF},\vec{k}}^{\left(1,-\right)}\delta\left(\omega+\Delta E_{\text{SF},\vec{k}}^{\left(1\right)}\right)\right.\nonumber \\
 &  & \left.\qquad+z_{\text{SF},\vec{k}}^{\left(2,+\right)}\delta\left(\omega-\Delta E_{\text{SF},\vec{k}}^{\left(2\right)}\right)-z_{\text{SF},\vec{k}}^{\left(2,-\right)}\delta\left(\omega+\Delta E_{\text{SF},\vec{k}}^{\left(2\right)}\right)\right\} ,\label{eq:SF G^(rho) equilibrium - 1}
\end{eqnarray}

\noindent where

\begin{eqnarray}
	z_{\text{SF},\vec{k}}^{\left(s,\pm\right)} & = & \left(-1\right)^{s+1}\frac{\left(\Delta E_{\text{SF},\vec{k}}^{\left(s\right)}\pm\Delta E_{\text{MI},\vec{k}}^{\left(+\right)}\right)\left(\Delta E_{\text{SF},\vec{k}}^{\left(s\right)}\mp\Delta E_{\text{MI},\vec{k}}^{\left(-\right)}\right)\left(\left\{ U+\mu\right\} \pm\Delta E_{\text{SF},\vec{k}}^{\left(s\right)}\right)}{2\Delta E_{\text{SF},\vec{k}}^{\left(s\right)}\left[\left(\Delta E_{\text{SF},\vec{k}}^{\left(1\right)}\right)^{2}-\left(\Delta E_{\text{SF},\vec{k}}^{\left(2\right)}\right)^{2}\right]}.\label{eq:z_SF - 1}
\end{eqnarray}

\noindent It is clear from Eq.~(\ref{eq:SF G^(rho) equilibrium - 1})
that $\Delta E_{\text{SF},\vec{k}}^{\left(s\right)}$ and $z_{\text{SF},\vec{k}}^{\left(s,\pm\right)}$
are the excitation energies and spectral weights respectively. Moreover,
for each branch the particle excitation energy is equal to the hole
excitation energy. Using Eq.~(\ref{eq:FDT - 1}) we have for the Keldysh
Green's function

\begin{eqnarray}
G_{\vec{k}}^{12,\left(K\right)}\left(\omega\right) & = & -2\pi i\left\{ z_{\text{SF},\vec{k}}^{\left(1,+\right)}\delta\left(\omega-\Delta E_{\text{SF},\vec{k}}^{\left(1\right)}\right)+z_{\text{SF},\vec{k}}^{\left(1,-\right)}\delta\left(\omega+\Delta E_{\text{SF},\vec{k}}^{\left(1\right)}\right)\right.\nonumber \\
 &  & \left.\qquad\qquad+z_{\text{SF},\vec{k}}^{\left(2,+\right)}\delta\left(\omega-\Delta E_{\text{SF},\vec{k}}^{\left(2\right)}\right)+z_{\text{SF},\vec{k}}^{\left(2,-\right)}\delta\left(\omega+\Delta E_{\text{SF},\vec{k}}^{\left(2\right)}\right)\right\} .\label{eq:SF G^(K) equilibrium - 1}
\end{eqnarray}

\subsubsection{$\vec{k}=0$\label{sub:SF k=00003D0 - 1}}

In the zero-quasi-momentum case, $G_{\vec{k}}^{12,\left(K\right)}\left(\omega\right)$
becomes

\begin{eqnarray}
G_{\vec{k}=0}^{12,\left(R\right)}\left(\omega\right) & = & \frac{\left(\omega^{+}+\Delta E_{\text{MI},\vec{k}=0}^{\left(+\right)}\right)\left(\omega^{+}-\Delta E_{\text{MI},\vec{k}=0}^{\left(-\right)}\right)\left(\omega^{+}+\left\{ U+\mu\right\} \right)}{\left(\omega^{+}-\Delta E_{\text{SF},\vec{k}=0}^{\left(1\right)}\right)\left(\omega^{+}+\Delta E_{\text{SF},\vec{k}=0}^{\left(1\right)}\right)\left(\omega^{+}\right)^{2}}.\label{eq:SF G^{R}_12 result - 2}
\end{eqnarray}

\noindent One cannot use the same Sokhotski-Plemelj formula as we
did above in deriving the spectral function, instead one must used
a generalized version of the formula

\begin{eqnarray}
\frac{f\left(x\right)}{\left(x+i0^{\pm}-x_{0}\right)^{n}} & = & \mp i\pi f^{\left(n-1\right)}\left(x_{0}\right)\delta\left(x-x_{0}\right)+\Gamma\left(n\right)\mathcal{P}\left\{ \frac{f\left(x\right)}{\left(x-x_{0}\right)^{n}}\right\} . \label{eq:Sokhotski-Plemelj theorem - 2}
\end{eqnarray}

\noindent Doing so yields the following spectral function

\begin{eqnarray}
G_{\vec{k}=0}^{12,\left(\rho\right)}\left(\omega\right) & = & 2\pi\left\{ z_{\text{SF},\vec{k}=0}^{\left(1,+\right)}\delta\left(\omega-\Delta E_{\text{SF},\vec{k}=0}^{\left(1\right)}\right)-z_{\text{SF},\vec{k}=0}^{\left(1,-\right)}\delta\left(\omega+\Delta E_{\text{SF},\vec{k}=0}^{\left(1\right)}\right)\right.\nonumber \\
 &  & \left.\qquad+\lim_{\vec{k}\to 0}\left[z_{\text{SF},\vec{k}}^{\left(2,+\right)}-z_{\text{SF},\vec{k}}^{\left(2,-\right)}\right]\delta\left(\omega\right)\right\} ,\label{eq:SF G^(rho) equilibrium - 2}
\end{eqnarray}

\noindent where

\begin{eqnarray}
\lim_{\vec{k}\to 0}\left[z_{\text{SF},\vec{k}}^{\left(2,+\right)}-z_{\text{SF},\vec{k}}^{\left(2,-\right)}\right] & = & \frac{\left(U+\mu\right)\left(\Delta E_{\text{MI},\vec{k}=0}^{\left(+\right)}-\Delta E_{\text{MI},\vec{k}=0}^{\left(-\right)}\right)-\Delta E_{\text{MI},\vec{k}=0}^{\left(+\right)}\Delta E_{\text{MI},\vec{k}=0}^{\left(-\right)}}{\left(\Delta E_{\text{SF},\vec{k}=0}^{\left(1\right)}\right)^{2}}.\label{eq:z_SF - 2}
\end{eqnarray}

\noindent In both $\vec{k}$ cases, $G_{\vec{k}}^{12,\left(\rho\right)}\left(\omega\right)$
is both properly normalized and signed \citep{SenguptaDupuis}.

In the case where $\vec{k}=0$, one needs to be careful when calculating
$G_{\vec{k}=0}^{12,\left(K\right)}\left(\omega\right)$ as the FDT
{[}Eq.~(\ref{eq:FDT - 1}){]} is ill-defined for $\omega=0$. Fortunately,
$G_{\vec{k}}^{12,\left(K\right)}\left(\omega\right)=0$ (see \ref{app:static limit of G^(K) - 1}
for a proof). Therefore we have for $G_{\vec{k}=0}^{12,\left(K\right)}\left(\omega\right)$

\begin{eqnarray}
G_{\vec{k}=0}^{12,\left(K\right)}\left(\omega=0\right) & = & 0,\label{eq:SF G^(K) equilibrium - 2}
\end{eqnarray}

\begin{eqnarray}
G_{\vec{k}=0}^{12,\left(K\right)}\left(\omega\neq0\right) & = & -2\pi i\left\{ z_{\text{SF},\vec{k}}^{\left(1,+\right)}\delta\left(\omega-\Delta E_{\text{SF},\vec{k}}^{\left(1\right)}\right)+z_{\text{SF},\vec{k}}^{\left(1,-\right)}\delta\left(\omega+\Delta E_{\text{SF},\vec{k}}^{\left(1\right)}\right)\right\} .\label{eq:SF G^(K) equilibrium - 3}
\end{eqnarray}

\subsubsection{Calculating $n_{\vec{k}}$ and $n$ \label{sub:SF calculating n_k and n - 1}}

\noindent One can calculate $n_{\vec{k}}$ from

\begin{eqnarray}
n_{\vec{k}} & = & \frac{1}{2}\left\langle iG_{\vec{k}}^{12,\left(K\right)}\left(s^{\prime}=0\right)+2\left\{ \left(2\pi\right)^{d}\delta_{\vec{k},\mathbf{0}}\right\} \left|\phi\right|^{2}-1\right\rangle ,\label{eq:SF n_k - G^(K) rel - 1}
\end{eqnarray}

\noindent where

\begin{eqnarray}
iG_{\vec{k}}^{12,\left(K\right)}\left(t^{\prime}=0\right) & = & \begin{cases}
z_{\text{SF},\vec{k}^{\prime}}^{\left(1,+\right)}+z_{\text{SF},\vec{k}^{\prime}}^{\left(1,-\right)}+z_{\text{SF},\vec{k}^{\prime}}^{\left(2,+\right)}+z_{\text{SF},\vec{k}^{\prime}}^{\left(2,-\right)} & \text{if }\vec{k}\neq0\\
z_{\text{SF},\vec{k}}^{\left(1,+\right)}+z_{\text{SF},\vec{k}}^{\left(1,-\right)} & \text{if }\vec{k}=0
\end{cases}.\label{eq:SF iG^(K) equal times - 1}
\end{eqnarray}

\begin{figure}
\begin{centering}
\includegraphics[scale=0.4]{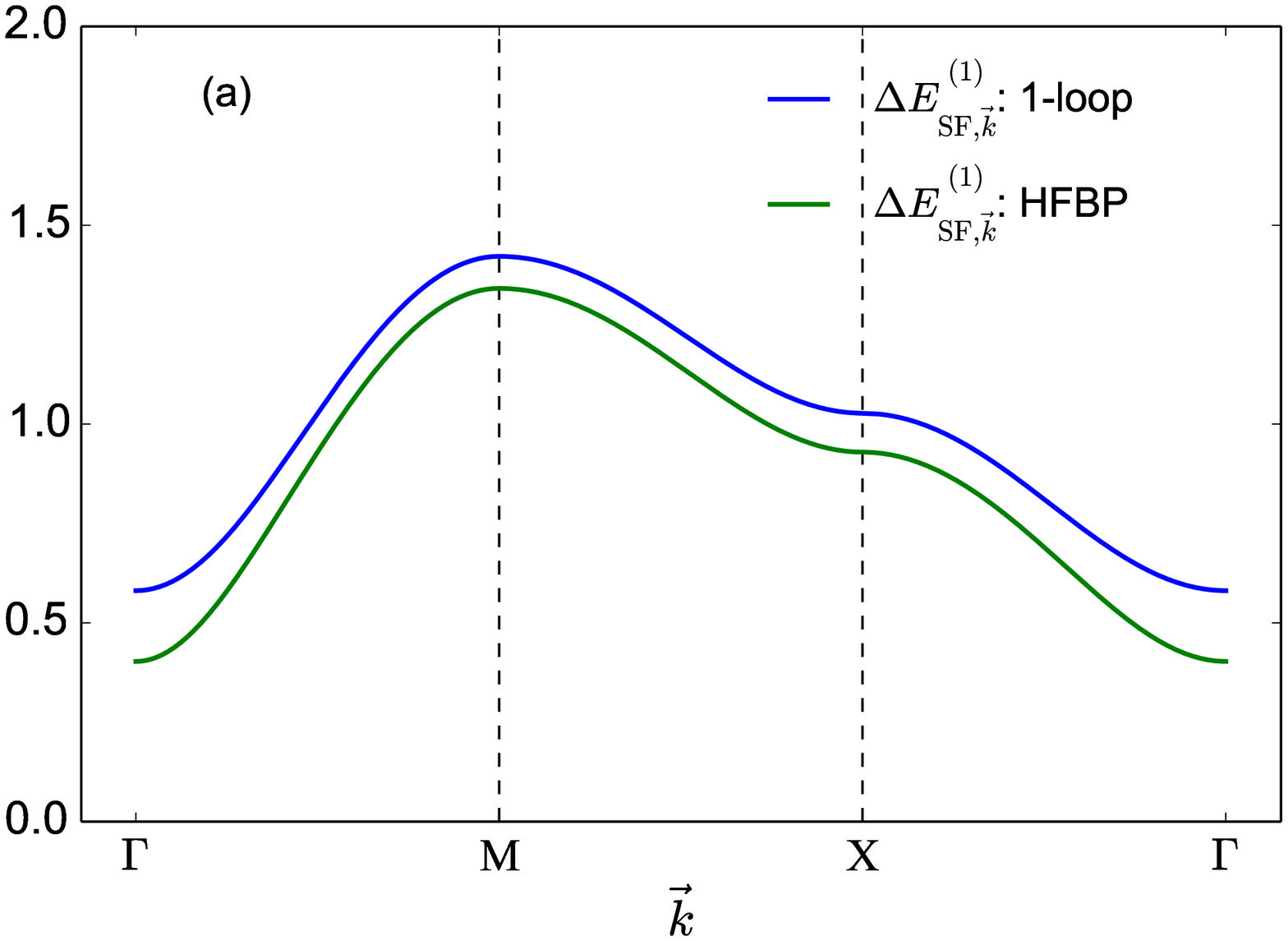}\includegraphics[scale=0.4]{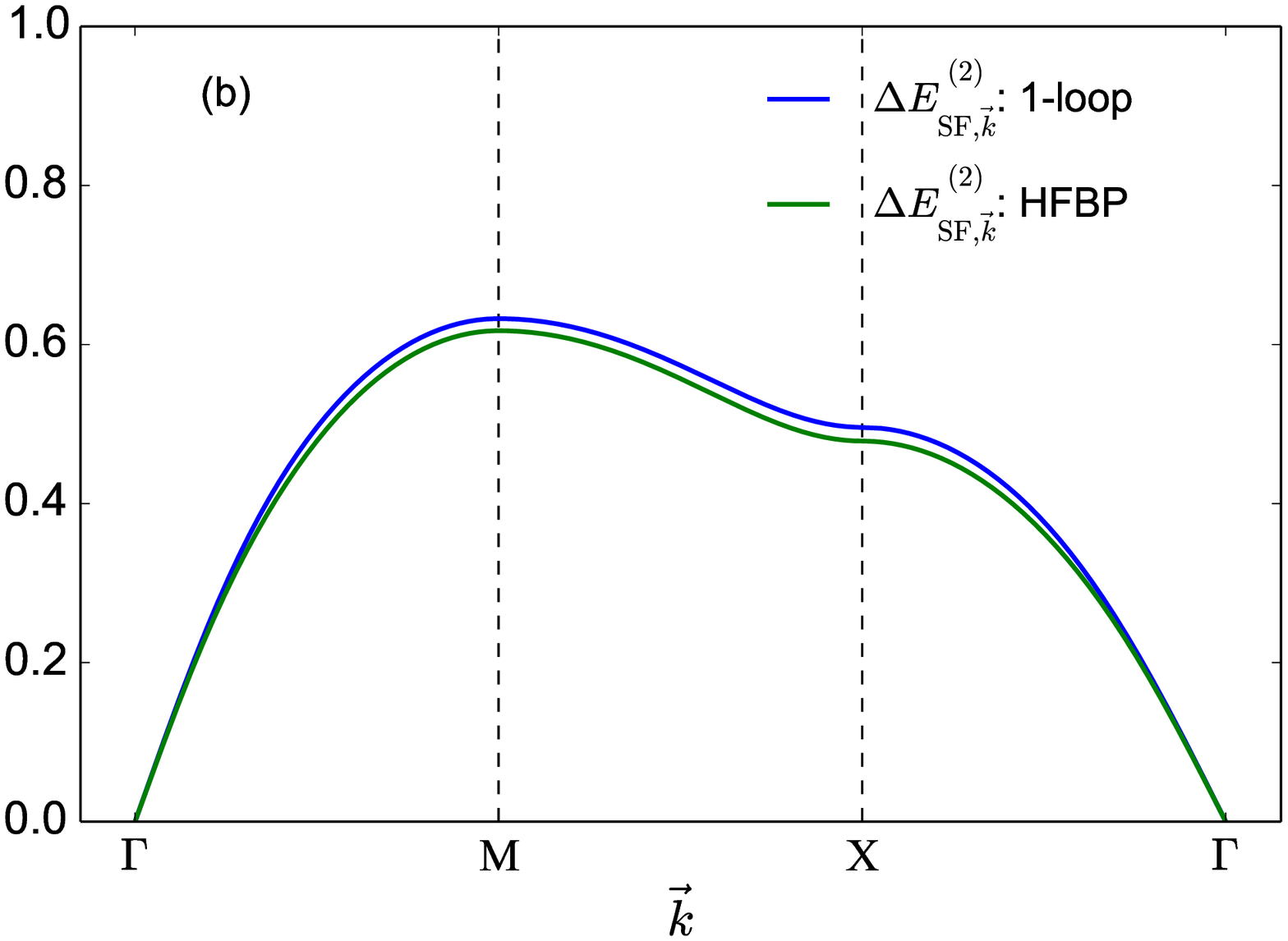}
\par\end{centering}

\begin{centering}
\includegraphics[scale=0.4]{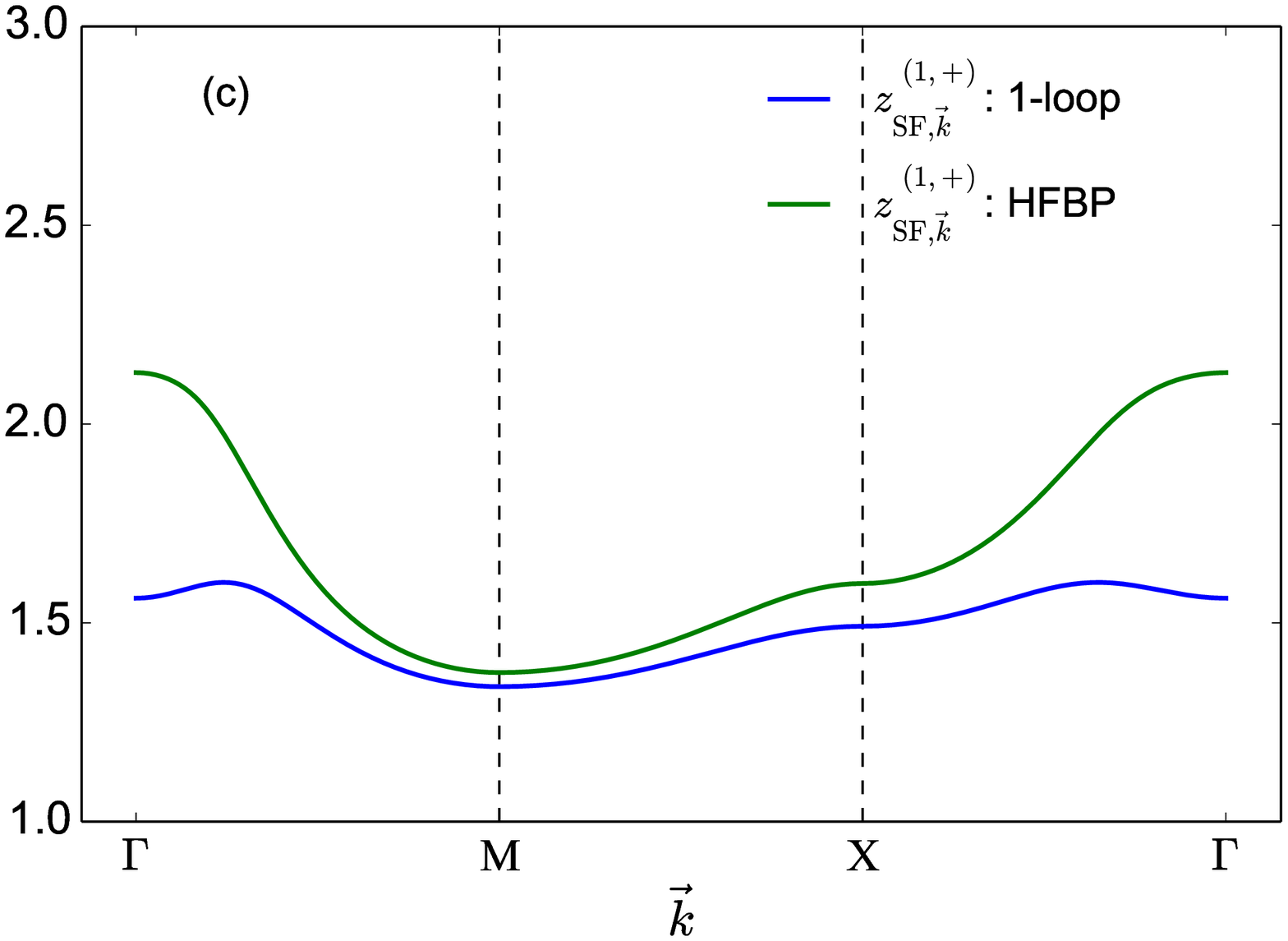}\includegraphics[scale=0.4]{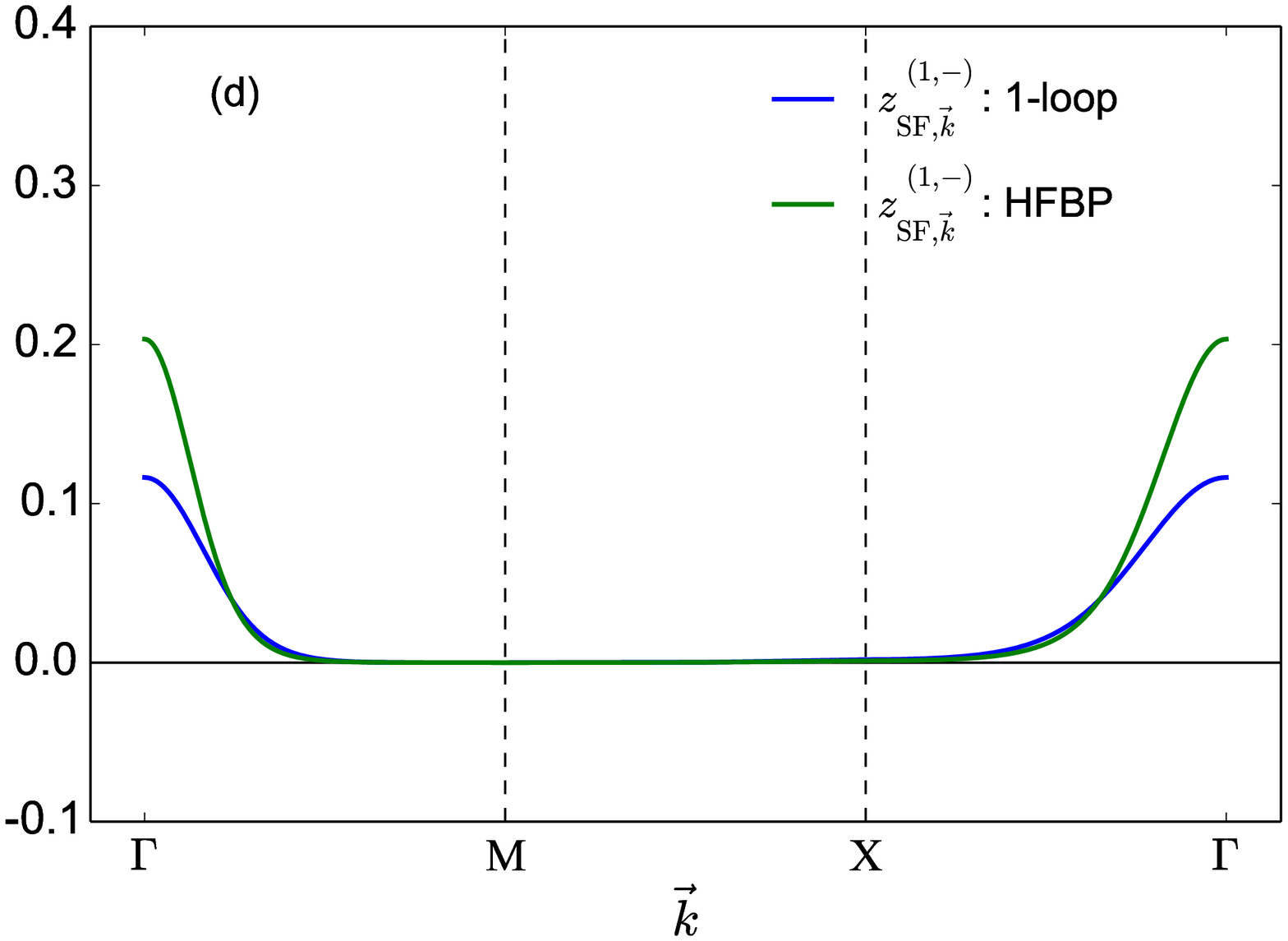}
\par\end{centering}

\begin{centering}
\includegraphics[scale=0.4]{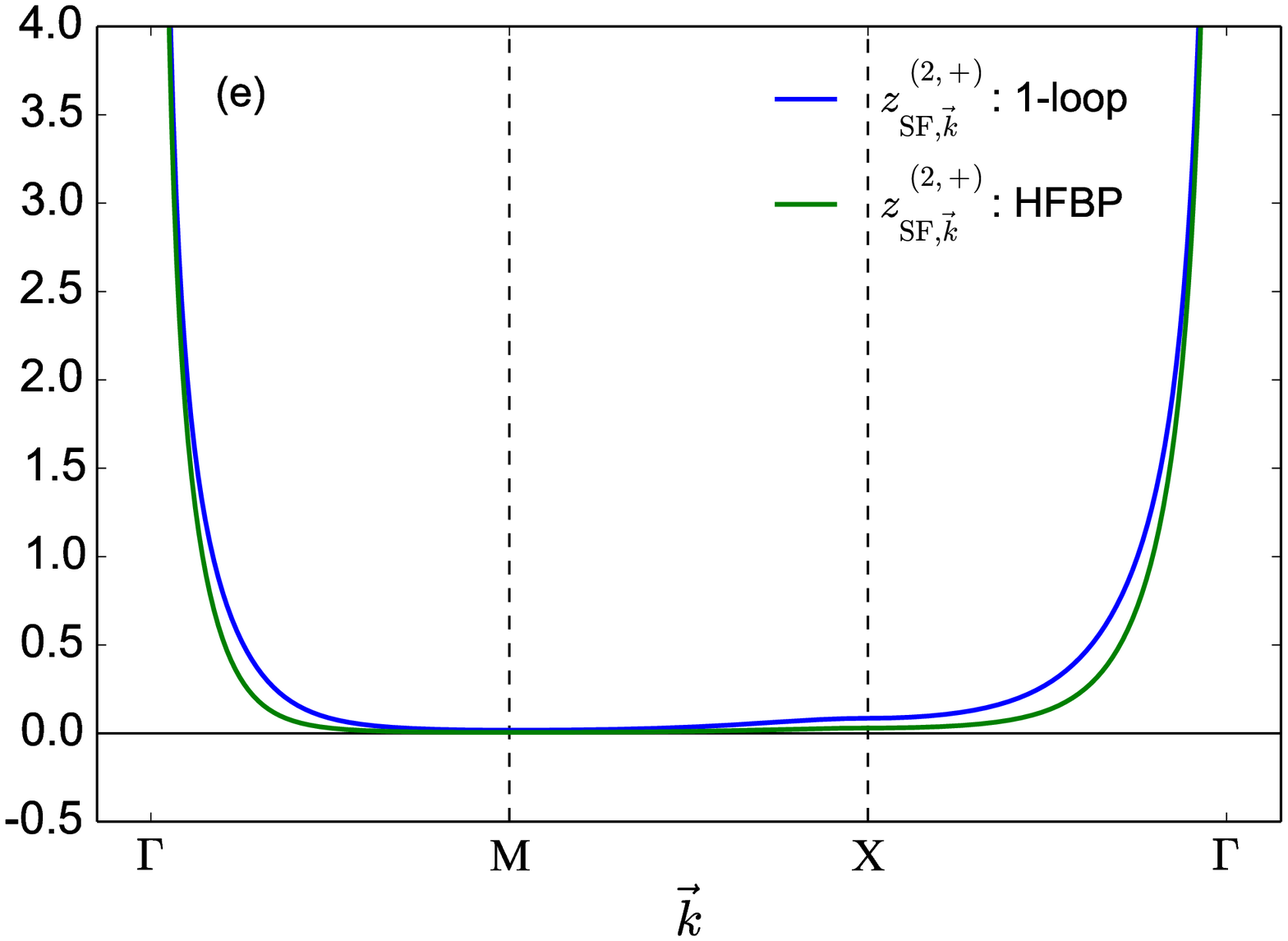}\includegraphics[scale=0.4]{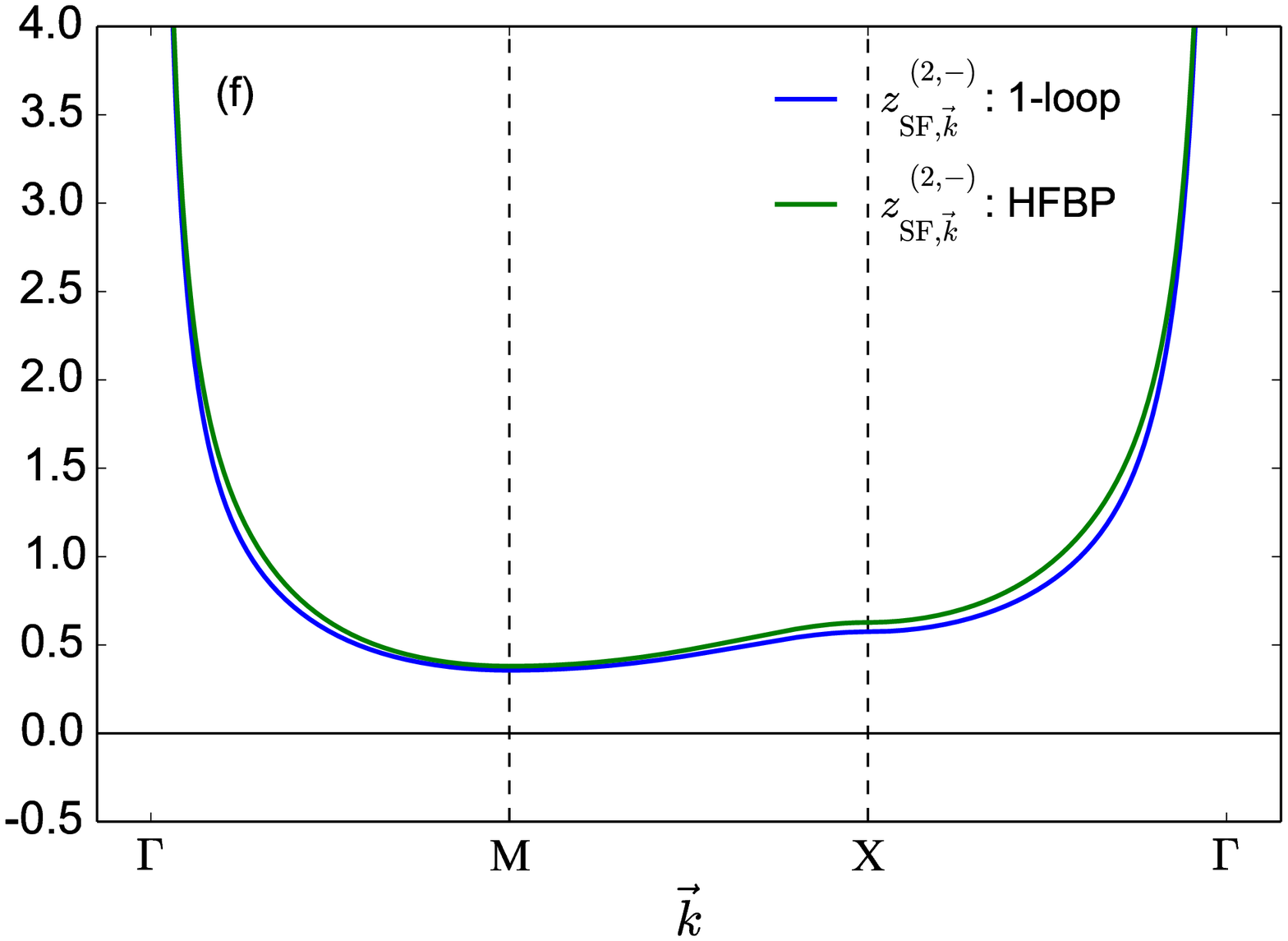}
\par\end{centering}

\caption{(Color online) Comparisons between the 1-loop and the HFBP equilibrium
solution in the superfluid phase. The parameters used were $d=2$,
$N_{s}=1000^{2}$, $\mu/U=0.36$, $J/U=0.07$, $\beta U=\infty$.
(a) The first particle/hole excitation energy branch $\Delta E_{\text{SF},\vec{k}}^{\left(1\right)}$,
(b) the second particle/hole excitation energy branch $\Delta E_{\text{SF},\vec{k}}^{\left(2\right)}$,
(c) the particle spectral weight $z_{\text{SF},\vec{k}}^{\left(1,+\right)}$for
the first branch, (d) the hole spectral weight $z_{\text{SF},\vec{k}}^{\left(1,-\right)}$for
the first branch, (e) the particle spectral weight $z_{\text{SF},\vec{k}}^{\left(2,+\right)}$for
the second branch, (f) the hole spectral weight $z_{\text{SF},\vec{k}}^{\left(2,-\right)}$for
the second branch. Note that $\Gamma=\left(0,0\right)$, $M=\left(\pi,\pi\right)$,
and $X=\left(\pi,0\right)$.\label{fig:fig7}}
\end{figure}

\noindent And lastly, the average particle density $n$ is calculated
using Eq.~(\ref{eq:n from n_k - 1}). Therefore, at the HFBP level, the
system can be solved self-consistently as follows:
\begin{enumerate}
\item Make an initial guess for $n$.
\item Use $n$ to calculate $\phi$ via Eq.~(\ref{eq:phi equilibrium eqn of motion - 2}).
\item Use $n$ and $\phi$ to calculate $\Sigma_{\vec{k}}^{12,\left(R\right)}$
and $\Sigma_{\vec{k}}^{22,\left(R\right)}$ via Eqs.~(\ref{eq:Sigma_12^(R) - 1}) and (\ref{eq:HFB-Popov Sigma_22^(RA) - 1}).
\item Use $\Sigma_{\vec{k}}^{12,\left(R\right)}$ to calculate $\Delta E_{\text{SF},\vec{k}}^{\left(s\right)}$
via Eqs. (\ref{eq:E_MI - 1})--(\ref{eq:C_k - 1}) and (\ref{eq:E_SF - 1})--(\ref{eq:C_k tilde defined - 1}).
\item Use $\Delta E_{\text{SF},\vec{k}}^{\left(s\right)}$ to calculate
$z_{\text{SF},\vec{k}}^{\left(s,\pm\right)}$ via Eqs.~(\ref{eq:z_SF - 1})
and (\ref{eq:z_SF - 2}).
\item Use $z_{\text{SF},\vec{k}}^{\left(s,\pm\right)}$ to calculate
$n_{k}$ via Eqs.~(\ref{eq:SF n_k - G^(K) rel - 1}) and (\ref{eq:SF iG^(K) equal times - 1}).
\item Use $n_{\vec{k}}$ to recalculate $n$ via Eq.~(\ref{eq:n from n_k - 1}).
\item Repeat steps 2 to 7 until self-consistency is reached.
\end{enumerate}
In Fig.~\ref{fig:fig7}, we compare the 1-loop and HFBP equilibrium
solutions in the superfluid phase by calculating the excitation energies
$\Delta E_{\text{SF},\vec{k}}^{\left(s\right)}$ and the spectral
weights $z_{\text{SF},\vec{k}}^{\left(s,\pm\right)}$ for a square
lattice system with $\mu/U=0.36$, $J/U=0.07$, and $\beta U=\infty$.
The 1-loop solution amounts to approximating the self-energy by $\Sigma_{\vec{k}}^{12,\left(R\right)}=\epsilon_{\vec{k}}+2u_{1}\left|\phi\right|^{2}$
and $\Sigma_{\vec{k}}^{22,\left(R\right)}=u_{1}\left(\phi\right)^{2}$
in the superfluid phase. We see that there is little qualitative change
in the excitation energies between the two approximations. Moreover,
the spectral weights in the second branch $s=2$ change very little
as well. We do observe appreciable differences in the spectral weights
for the first branch $s=1$ in the long-wavelength limit, similar
to the Mott-insulator case. As was argued for in the Mott-insulator case, 
since the HFBP calculation yields a more accurate phase boundary, we
believe this method will also yield a more accurate result for $z_{\text{SF},\vec{k}}^{\left(1,\pm\right)}$
in the long-wavelength limit as compared to the 1-loop result.

\subsection{Phase boundary\label{sub:phase boundary - 1}}

To calculate the phase boundary, we make a slight modification
to our solution scheme for the MI phase. The modification comes from
the extra step of calculating the critical hopping $J_{c}$. Consider
again the $\phi$-equation Eq. (\ref{eq:phi equilibrium eqn of motion - 2}).
At the boundary, $\phi=G_{\vec{r}=\mathbf{0}}^{22,\left(K\right)}\left(s^{\prime}=0\right)=0$.
Solving for $J$ we get

\begin{eqnarray}
J_{c} & = & \frac{1}{2d}\left\{ 2u_{1}\left(n-n_{0}\right)-\left\{ \mathcal{G}^{12,\left(R\right)}\left(\omega^{\prime}=0\right)\right\} ^{-1}\right\} .\label{eq:phi equation of motion at boundary - 1}
\end{eqnarray}

\noindent With this established, we can outline the phase boundary
solution as follows
\begin{enumerate}
\item Make an initial guess for the average particle density $n$
\item Use $n$ to calculate the hopping $J_{c}$, see Eq.~(\ref{eq:phi equation of motion at boundary - 1})
\item Use $n$ and $J_{c}$ to calculate the self-energy $\Sigma_{\vec{k}}^{12,\left(R\right)}$,
see Eq.~(\ref{eq:Sigma_12^(R) - 1})
\item Use $\Sigma_{\vec{k}}^{12,\left(R\right)}$ to calculate $\Delta E_{\text{MI},\vec{k}}^{\left(\pm\right)}$
via Eqs.~(\ref{eq:E_MI - 1})--(\ref{eq:C_k - 1}).
\item Use $\Delta E_{\text{MI},\vec{k}}^{\left(\pm\right)}$ to calculate
$z_{\text{MI},\vec{k}}^{\left(\pm\right)}$ via Eq.~(\ref{eq:z_MI - 1}).
\item Use $z_{\text{MI},\vec{k}}^{\left(\pm\right)}$ to calculate $n_{k}$
via Eq.~(\ref{eq:MI n_k - 1}).
\item Use $n_{\vec{k}}$ to recalculate $n$ via Eq.~(\ref{eq:n from n_k - 1}).
\item Repeat steps 2 to 7 until self-consistency is reached.
\end{enumerate}
This calculation ends up reproducing the phase boundary found from the Mott insulating side 
since the anomalous Green's functions vanish at the phase boundary.

\section{Discussion and Conclusions\label{sec:discussion and conclusions - 1}}

The ability to address single sites in cold atom experiments \citep{Bakr}
has allowed for experimental exploration of spatio-temporal correlations
in the BHM \citep{Cheneau}. This has led to theoretical investigations
of these correlations in both one \citep{Barmettler} and higher dimensions
\citep{Carleo,Natu2,Yanay,Krutitsky} in the presence of a quench.
In dimensions higher than one, where numerical approaches are limited,
a theoretical challenge has been to develop a framework which can
treat correlations in both the superfluid and Mott insulating phases
over the course of a quench. An important result in this paper is
that we have developed a formalism that allows for the description
of the space and time dependence of correlations in both phases during
a quench. The specific approach we took was to derive a 2PI effective
action for the BHM using the contour-time technique building on the
1PI real-time strong-coupling theory developed in Ref.~\citep{Kennett}
which generalized the imaginary-time theory developed in Ref.~\citep{SenguptaDupuis}.
From this 2PI effective action we were able to derive equations of
motion that treat the superfluid order parameter and the full two-point
Green's functions on equal footing. We emphasise that our formalism
is applicable even in the limit of low occupation number per site.

Even at the level of the 1PI real-time theory, the quartic coupling
becomes non-local in time, which in the 2PI theory leads to complicated expressions in the
equations of motion, involving up to four time integrals, even at the first
order in the interaction vertices. We showed
that by taking a low frequency approximation, this complexity can
be reduced to at most a single time integral. The equations of motion
obtained at this point are somewhat similar to previous 2PI studies
of the out of equilibrium dynamics of interacting bosons \citep{Rey1,Rey2,Aarts1,Aarts2,Aarts3,Aarts4}.
However, in contrast to these previous studies, the equations of motion
we obtain are a series of integral equations rather than integro-differential
equations. 

We showed that taking a HFB(P) approximation of the 2PI effective
action yields significant improvements to the calculation of the particle
density and phase boundary when compared to the 1-loop approximation
considered in Ref.~\citep{SenguptaDupuis}. Our results also suggest
that the HFB(P) approximation gives a better account of the spectral
weights in the long-wavelength limit. These improvements in the equilibrium
case suggest that our formalism should be suitable for accurately
describing spatio-temporal correlations in nonequilibrium scenarios.

The space and time dependence of correlations after a quantum quench
give insight into the propagation of excitations generated by that
quench, and hence we hope that the formalism we have developed here
will allow further theoretical investigation of the excitations after
quenches in the BHM, to complement experimental efforts in the same
direction. In future work we plan to investigate a broad range of
quench protocols, including quenches in the Mott phase where one can
study the light-cone-like spreading of single-particle correlations.
Other quench protocols of interests are those beginning in the superfluid
phase and then ending in the Mott phase. In such scenarios, one may
be interested in studying for example the possibility of aging-like
phenomena. Lastly, we plan to investigate generalizations such as
the inclusion of a harmonic trap, coupling to a bath \citep{Robertson,Dalidovich}
or a multicomponent BHM.

\section*{Acknowledgements\label{sec:acknowledgements - 1}}

The authors thank N. Dupuis, T. Gasenzer, A. M. Rey, and A. Pelster
for helpful discussions and communications. This work was supported by NSERC. 

\appendix

\section{Deriving the strong-coupling effective theory\label{app:deriving effective theory - 1}}

In this appendix, we briefly review the derivation of the effective
theory for the BHM {[}Eq.~\ref{eq:Effective theory of BHM - 1}{]}
and make note of some minor mistakes in Ref.~\citep{Kennett} (all
of these mistakes relate to mislabelling of Keldysh indices -- numerical
results in Ref.~\citep{Kennett} are unaffected). The derivation given
in Ref.~\citep{Kennett} was for the case of the Schwinger-Keldysh
contour, here we extend the derivation to the more general contour
illustrated in Fig.~\ref{fig:fig1}. We make use of the compact notation
introduced in Section \ref{sub:2PI Formalism and the effective action - 1}
when it is helpful.

We start with the generating functional $\mathcal{Z}\left[f\right]$

\begin{eqnarray}
\mathcal{Z}\left[f\right] & = & \int\left[\mathcal{D}a^{a}\right]e^{\frac{i}{2!}\sum_{\vec{r}_{1}\vec{r}_{2}}\left(2J_{\vec{r}_{1}\vec{r}_{2},\tau_{1}\tau_{2}}^{\overline{a_{1}}\overline{a_{2}}}\right)a_{\vec{r}_{1},\tau_{1}}^{a_{1}}a_{\vec{r}_{2},\tau_{2}}^{a_{2}}+iS_{0}\left[a\right]+iS_{f}\left[a\right]},\label{eq:Path integral form of Z - 2}
\end{eqnarray}

\noindent where $J_{\vec{r}_{1}\vec{r}_{2},\tau_{1}\tau_{2}}^{a_{1}a_{2}}$
is defined in Eq.~(\ref{eq:J tensor defined - 1}), $S_{f}\left[a\right]$
is defined in Eq.~(\ref{eq:S_f in qcT basis - 1}), and

\begin{eqnarray}
	S_{0} & = & \frac{1}{2}\int_{0}^{s_{\alpha_{1}\alpha_{2}}^{f}}ds\sum_{\vec{r}}\left[a_{\vec{r},\alpha_{1}}^{a_{1}}\left(s\right)\left(\left[\tau^{0}\right]_{\alpha_{1}\alpha_{3}}^{\dagger}\tau_{\alpha_{3}\alpha_{2}}^{1}\sigma_{2}^{a_{1}a_{2}}\partial_{s}\right)a_{\vec{r},\alpha_{2}}^{a_{2}}\left(s\right)\right]+S_{U}\left[a\right],\label{eq:S_0 defined - 1}
\end{eqnarray}

\noindent is the atomic part of the BHM action. Next we introduce
an auxiliary field $\psi$ via a complex Hubbard-Stratonovich transformation
\citep{SenguptaDupuis,Kennett} so the generating functional $\mathcal{Z}\left[f\right]$
takes the form

\begin{eqnarray}
\mathcal{Z}\left[f\right] & = & \int\left[\mathcal{D}\psi^{a}\right]\int\left[\mathcal{D}a^{a}\right]e^{-\frac{i}{2!}\sum_{\vec{r}_{1}\vec{r}_{2}}\left(\frac{1}{2}\left[J^{-1}\right]_{\vec{r}_{1}\vec{r}_{2},\tau_{1}\tau_{2}}^{\overline{a_{1}}\overline{a_{2}}}\right)\psi_{\vec{r}_{1},\tau_{1}}^{a_{1}}\psi_{\vec{r}_{2},\tau_{2}}^{a_{2}}-iS_{\psi}\left[a\right]+iS_{0}\left[a\right]+iS_{f}\left[a\right]},\label{eq:Path integral form of Z - 3}
\end{eqnarray}

\noindent where

\begin{eqnarray}
S_{\psi}\left[a\right] & = & \sum_{\vec{r}}\psi_{\vec{r},\tau}^{\overline{a}}a_{\vec{r},\tau}^{a}.\label{eq:S_psi - 1}
\end{eqnarray}

We can eliminate the $iS_{f}$ term in Eq. (\ref{eq:Path integral form of Z - 3})
by making a field substitution, $\psi_{\vec{r},\tau}^{a}\to-\psi_{\vec{r},\tau}^{a}+f_{\vec{r},\tau}^{a}$,
which gives

\begin{eqnarray}
\mathcal{Z}\left[f\right] & = & \int\left[\mathcal{D}\psi^{a}\right]e^{-\frac{i}{2!}\sum_{\vec{r}_{1}\vec{r}_{2}}\left(\frac{1}{2}\left[J^{-1}\right]_{\vec{r}_{1}\vec{r}_{2},\tau_{1}\tau_{2}}^{\overline{a_{1}}\overline{a_{2}}}\right)\left(\psi_{\vec{r}_{1},\tau_{1}}^{a_{1}}-f_{\vec{r}_{1},\tau_{1}}^{a_{1}}\right)\left(\psi_{\vec{r}_{2},\tau_{2}}^{a_{2}}-f_{\vec{r}_{2},\tau_{2}}^{a_{2}}\right)+iW_{0}\left[\psi\right]},\label{eq:Path integral form of Z - 4}
\end{eqnarray}

\noindent where

\begin{eqnarray}
e^{iW_{0}\left[\psi\right]} & = & \frac{1}{\mathcal{N}_{0}}\int\left[\mathcal{D}a^{a}\right]e^{iS_{0}\left[a\right]+iS_{\psi}\left[a\right]},\label{eq:introducing W_0 - 1}\\
\mathcal{N}_{0} & = & \int\left[\mathcal{D}a^{a}\right]e^{iS_{0}\left[a\right]},\label{eq:W_0 normalization factor - 1}
\end{eqnarray}

\noindent In obtaining Eq.~(\ref{eq:Path integral form of Z - 4})
we absorbed a factor of $\mathcal{N}_{0}$ into the $\psi$-measure
$\int\left[\mathcal{D}\psi^{a}\right]$. Comparing Eq.~(\ref{eq:introducing W_0 - 1})
with Eq.~(\ref{eq:Path integral form of Z - 1}), we see that $W_{0}\left[\psi\right]$
is the generator of atomic CCOGFs $\mathcal{G}^{c}$ for the bosonic
field $a$. The CCOGFs considered explicitly by the authors in Ref.~\citep{Kennett} were

\begin{eqnarray}
 &  & \mathcal{G}_{\vec{r},\alpha_{1}\ldots\alpha_{n}\alpha_{1}^{\prime}\ldots\alpha_{n}^{\prime}}^{n,c}\left(s_{1},\ldots,s_{n},s_{1}^{\prime}\ldots,s_{n}^{\prime}\right)\nonumber \\
 &  & \quad\equiv\mathcal{G}_{\underbrace{\vec{r}\ldots\vec{r}}_{2n\text{ terms}},\alpha_{1}\ldots\alpha_{n}\alpha_{1}^{\prime}\ldots\alpha_{n}^{\prime}}^{\overbrace{1\ldots1}^{n\text{ terms}}\overbrace{2\ldots2}^{n\text{ terms}},c}\left(s_{1},\ldots,s_{n},s_{1}^{\prime}\ldots,s_{n}^{\prime}\right)\nonumber \\
 &  & \quad=\left(-1\right)\left(\left[\tau^{1}\right]_{\alpha_{1}\alpha_{1}^{\prime\prime}}^{\dagger}\ldots\left[\tau^{1}\right]_{\alpha_{n}\alpha_{n}^{\prime\prime}}^{\dagger}\left[\tau^{1}\right]_{\alpha_{1}^{\prime}\alpha_{1}^{\prime\prime\prime}}^{\dagger}\ldots\left[\tau^{1}\right]_{\alpha_{n}^{\prime}\alpha_{n}^{\prime\prime\prime}}^{\dagger}\right)\nonumber \\
 &  & \quad\phantom{=}\quad\times\left.\frac{\delta^{2n}W_{0}\left[\psi\right]}{\delta f_{\vec{r},\alpha_{1}^{\prime\prime}}^{*}\left(s_{1}\right)\ldots\delta f_{\vec{r},\alpha_{n}^{\prime\prime}}^{*}\left(s_{n}\right)\delta f_{\vec{r},\alpha_{1}^{\prime\prime\prime}}\left(s_{1}^{\prime}\right)\ldots\delta f_{\vec{r},\alpha_{n}^{\prime\prime\prime}}\left(s_{n}^{\prime}\right)}\right|_{\psi=0}\nonumber \\
 &  & \quad=i\left(-1\right)^{n}\left\langle a_{\vec{r},\alpha_{1}}\left(s_{1}\right)\ldots a_{\vec{r},\alpha_{n}}\left(s_{n}\right)a_{\vec{r},\alpha_{1}^{\prime}}^{*}\left(s_{1}^{\prime}\right)\ldots a_{\vec{r},\alpha_{n}^{\prime}}^{*}\left(s_{n}^{\prime}\right)\right\rangle _{S_{0}}^{c}.\label{eq:G0s from Kennett and Dalidovich - 1}
\end{eqnarray}

\noindent Note that Eq.~(\ref{eq:G0s from Kennett and Dalidovich - 1})
corrects Eq.~(6) in Ref.~\citep{Kennett}. Moreover, note that for
the uniform BHM as considered here, the atomic CCOGFs are independent
of site index, and so we drop these indices when they do not affect
the clarity of the exposition in this paper. 

Inverting Eq.~(\ref{eq:CCOGFs from W - 1}), with $G^{c}\to\mathcal{G}^{c}$,
we may rewrite $W_{0}$ as

\begin{eqnarray}
W_{0}\left[\psi\right] & = & -\sum_{\vec{r}}\sum_{n=1}^{\infty}\frac{1}{\left(2n\right)!}\mathcal{G}_{\tau_{1}\ldots\tau_{2n}}^{\overline{a_{1}}\ldots\overline{a_{2n}},c}\psi_{\vec{r},\tau_{1}}^{a_{1}}\ldots\psi_{\vec{r},\tau_{2n}}^{a_{2n}},\label{eq:W_0 rewritten - 1}
\end{eqnarray}

\noindent which corrects Eq.~(7) in Ref.~\citep{Kennett} by a factor
of $-\left(-1\right)^{n}$, and so

\begin{eqnarray}
e^{iW_{0}\left[\psi\right]} & = & e^{i\sum_{n=1}^{\infty}S_{\text{int}}^{n}\left[\psi\right]},\label{eq:W_0 rewritten - 2}
\end{eqnarray}

\noindent where

\begin{eqnarray}
S_{\text{int}}^{n}\left[\psi\right] & = & -\sum_{\vec{r}}\frac{1}{\left(2n\right)!}\mathcal{G}_{\tau_{1}\ldots\tau_{2n}}^{\overline{a_{1}}\ldots\overline{a_{2n}},c}\psi_{\vec{r},\tau_{1}}^{a_{1}}\ldots\psi_{\vec{r},\tau_{2n}}^{a_{2n}},\label{eq:S_int^n defined - 1}
\end{eqnarray}

\noindent which corrects Eq.~(8) in Ref.~\citep{Kennett} by the same
factor of $-\left(-1\right)^{n}$.

Truncating $W_{0}\left[\psi\right]$ to quartic order in the $\psi$
fields and setting the source currents $f$ to zero in Eq.~(\ref{eq:Path integral form of Z - 4}),
the action from Eq.~(\ref{eq:Path integral form of Z - 4}) is found
to be

\begin{eqnarray}
S_{\text{eff}}\left[\psi\right] & = & -\frac{1}{2!}\sum_{\vec{r}_{1}\vec{r}_{2}}\left(\frac{1}{2}\left[J^{-1}\right]_{\vec{r}_{1}\vec{r}_{2},\tau_{1}\tau_{2}}^{\overline{a_{1}}\overline{a_{2}}}+\mathcal{G}_{\vec{r}_{1}\vec{r}_{2},\tau_{1}\tau_{2}}^{\overline{a_{1}}\overline{a_{2}},c}\right)\psi_{\vec{r}_{1},\tau_{1}}^{a_{1}}\psi_{\vec{r}_{2},\tau_{2}}^{a_{2}}\nonumber \\
 &  & -\frac{1}{4!}\sum_{\vec{r}}\mathcal{G}_{\tau_{1}\tau_{2}\tau_{3}\tau_{4}}^{\overline{a_{1}}\overline{a_{2}}\overline{a_{3}}\overline{a_{4}},c}\psi_{\vec{r},\tau_{1}}^{a_{1}}\psi_{\vec{r},\tau_{2}}^{a_{2}}\psi_{\vec{r},\tau_{3}}^{a_{3}}\psi_{\vec{r},\tau_{4}}^{a_{4}}.\label{eq:new action after first HS - 1}
\end{eqnarray}

As pointed out in Ref.~\citep{SenguptaDupuis}, the quadratic terms
in the equilibrium action of the form in Eq.~(\ref{eq:new action after first HS - 1})
allow one to calculate the mean-field phase boundary, however it yields
an unphysical excitation spectrum in the superfluid regime. This issue
is circumvented by performing a second Hubbard-Stratonovich transformation
\citep{SenguptaDupuis,Kennett}. Starting from Eq.~(\ref{eq:Path integral form of Z - 4})
(keeping the source currents $f$ this time), we introduce a second
field $z$ such that

\begin{eqnarray}
\mathcal{Z}\left[f\right] & = & \int\left[\mathcal{D}z^{a}\right]e^{\frac{i}{2!}\sum_{\vec{r}_{1}\vec{r}_{2}}\left(2J_{\vec{r}_{1}\vec{r}_{2},\tau_{1}\tau_{2}}^{\overline{a_{1}}\overline{a_{2}}}\right)z_{\vec{r}_{1},\tau_{1}}^{a_{1}}z_{\vec{r}_{2},\tau_{2}}^{a_{2}}+i\widetilde{W}\left[z\right]+iS_{f}\left[z\right]},\label{eq:Z after second HS - 1}
\end{eqnarray}

\noindent where

\begin{eqnarray}
S_{f}\left[z\right] & = & \sum_{\vec{r}}f_{\vec{r},\tau}^{\overline{a}}z_{\vec{r},\tau}^{a},\label{eq:S_f with z fields  - 1}\\
e^{i\widetilde{W}\left[z\right]} & = & \frac{1}{\mathcal{N}_{\psi}}\int\left[\mathcal{D}\psi^{a}\right]e^{iW_{0}\left[a\right]+iS_{z}\left[\psi\right]},\label{eq:W tilde introduced - 1}\\
\mathcal{N}_{\psi} & = & \int\left[\mathcal{D}\psi^{a}\right]e^{iS_{\text{int}}^{1}\left[\psi\right]},\label{eq:normalization constant for W tilde - 1}\\
S_{z}\left[\psi\right] & = & \sum_{\vec{r}}z_{\vec{r},\tau}^{\overline{a}}\psi_{\vec{r},\tau}^{a}.\label{eq:S_z for psi fields - 1}
\end{eqnarray}

\noindent By comparing Eq.~(\ref{eq:Z after second HS - 1}) to Eq.~(\ref{eq:Path integral form of Z - 1}), we can see that the COGFs
of the $z$ field generated by $\mathcal{Z}\left[f\right]$ are identical
to those of the bosonic field $a$. The last step is to perform a
cumulant expansion of $\widetilde{W}\left[z\right]$ \citep{Kennett,SenguptaDupuis,Dupuis}.
Upon doing this, we can write the generating functional $\mathcal{Z}\left[f\right]$
as

\begin{eqnarray}
\mathcal{Z}\left[f\right] & = & \int\left[\mathcal{D}z^{a}\right]e^{iS_{\text{BHM}}\left[z\right]+iS_{f}\left[z\right]},\label{eq:final Z after 2 HS - 1}
\end{eqnarray}

\noindent where $S_{\text{BHM}}\left[z\right]$ is given by

\begin{eqnarray}
S_{\text{BHM}}\left[z\right] & = & \frac{1}{2!}\sum_{\vec{r}_{1}\vec{r}_{2}}\left(2J_{\vec{r}_{1}\vec{r}_{2},\tau_{1}\tau_{2}}^{\overline{a_{1}}\overline{a_{2}}}+\left[\mathcal{G}^{-1}\right]_{\vec{r}_{1}\vec{r}_{2},\tau_{1}\tau_{2}}^{\overline{a_{1}}\overline{a_{2}},c}+\delta_{\vec{r}_{1}\vec{r}_{2}}\tilde{u}_{\tau_{1}\tau_{2}}^{\overline{a_{1}}\overline{a_{2}}}\right)z_{\vec{r}_{1},\tau_{1}}^{a_{1}}z_{\vec{r}_{2},\tau_{2}}^{a_{2}}\nonumber \\
 &  & +\sum_{\vec{r}}\sum_{n=2}^{\infty}\frac{1}{\left(2n\right)!}\left(u_{\tau_{1}\ldots\tau_{2n}}^{\overline{a_{1}}\ldots\overline{a_{2n}}}+\tilde{u}_{\tau_{1}\ldots\tau_{2n}}^{\overline{a_{1}}\ldots\overline{a_{2n}}}\right)z_{\vec{r},\tau_{1}}^{a_{1}}\ldots z_{\vec{r},\tau_{2n}}^{a_{2n}},\label{eq:z field theory - 1}
\end{eqnarray}

\noindent with

\begin{eqnarray}
u_{\tau_{1}\ldots\tau_{2n}}^{a_{1}\ldots a_{2n}} & = & -\prod_{m=1}^{n}\left(\left[\mathcal{G}^{-1}\right]_{\tau_{2m-1}\tau_{2m-1}^{\prime}}^{a_{2m-1}a_{2m-1}^{\prime},c}\left[\mathcal{G}^{-1}\right]_{\tau_{2m}\tau_{2m}^{\prime}}^{a_{2m}a_{2m}^{\prime},c}\right)\mathcal{G}_{\tau_{1}\ldots\tau_{2n}}^{\overline{a_{1}}\ldots\overline{a_{2n}},c},\label{eq:2n-point u vertices - 1}
\end{eqnarray}

\noindent and the $\tilde{u}$ vertices contain an infinite set of
``anomalous'' diagrams, i.e. diagrams that contain internal inverse
bare propagator lines. Such diagrams have no physical meaning and
should not contribute to the physical quantities \citep{Dupuis}.
It should be noted that in addition to the physical diagrams, the
$u$ vertices also generate ``anomalous'' terms. In \ref{app:cancellation of anomalous diagrams - 1},
we show that these anomalous terms cancel one another out when calculating
the superfluid order parameter $\phi$ and the full two-point CCOGF.
That being said, the action in Eq.~(\ref{eq:z field theory - 1})
contains an infinite sum, therefore one will eventually have to truncate
said action which will ultimately lead to only certain subclasses
of ``anomalous'' terms cancelling out. 

In this paper, we truncate the action to quartic order in the $z$
fields

\begin{eqnarray}
S_{\text{BHM}}\left[z\right] & = & \frac{1}{2!}\sum_{\vec{r}_{1}\vec{r}_{2}}\left(2J_{\vec{r}_{1}\vec{r}_{2},\tau_{1}\tau_{2}}^{\overline{a_{1}}\overline{a_{2}}}+\left[\mathcal{G}^{-1}\right]_{\vec{r}_{1}\vec{r}_{2},\tau_{1}\tau_{2}}^{\overline{a_{1}}\overline{a_{2}},c}+\delta_{\vec{r}_{1}\vec{r}_{2}}\tilde{u}_{\tau_{1}\tau_{2}}^{\overline{a_{1}}\overline{a_{2}}}\right)z_{\vec{r}_{1},\tau_{1}}^{a_{1}}z_{\vec{r}_{2},\tau_{2}}^{a_{2}}\nonumber \\
 &  & +\sum_{\vec{r}}\frac{1}{4!}\left(u_{\tau_{1}\tau_{2}\tau_{3}\tau_{4}}^{\overline{a_{1}}\overline{a_{2}}\overline{a_{3}}\overline{a_{4}}}+\tilde{u}_{\tau_{1}\tau_{2}\tau_{3}\tau_{4}}^{\overline{a_{1}}\overline{a_{2}}\overline{a_{3}}\overline{a_{4}}}\right)z_{\vec{r},\tau_{1}}^{a_{1}}z_{\vec{r},\tau_{2}}^{a_{2}}z_{\vec{r},\tau_{3}}^{a_{3}}z_{\vec{r},\tau_{4}}^{a_{4}},\label{eq:z field theory truncated - 1}
\end{eqnarray}

\noindent where we approximate $\tilde{u}^{\left(2\right)}$ by

\begin{eqnarray}
\tilde{u}_{\tau_{1}\tau_{2}}^{a_{1}a_{2}} & = & -\frac{1}{2!}u_{\tau_{1}\tau_{2}\tau_{3}\tau_{4}}^{a_{1}a_{2}a_{3}a_{4},c}\left(i\mathcal{G}_{\tau_{3}\tau_{4}}^{\overline{a_{3}}\overline{a_{4}},c}\right),\label{eq:approximation of 2-point u tilde - 2}
\end{eqnarray}

\noindent and neglect any contributions from $\tilde{u}^{\left(4\right)}$.
In Refs.~\citep{SenguptaDupuis,Kennett}, all $\tilde{u}$ terms were
neglected. By including the $\tilde{u}$ term given in Eq.~(\ref{eq:approximation of 2-point u tilde - 2}),
one obtains equations of motion which are accurate to first order
in $\mathcal{G}^{\left(4\right),c}$, which is not the case in Refs.~\citep{SenguptaDupuis,Kennett}. Lastly, we stress that this approach
leads to a strong-coupling theory that is not simply an expansion
order by order in $J/U$.

\section{Cancellation of anomalous diagrams\label{app:cancellation of anomalous diagrams - 1}}

In this appendix, we show that the anomalous terms introduced in \ref{app:deriving effective theory - 1}
do not contribute when calculating the mean field $\phi$ and the
two-point CCOGF $G^{c}$ of the original field $a$. For the sake of economy
in writing, we adopt the notation introduced in Section \ref{sub:2PI Formalism and the effective action - 1}
and condense it even further such that

\begin{eqnarray}
X_{x_{1}\ldots x_{n}} & \equiv & X_{\vec{r}_{1}\ldots\vec{r}_{n},\tau_{1}\ldots\tau_{n}}^{a_{1}\ldots a_{n}},\label{eq:compact notation for 2PI calculations - 2}\\
X_{x}Y_{x} & = & \sum_{\vec{r}}X_{\vec{r},\tau}^{a}Y_{\vec{r},\tau}^{\overline{a}}.\label{eq:Einstein summation convention for tau - 2}
\end{eqnarray}

\noindent We start with Eq.~(\ref{eq:Path integral form of Z - 3})

\begin{eqnarray}
\mathcal{Z}\left[f\right] & = & \int\left[\mathcal{D}\psi^{a}\right]\int\left[\mathcal{D}a^{a}\right]e^{-\frac{i}{2!}\left(\frac{1}{2}\left[J^{-1}\right]_{x_{1}x_{2}}\right)\psi_{x_{1}}\psi_{x_{2}}-iS_{\psi}\left[a\right]+iS_{0}\left[a\right]+iS_{f}\left[a\right]}\nonumber \\
 & = & \int\left[\mathcal{D}\psi^{a}\right]\int\left[\mathcal{D}a^{a}\right]e^{\frac{i}{2!}\left(-\frac{1}{2}\left[J^{-1}\right]_{x_{1}x_{2}}\right)\psi_{x_{1}}\psi_{x_{2}}}\left\langle e^{i\left(S_{\psi}\left[a\right]+S_{f}\left[a\right]\right)}\right\rangle _{S_{0}},\label{eq:Path integral form of Z - 5}
\end{eqnarray}

\noindent where we performed the field substitution $\psi_{x}\to-\psi_{x}$.
We first establish a relationship between the expectation values of
the $a$-field, $\phi_{x}$, and of the $\psi$-field, $\mathcal{V}_{x}$.
To do this, we start by calculating $\phi_{x_{1}}=\left\langle a_{x_{1}}\right\rangle $
as follows

\begin{eqnarray}
\phi_{x_{1}} & = & \left\langle a_{x_{1}}\right\rangle \nonumber \\
 & = & -i\lim_{f\to0}\frac{1}{\mathcal{Z}\left[f\right]}\frac{\delta\mathcal{Z}\left[f\right]}{\delta f_{x_{1}}}\nonumber \\
 & = & -i\lim_{f\to0}\frac{1}{\mathcal{Z}\left[f\right]}\int\left[\mathcal{D}\psi^{a}\right]e^{\frac{i}{2!}\left(-\frac{1}{2}\left[J^{-1}\right]_{x_{2}x_{3}}\right)\psi_{x_{2}}\psi_{x_{3}}}\frac{\delta}{\delta f_{x_{1}}}\left\{ \left\langle e^{i\left(S_{\psi}\left[a\right]+S_{f}\left[a\right]\right)}\right\rangle _{S_{0}}\right\} \nonumber \\
 & = & -i\lim_{f\to0}\frac{1}{\mathcal{Z}\left[f\right]}\int\left[\mathcal{D}\psi^{a}\right]e^{\frac{i}{2!}\left(-\frac{1}{2}\left[J^{-1}\right]_{x_{2}x_{3}}\right)\psi_{x_{2}}\psi_{x_{3}}}\frac{\delta}{\delta\psi_{x_{1}}}\left\{ \left\langle e^{i\left(S_{\psi}\left[a\right]+S_{f}\left[a\right]\right)}\right\rangle _{S_{0}}\right\} ,\label{eq:calc phi from psi-theory - 1}
\end{eqnarray}

\noindent and then integrate by parts to get

\begin{eqnarray}
 & = & i\lim_{f\to0}\frac{1}{\mathcal{Z}\left[f\right]}\int\left[\mathcal{D}\psi^{a}\right]\frac{\delta}{\delta\psi_{x_{1}}}\left\{ e^{\frac{i}{2!}\left(-\frac{1}{2}\left[J^{-1}\right]_{x_{2}x_{3}}\right)\psi_{x_{2}}\psi_{x_{3}}}\right\} \left\langle e^{i\left(S_{\psi}\left[a\right]+S_{f}\left[a\right]\right)}\right\rangle _{S_{0}}\nonumber \\
 & = & \frac{1}{2}\left[J^{-1}\right]_{x_{1}x_{2}}\left(\lim_{f\to0}\frac{1}{\mathcal{Z}\left[f\right]}\int\left[\mathcal{D}\psi^{a}\right]\psi_{x_{2}}e^{\frac{i}{2!}\left(-\frac{1}{2}\left[J^{-1}\right]_{x_{3}x_{4}}\right)\psi_{x_{3}}\psi_{x_{4}}+iW_{0}\left[\psi+f\right]}\right)\nonumber \\
 & = & \frac{1}{2}\left[J^{-1}\right]_{x_{1}x_{2}}\mathcal{V}_{x_{2}},\label{eq:calc phi from psi-theory - 2}
\end{eqnarray}

\noindent which establishes a relation between $\phi_{x}$ and $\mathcal{V}_{x}$.
Note that

\begin{eqnarray}
\frac{\delta}{\delta\Phi_{x}}\left(\ldots\right) & \equiv & \frac{\delta}{\delta\Phi_{\vec{r},\tau}^{\overline{a}}}\left(\ldots\right),\label{eq:condensed functional derivative notation - 1}
\end{eqnarray}

\noindent where $\Phi$ is some arbitrary field. By similar calculation,
one can show that

\begin{eqnarray}
G_{x_{1}x_{2}}^{c} & = & \frac{1}{2}\left[J^{-1}\right]_{x_{1}x_{2}}+\left(\frac{1}{2}\left[J^{-1}\right]_{x_{1}x_{3}}\right)\left(\frac{1}{2}\left[J^{-1}\right]_{x_{2}x_{4}}\right)\mathcal{V}_{x_{3}x_{4}}^{c},\label{eq:calc G from psi-theory - 1}
\end{eqnarray}

\noindent where $\mathcal{V}_{x_{1}x_{2}}^{c}$ is the two-point CCGOF
for the field $\psi$. Taking the inverses of the above relations
yields

\begin{eqnarray}
\mathcal{V}_{x_{1}} & = & \left(2J_{x_{1}x_{2}}\right)\phi_{x_{2}},\label{eq:1-pt V relation to phi - 1}\\
\mathcal{V}_{x_{1}x_{2}}^{c} & = & -\left(2J_{x_{1}x_{2}}\right)+\left(2J_{x_{1}x_{3}}\right)\left(2J_{x_{2}x_{4}}\right)G_{x_{3}x_{4}}^{c}.\label{eq:2-pt V relation to G - 1}
\end{eqnarray}

We now use the $\psi$ theory to calculate the 2PI equations of motion
for $\mathcal{V}_{x_{1}}$ and $\mathcal{V}_{x_{1}x_{2}}^{c}$. The action
$S_{\text{aux}}\left[\psi\right]$ for the auxiliary field $\psi$
can be expressed as

\begin{eqnarray}
S_{\text{aux}}\left[\psi\right] & = & \frac{1}{2!}\left(-\frac{1}{2}\left[J^{-1}\right]_{x_{1}x_{2}}\right)\psi_{x_{1}}\psi_{x_{2}}-\sum_{n=1}^{\infty}\frac{1}{\left(2n\right)!}\mathcal{G}_{x_{1}\ldots x_{2n}}^{c}\psi_{x_{1}}\ldots\psi_{x_{2n}},\label{eq:S_aux - 1}
\end{eqnarray}

\noindent and hence using this action in Eqs.~(\ref{eq:phi eqn of motion - 1})
and (\ref{eq:G eqn of motion - 1}) and rearranging terms, we obtain
the following relations

\begin{eqnarray}
\mathcal{V}_{x_{1}} & = & -\left(2J_{x_{1}x_{2}}\right)\mathcal{G}_{x_{2}x_{3}}^{c}\mathcal{V}_{x_{3}}\nonumber \\
 &  & -\left(2J_{x_{1}x_{2}}\right)\sum_{n=2}^{\infty}\frac{1}{\left(2n-3\right)!}\mathcal{G}_{x_{2}x_{3}x_{4}x_{5}\ldots x_{2n+1}}^{c}\nonumber \\
 &  & \phantom{-\left(2J_{x_{1}x_{2}}\right)\sum_{n=2}^{\infty}}\times\left\{ \frac{1}{\left(2n-1\right)\left(2n-2\right)}\mathcal{V}_{x_{3}}\mathcal{V}_{x_{4}}+\frac{1}{2}\left(i\mathcal{V}_{x_{3}x_{4}}^{c}\right)\right\} \mathcal{V}_{x_{5}}\ldots\mathcal{V}_{x_{2n+1}}\nonumber \\
 &  & +\left(2J_{x_{1}x_{2}}\right)\Xi_{x_{2}}\left[\mathcal{G}^{\left(2n\right),c},\mathcal{V}^{\left(1\right)},\mathcal{V}^{\left(2\right),c}\right],\label{eq:1-pt V equation of motion - 1}\\
\mathcal{V}_{x_{1}x_{2}}^{c} & = & -\left(2J_{x_{1}x_{2}}\right)-\left(2J_{x_{1}x_{3}}\right)\mathcal{G}_{x_{3}x_{4}}^{c}\mathcal{V}_{x_{4}x_{2}}^{c}\nonumber \\
 &  & -\left(2J_{x_{1}x_{3}}\right)\left\{ \sum_{n=2}^{\infty}\frac{1}{\left(2n-2\right)!}\mathcal{G}_{x_{3}x_{4}x_{5}\ldots x_{2n+2}}^{c}\mathcal{V}_{x_{5}}\ldots\mathcal{V}_{x_{2n+2}}\right\} \mathcal{V}_{x_{4}x_{2}}^{c}\nonumber \\
 &  & -\left(2J_{x_{1}x_{3}}\right)\Sigma_{x_{3}x_{4}}^{\text{aux}}\left[\mathcal{G}^{\left(2n\right),c},\mathcal{V}^{\left(1\right)},\mathcal{V}^{\left(2\right),c}\right]\mathcal{V}_{x_{4}x_{2}}^{c},\label{eq:2-pt V equation of motion - 1}
\end{eqnarray}

\noindent where $\Xi$ and $\Sigma$ are obtained from the corresponding
$\Gamma_{2}$. Next, we apply Eqs.~(\ref{eq:1-pt V relation to phi - 1})
and (\ref{eq:2-pt V relation to G - 1}) to obtain recursive expressions
for $\phi$ and $G^{c}$

\begin{eqnarray}
\phi_{x_{1}} & = & -\mathcal{G}_{x_{1}x_{2}}^{c}\left(2J_{x_{2}x_{2}^{\prime}}\right)\phi_{x_{2}^{\prime}}\nonumber \\
 &  & -\sum_{n=2}^{\infty}\frac{1}{\left(2n-3\right)!}\mathcal{G}_{x_{1}x_{2}x_{3}x_{4}\ldots x_{2n}}^{c}\left\{ \frac{1}{\left(2n-1\right)\left(2n-2\right)}\left(2J_{x_{2}x_{2}^{\prime}}\right)\left(2J_{x_{3}x_{3}^{\prime}}\right)\phi_{x_{2}^{\prime}}\phi_{x_{3}^{\prime}}\right.\nonumber \\
 &  & \phantom{-\sum_{n=2}^{\infty}\frac{1}{\left(2n-3\right)!}\mathcal{G}_{x_{1}x_{2}x_{3}x_{4}\ldots x_{2n}}^{c}}\quad\left.+\frac{i}{2}\left[-\left(2J_{x_{2}x_{3}}\right)+\left(2J_{x_{2}x_{2}^{\prime}}\right)\left(2J_{x_{3}x_{3}^{\prime}}\right)G_{x_{2}^{\prime}x_{3}^{\prime}}^{c}\right]\right\} \nonumber \\
 &  & \phantom{-\sum_{n=2}^{\infty}}\quad\times\left(2J_{x_{4}x_{4}^{\prime}}\right)\ldots\left(2J_{x_{2n}x_{2n}^{\prime}}\right)\phi_{x_{4}^{\prime}}\ldots\phi_{x_{2n}^{\prime}}\nonumber \\
 &  & +\Xi_{x_{1}}\left[\mathcal{G}^{\left(2n\right),c},\left(2J_{xx^{\prime}}\right)\phi_{x^{\prime}},-\left(2J_{xy}\right)+\left(2J_{xx^{\prime}}\right)\left(2J_{yy^{\prime}}\right)G_{x^{\prime}y^{\prime}}^{c}\right],\label{eq:phi equation of motion from S_aux - 1}
\end{eqnarray}

\begin{eqnarray}
G_{x_{1}x_{2}}^{c} & = & \left\{ \mathcal{G}_{x_{1}x_{3}}^{c}+\sum_{n=2}^{\infty}\frac{1}{\left(2n-2\right)!}\mathcal{G}_{x_{1}x_{3}x_{4}\ldots x_{2n+1}}^{c}\left(2J_{x_{4}x_{4}^{\prime}}\phi_{x_{4}^{\prime}}\right)\ldots\left(2J_{x_{2n+1}x_{2n+1}^{\prime}}\phi_{x_{2n+1}^{\prime}}\right)\right.\nonumber \\
 &  & \left.\quad+\Sigma_{x_{1}x_{3}}^{\text{aux}}\left[\mathcal{G}^{\left(2n\right),c},\left(2J_{xx^{\prime}}\right)\phi_{x^{\prime}},-\left(2J_{xy}\right)+\left(2J_{xx^{\prime}}\right)\left(2J_{yy^{\prime}}\right)G_{x^{\prime}y^{\prime}}^{c}\right]\right\} \nonumber \\
 &  & \quad\times\left\{ \delta_{x_{3}x_{2}}-\left(2J_{x_{3}x_{3}^{\prime}}\right)G_{x_{3}^{\prime}x_{2}}^{c}\right\} .\label{eq:G equation of motion from S_aux - 1}
\end{eqnarray}

We now derive recursive relations for $\phi$ and $G^{c}$ by an alternative
approach: we apply the 2PI approach to the theory of the $z$-fields,
allowing for anomalous terms, which is given by Eq.~(\ref{eq:z field theory - 1})
and written again here in compact form

\begin{eqnarray}
S_{\text{BHM}} & = & \frac{1}{2!}\left(\left[\mathcal{G}^{-1}\right]_{x_{1}x_{2}}^{c}+\tilde{u}_{x_{1}x_{2}}\right)z_{x_{1}}z_{x_{2}}+\frac{1}{2!}\left(2J_{x_{1}x_{2}}\right)z_{x_{1}}z_{x_{2}}\nonumber \\
 &  & +\sum_{n=2}^{\infty}\frac{1}{\left(2n\right)!}\left(u_{x_{1}\ldots x_{2n}}+\tilde{u}_{x_{1}\ldots x_{2n}}\right)z_{x_{1}}\ldots z_{x_{2n}}.\label{eq:z-field theory compact form - 1}
\end{eqnarray}

\noindent As noted in \ref{app:deriving effective theory - 1}, the
Green's functions for the $z$-fields are the same as those for the
$a$-fields. Similarly to the calculations leading to the recursive
relations $\mathcal{V}_{x_{1}}$ and $\mathcal{V}_{x_{1}x_{2}}^{c}$,
we calculate the following recursive 2PI relations for $\phi$ and
$G^{c}$

\begin{eqnarray}
\phi_{x_{1}} & = & -\mathcal{G}_{x_{1}x_{2}}^{c}\left(2J_{x_{2}x_{3}}\right)\phi_{x_{3}}-\mathcal{G}_{x_{1}x_{2}}^{c}\tilde{u}_{x_{2}x_{3}}\phi_{x_{3}}\nonumber \\
 &  & -\mathcal{G}_{x_{1}x_{2}}^{c}\sum_{n=2}^{\infty}\frac{1}{\left(2n-3\right)!}\left\{ u_{x_{2}x_{3}x_{4}x_{5}\ldots x_{2n+1}}+\tilde{u}_{x_{2}x_{3}x_{4}x_{5}\ldots x_{2n+1}}\right\} \nonumber \\
 &  & \phantom{-\mathcal{G}_{x_{1}x_{2}}^{c}\sum_{n=2}^{\infty}}\quad\times\left\{ \frac{1}{\left(2n-1\right)\left(2n-2\right)}\phi_{x_{3}}\phi_{x_{4}}+\frac{i}{2}G_{x_{3}x_{4}}^{c}\right\} \phi_{x_{5}}\ldots\phi_{x_{2n+1}}\nonumber \\
 &  & -\mathcal{G}_{x_{1}x_{2}}^{c}\Xi_{x_{2}}\left[-u^{\left(2n\right)}-\tilde{u}^{\left(2n\right)},\phi,G^{c}\right],\label{eq:phi equation of motion from z-theory - 1}
\end{eqnarray}

\begin{eqnarray}
G_{x_{1}x_{2}}^{c} & = & \mathcal{G}_{x_{1}x_{2}}^{c}-\mathcal{G}_{x_{1}x_{3}}^{c}\left(2J_{x_{3}x_{4}}\right)G_{x_{4}x_{2}}^{c}-\mathcal{G}_{x_{1}x_{3}}^{c}\tilde{u}_{x_{3}x_{4}}G_{x_{4}x_{2}}^{c}\nonumber \\
 &  & -\mathcal{G}_{x_{1}x_{3}}^{c}\left(\sum_{n=2}^{\infty}\frac{1}{\left(2n-2\right)!}\left\{ u_{x_{3}x_{4}x_{5}\ldots x_{2n+2}}+\tilde{u}_{x_{3}x_{4}x_{5}\ldots x_{2n+2}}\right\} \phi_{x_{5}}\ldots\phi_{x_{2n+2}}\right)G_{x_{4}x_{2}}^{c}\nonumber \\
 &  & +\mathcal{G}_{x_{1}x_{3}}^{c}\Sigma_{x_{3}x_{4}}^{\text{aux}}\left[-u^{\left(2n\right)}-\tilde{u}^{\left(2n\right)},\phi,G^{c}\right]G_{x_{4}x_{2}}^{c}.\label{eq:G equation of motion from z-theory - 1}
\end{eqnarray}

\noindent We momentarily drop the terms containing $\tilde{u}$ and
focus on the remaining terms in the recursive expressions

\begin{eqnarray}
\phi_{x_{1}} & = & -\mathcal{G}_{x_{1}x_{2}}^{c}\left(2J_{x_{2}x_{3}}\right)\phi_{x_{3}}\nonumber \\
 &  & -\mathcal{G}_{x_{1}x_{2}}^{c}\sum_{n=2}^{\infty}\frac{1}{\left(2n-3\right)!}u_{x_{2}x_{3}x_{4}x_{5}\ldots x_{2n+1}}\nonumber \\
 &  & \phantom{-\mathcal{G}_{x_{1}x_{2}}^{c}\sum_{n=2}^{\infty}}\quad\times\left\{ \frac{1}{\left(2n-1\right)\left(2n-2\right)}\phi_{x_{3}}\phi_{x_{4}}+\frac{1}{2}\left(iG_{x_{3}x_{4}}^{c}\right)\right\} \phi_{x_{5}}\ldots\phi_{x_{2n+1}}\nonumber \\
 &  & -\mathcal{G}_{x_{1}x_{2}}^{c}\Xi_{x_{2}}\left[-u^{\left(2n\right)},\phi,G^{c}\right]+\ldots,\label{eq:phi equation of motion from z-theory - 2}
\end{eqnarray}

\begin{eqnarray}
G_{x_{1}x_{2}} & = & \mathcal{G}_{x_{1}x_{2}}^{c}-\mathcal{G}_{x_{1}x_{3}}^{c}\left(2J_{x_{3}x_{4}}\right)G_{x_{4}x_{2}}^{c}\nonumber \\
 &  & -\mathcal{G}_{x_{1}x_{3}}^{c}\left(\sum_{n=2}^{\infty}\frac{1}{\left(2n-2\right)!}u_{x_{3}x_{4}x_{5}\ldots x_{2n+2}}\phi_{x_{5}}\ldots\phi_{x_{2n+2}}\right)G_{x_{4}x_{2}}^{c}\nonumber \\
 &  & +\mathcal{G}_{x_{1}x_{3}}^{c}\Sigma_{x_{3}x_{4}}^{\text{aux}}\left[-u^{\left(2n\right)},\phi,G^{c}\right]G_{x_{4}x_{2}}^{c}+\ldots\,.\label{eq:G equation of motion from z-theory - 2}
\end{eqnarray}

\noindent We now iterate the recursive expressions: for every additive
term in Eqs.~(\ref{eq:phi equation of motion from z-theory - 2})
and (\ref{eq:G equation of motion from z-theory - 2}) that contains
at least one $u$ vertex, we apply the recursion relations to each
$\phi$ and $G^{c}$, and keep explicitly the following (infinite) subsets
of terms respectively

\begin{eqnarray}
\phi_{x_{1}} & \to & -\mathcal{G}_{x_{1}x_{2}}^{c}\left(2J_{x_{2}x_{3}}\right)\phi_{x_{3}},\label{eq:recursive insertions - 1}\\
G_{x_{1}x_{2}}^{c} & \to & -\mathcal{G}_{x_{1}x_{3}}^{c}\left(2J_{x_{3}x_{3}^{\prime}}\right)\mathcal{G}_{x_{3}^{\prime}x_{2}}^{c}+\mathcal{G}_{x_{1}x_{3}}^{c}\left(2J_{x_{3}x_{4}}\right)G_{x_{4}x_{4}^{\prime}}^{c}\left(2J_{x_{4}^{\prime}x_{3}^{\prime}}\right)\mathcal{G}_{x_{3}^{\prime}x_{2}}^{c}\quad\text{(internal lines)},\label{eq:recursive insertions - 2}\\
G_{x_{1}x_{2}}^{c} & \to & \mathcal{G}_{x_{1}x_{2}}^{c}-\mathcal{G}_{x_{1}x_{3}}^{c}\left(2J_{x_{3}x_{3}^{\prime}}\right)G_{x_{3}^{\prime}x_{2}}^{c}\quad\text{(external lines)},\label{eq:recursive insertions - 3}
\end{eqnarray}

\noindent which yields

\begin{eqnarray}
\phi_{x_{1}} & = & -\mathcal{G}_{x_{1}x_{2}}^{c}\left(2J_{x_{2}x_{2}^{\prime}}\right)\phi_{x_{2}^{\prime}}\nonumber \\
 &  & -\sum_{n=2}^{\infty}\frac{1}{\left(2n-3\right)!}\mathcal{G}_{x_{1}x_{2}x_{3}x_{4}\ldots x_{2n}}^{c}\left\{ \frac{1}{\left(2n-1\right)\left(2n-2\right)}\left(2J_{x_{2}x_{2}^{\prime}}\right)\left(2J_{x_{3}x_{3}^{\prime}}\right)\phi_{x_{2}^{\prime}}\phi_{x_{3}^{\prime}}\right.\nonumber \\
 &  & \phantom{-\sum_{n=2}^{\infty}\frac{1}{\left(2n-3\right)!}\mathcal{G}_{x_{1}x_{2}x_{3}x_{4}\ldots x_{2n}}^{c}}\quad\left.+\frac{i}{2}\left[-\left(2J_{x_{2}x_{3}}\right)+\left(2J_{x_{2}x_{2}^{\prime}}\right)\left(2J_{x_{3}x_{3}^{\prime}}\right)G_{x_{2}^{\prime}x_{3}^{\prime}}^{c}\right]\right\} \nonumber \\
 &  & \phantom{-\sum_{n=2}^{\infty}}\quad\times\left(2J_{x_{4}x_{4}^{\prime}}\right)\ldots\left(2J_{x_{2n}x_{2n}^{\prime}}\right)\phi_{x_{4}^{\prime}}\ldots\phi_{x_{2n}^{\prime}}\nonumber \\
 &  & +\Xi_{x_{1}}\left[\mathcal{G}^{\left(2n\right),c},\left(2J_{xx^{\prime}}\right)\phi_{x^{\prime}},-\left(2J_{xy}\right)+\left(2J_{xx^{\prime}}\right)\left(2J_{yy^{\prime}}\right)G_{x^{\prime}y^{\prime}}^{c}\right]+F^{\phi}\left[\left(\mathcal{G}^{c}\right)^{-1}\right],\label{eq:phi equation of motion from z-theory - 3}
\end{eqnarray}

\begin{eqnarray}
G_{x_{1}x_{2}}^{c} & = & \left\{ \mathcal{G}_{x_{1}x_{3}}^{c}+\sum_{n=2}^{\infty}\frac{1}{\left(2n-2\right)!}\mathcal{G}_{x_{1}x_{3}x_{4}\ldots x_{2n+1}}^{c}\left(2J_{x_{4}x_{4}^{\prime}}\right)\ldots\left(2J_{x_{2n+1}x_{2n+1}^{\prime}}\right)\phi_{x_{4}^{\prime}}\ldots\phi_{x_{2n+1}^{\prime}}\right.\nonumber \\
 &  & \left.\quad+\Sigma_{x_{1}x_{3}}^{\text{aux}}\left[\mathcal{G}^{\left(2n\right),c},\left(2J_{xx^{\prime}}\right)\phi_{x^{\prime}},-\left(2J_{xy}\right)+\left(2J_{xx^{\prime}}\right)\left(2J_{yy^{\prime}}\right)G_{x^{\prime}y^{\prime}}^{c}\right]\right\} \nonumber \\
 &  & \quad\times\left\{ \delta_{x_{3}x_{2}}-\left(2J_{x_{3}x_{3}^{\prime}}\right)G_{x_{3}^{\prime}x_{2}}^{c}\right\} +F^{G^{c}}\left[\left(\mathcal{G}^{c}\right)^{-1}\right],\label{eq:G equation of motion from z-theory - 3}
\end{eqnarray}

\noindent where the $F^{\phi,G^{c}}\left[\left(\mathcal{G}^{c}\right)^{-1}\right]$ terms contain
an (infinite) set of terms with internal inverse atomic propagator
lines $\left(\mathcal{G}^{c}\right)^{-1}$. These are the anomalous terms we made reference
to in \ref{app:deriving effective theory - 1}. Note that in obtaining
Eqs. (\ref{eq:phi equation of motion from z-theory - 3}) and (\ref{eq:G equation of motion from z-theory - 3})
we made use of the following facts

\begin{eqnarray}
\Xi_{x_{1}}\left[\mathcal{G}^{\left(2n\right),c},-A,B\right] & = & -\Xi_{x_{1}}\left[\mathcal{G}^{\left(2n\right),c},A,B\right],\label{eq:Xi sign property - 1}\\
\Sigma_{x_{1}x_{2}}^{\text{aux}}\left[\mathcal{G}^{\left(2n\right),c},-A,B\right] & = & \Sigma_{x_{1}x_{2}}^{\text{aux}}\left[\mathcal{G}^{\left(2n\right),c},A,B\right].\label{eq:Sigma^aux sign property - 1}
\end{eqnarray}

\noindent Equations~(\ref{eq:Xi sign property - 1}) and (\ref{eq:Sigma^aux sign property - 1})
can be proven straightforwardly. First, note that diagrammatically,
\\$\Xi_{x_{1}}\left[\mathcal{G}^{\left(2n\right),c},A,B\right]$ and $\Sigma_{x_{1}x_{2}}^{\text{aux}}\left[\mathcal{G}^{\left(2n\right),c},A,B\right]$
are represented by infinite sums of diagrams, where each diagram is
made up of vertices $\mathcal{G}^{\left(2n\right),c}$, each of which
contain an even number of state-labels. Therefore, the total number
of vertex state-labels for each diagram is an even number. Each state-label
will either contract with a one-point propagator $A$, contract with
a two-point propagator $B$ (along with another state-label), or represent
an external state-label. Keeping in mind that each internal line $B$
contracts with two vertex state-labels, we must have that each diagram
in $\Xi_{x_{1}}\left[\mathcal{G}^{\left(2n\right),c},A,B\right]$ and
$\Sigma_{x_{1}x_{2}}^{\text{aux}}\left[\mathcal{G}^{\left(2n\right),c},A,B\right]$
contain an odd and even number of $A$ factors respectively, since
the former contains an odd number of external vertex state-labels
and the latter contains an even number. Eqs.~(\ref{eq:Xi sign property - 1})
and (\ref{eq:Sigma^aux sign property - 1}) immediately follow from
this observation.

Comparing Eqs.~(\ref{eq:phi equation of motion from z-theory - 3})
and (\ref{eq:G equation of motion from z-theory - 3}) to Eqs.~(\ref{eq:phi equation of motion from S_aux - 1})
and (\ref{eq:G equation of motion from S_aux - 1}), we see that these
are only consistent if all anomalous terms i.e. $\tilde{u}$ and $F^{\phi,G^{c}}\left[\left(\mathcal{G}^{c}\right)^{-1}\right]$
terms are omitted from the 2PI equations of motion. This completes
the proof that the anomalous terms cancel one another out when calculating
$\phi$ and $G^{c}$.

\section{Keldysh components of $\mathcal{G}^{c}$\label{app:keldysh components of G0 - 1}}

The Keldysh components of the atomic Green's function $\mathcal{G}^{c}$
can be expressed as follows

\begin{eqnarray}
\mathcal{G}^{12,\left(R\right)}\left(s_{1},s_{2}\right) & = & -\frac{i}{\mathcal{Z}_{0}}\Theta\left(s_{1}-s_{2}\right)\sum_{n=0}^{\infty}e^{-\beta\left(\mathcal{E}_{n}-\mathcal{E}_{n_{0}}\right)}\left\{ \left(n+1\right)e^{-i\left(\mathcal{E}_{n+1}-\mathcal{E}_{n}\right)\left(s_{1}-s_{2}\right)}\right. \nonumber \\
 &  & \phantom{-\frac{i}{\mathcal{Z}_{0}}\Theta\left(s_{1}-s_{2}\right)\sum_{n=0}^{\infty}e^{-\beta\left(\mathcal{E}_{n}-\mathcal{E}_{n_{0}}\right)}}\left.\quad-ne^{i\left(\mathcal{E}_{n-1}-\mathcal{E}_{n}\right)\left(s_{1}-s_{2}\right)}\right\} ,\label{eq:G0^(R) - 1}\\
\mathcal{G}^{12,\left(A\right)}\left(s_{1},s_{2}\right) & = & \frac{i}{\mathcal{Z}_{0}}\Theta\left(s_{2}-s_{1}\right)\sum_{n=0}^{\infty}e^{-\beta\left(\mathcal{E}_{n}-\mathcal{E}_{n_{0}}\right)}\left\{ \left(n+1\right)e^{-i\left(\mathcal{E}_{n+1}-\mathcal{E}_{n}\right)\left(s_{1}-s_{2}\right)}\right.\nonumber \\
 &  & \phantom{\frac{i}{\mathcal{Z}_{0}}\Theta\left(s_{2}-s_{1}\right)\sum_{n=0}^{\infty}e^{-\beta\left(\mathcal{E}_{n}-\mathcal{E}_{n_{0}}\right)}}\left.\quad-ne^{i\left(\mathcal{E}_{n-1}-\mathcal{E}_{n}\right)\left(s_{1}-s_{2}\right)}\right\} ,\label{eq:G0^(A) - 1}\\
\mathcal{G}^{12,\left(K\right)}\left(s_{1},s_{2}\right) & = & -\frac{i}{\mathcal{Z}_{0}}\sum_{n=0}^{\infty}e^{-\beta\left(\mathcal{E}_{n}-\mathcal{E}_{n_{0}}\right)}\left\{ \left(n+1\right)e^{-i\left(\mathcal{E}_{n+1}-\mathcal{E}_{n}\right)\left(s_{1}-s_{2}\right)}\right.\nonumber \\
 &  & \phantom{-\frac{i}{\mathcal{Z}_{0}}\sum_{n=0}^{\infty}e^{-\beta\left(\mathcal{E}_{n}-\mathcal{E}_{n_{0}}\right)}}\left.\quad+ne^{i\left(\mathcal{E}_{n-1}-\mathcal{E}_{n}\right)\left(s_{1}-s_{2}\right)}\right\} ,\label{eq:G0^(K) - 1}
\end{eqnarray}
\begin{eqnarray}
\mathcal{G}^{12,\left(\lceil\right)}\left(s_{1},s_{2}\right) & = & -\frac{i}{\mathcal{Z}_{0}}\sum_{n=0}^{\infty}\left(n+1\right)e^{-\beta\left(\mathcal{E}_{n}-\mathcal{E}_{n_{0}}\right)}e^{i\left(\mathcal{E}_{n+1}-\mathcal{E}_{n}\right)s_{2}}e^{-\left(\mathcal{E}_{n+1}-\mathcal{E}_{n}\right)s_{1}},\label{eq:G0^(left) - 1}\\
\mathcal{G}^{12,\left(\rceil\right)}\left(s_{1},s_{2}\right) & = & -\frac{i}{\mathcal{Z}_{0}}\sum_{n=0}^{\infty}ne^{-\beta\left(\mathcal{E}_{n}-\mathcal{E}_{n_{0}}\right)}e^{i\left(\mathcal{E}_{n-1}-\mathcal{E}_{n}\right)s_{1}}e^{-\left(\mathcal{E}_{n-1}-\mathcal{E}_{n}\right)s_{2}},\label{eq:G0^(right) - 1}\\
\mathcal{G}^{12,\left(M\right)}\left(s_{1},s_{2}\right) & = & -\frac{1}{\mathcal{Z}_{0}}\sum_{n=0}^{\infty}e^{-\beta\left(\mathcal{E}_{n}-\mathcal{E}_{n_{0}}\right)}\left\{ \Theta\left(s_{1}-s_{2}\right)\left(n+1\right)e^{-\left(\mathcal{E}_{n+1}-\mathcal{E}_{n}\right)\left(s_{1}-s_{2}\right)}\right.\nonumber \\
 &  & \phantom{-\frac{1}{\mathcal{Z}_{0}}\sum_{n=0}^{\infty}e^{-\beta\left(\mathcal{E}_{n}-\mathcal{E}_{n_{0}}\right)}}\left.\quad+\Theta\left(s_{2}-s_{1}\right)ne^{\left(\mathcal{E}_{n-1}-\mathcal{E}_{n}\right)\left(s_{1}-s_{2}\right)}\right\} ,\label{eq:G0^(M) - 1}
\end{eqnarray}

\noindent where $\mathcal{Z}_{0}$ is the atomic partition function

\begin{eqnarray}
\mathcal{Z}_{0} & \equiv & \sum_{n=0}^{\infty}e^{-\beta\left(\mathcal{E}_{n}-\mathcal{E}_{n_{0}}\right)},\label{eq:atomic partition function - 1}
\end{eqnarray}

\noindent and $n_{0}$ and $\mathcal{E}_{n}$ are given by Eqs.~(\ref{eq:n0 - 1})
and (\ref{eq:E_atomic - 1}) respectively.

Given that the Fourier transforms $\mathcal{G}^{12,\left(R,K\right)}\left(\omega\right)$
are used throughout this paper, it is worth explicitly writing out
the expressions for these particular Keldysh components

\begin{eqnarray}
\mathcal{G}^{12,\left(R\right)}\left(\omega\right) & = & \frac{1}{\mathcal{Z}_{0}}\sum_{n=0}^{\infty}e^{-\beta\left(\mathcal{E}_{n}-\mathcal{E}_{n_{0}}\right)}\left\{ \frac{\left(n+1\right)}{\left(\omega-\left[\mathcal{E}_{n+1}-\mathcal{E}_{n}\right]\right)+i0^{+}}-\frac{n}{\left(\omega+\left[\mathcal{E}_{n-1}-\mathcal{E}_{n}\right]\right)+i0^{+}}\right\} ,\label{eq:G0^(R)(w) - 1}\\
\mathcal{G}^{12,\left(K\right)}\left(\omega\right) & = & -\frac{2\pi i}{\mathcal{Z}_{0}}\sum_{n=0}^{\infty}e^{-\beta\left(\mathcal{E}_{n}-\mathcal{E}_{n_{0}}\right)}\left\{ \left(n+1\right)\delta\left(\omega-\left[\mathcal{E}_{n+1}-\mathcal{E}_{n}\right]\right)+n\delta\left(\omega+\left[\mathcal{E}_{n-1}-\mathcal{E}_{n}\right]\right)\right\} .\label{eq:G0^(K)(w) - 1}
\end{eqnarray}

\section{Low frequency approximation to four-point vertex $u^{\left(4\right)}$\label{app:low frequency approximation - 1}}

To calculate the low frequency approximation to the four-point vertex
$u_{\alpha_{1}\alpha_{2}\alpha_{3}\alpha_{4}}^{a_{1}a_{2}a_{3}a_{4}}\left(s_{1},s_{2},s_{3},s_{4}\right)$,
we begin with Eq.~(\ref{eq:u-vertex defined - 1}). We make use of
the time-translational invariance of the atomic two-point Green's
function and take the low-frequency approximation, which gives (noting
that there is no contribution from the Keldysh Green's function except
at points where the Mott lobes are degenerate) \citep{Kennett}

\begin{eqnarray}
 &  & u_{\alpha_{1}\alpha_{2}\alpha_{3}\alpha_{4}}^{a_{1}a_{2}a_{3}a_{4}}\left(s_{1},s_{2},s_{3},s_{4}\right)\nonumber \\
 &  & \quad=-\left\{ \mathcal{G}^{12,\left(R\right)}\left(\omega^{\prime}=0\right)\right\} ^{-4}\prod_{m=1}^{4}\left(\int_{-\infty}^{\infty}\frac{d\omega_{m}}{2\pi}e^{-i\omega_{m}s_{m}}\right)\nonumber \\
 &  & \phantom{\quad=}\quad\times\begin{cases}
\mathcal{G}_{\alpha_{1}\alpha_{2}\alpha_{3}\alpha_{4}}^{a_{1}a_{2}a_{3}a_{4},c}\left(\omega_{1},\omega_{2},\omega_{3},\omega_{4}\right), & \begin{aligned} & \text{if }\alpha_{m}=q\text{ or }c\text{ for }m=1,\ldots4 , \\
 & \text{or if }\left\{ \alpha_{m}\right\} _{m=1}^{4}=\left\{ \mathcal{T},\mathcal{T},\mathcal{T},\mathcal{T}\right\}, 
\end{aligned}
\\
\\
0, & \begin{aligned} & \text{otherwise},\end{aligned}
\end{cases}\label{eq:u vertex low frequency approx - 1}
\end{eqnarray}

\noindent where $\mathcal{G}^{12,\left(R\right)}\left(\omega^{\prime}=0\right)$
is easily determined from Eq.~(\ref{eq:G0^(R)(w) - 1}) to be

\begin{eqnarray}
\mathcal{G}^{12,\left(R\right)}\left(\omega^{\prime}=0\right) & = & -\frac{1}{\mathcal{Z}_{0}}\sum_{n=0}^{\infty}e^{-\beta\left(\mathcal{E}_{n}-\mathcal{E}_{n_{0}}\right)}\left\{ \frac{\left(n+1\right)}{\mathcal{E}_{n+1}-\mathcal{E}_{n}}+\frac{n}{\mathcal{E}_{n-1}-\mathcal{E}_{n}}\right\} .\label{eq:static limit of G0^(R) - 1}
\end{eqnarray}

\noindent Explicit calculation of $\mathcal{G}_{\alpha_{1}\alpha_{2}\alpha_{3}\alpha_{4}}^{a_{1}a_{2}a_{3}a_{4},c}\left(\omega_{1},\omega_{2},\omega_{3},\omega_{4}\right)$
followed by taking the low frequency limit leads to the two constants
introduced in Eq.~(\ref{eq:static lim of u^(4) - 1}):

\begin{eqnarray}
u_{1} & = & -\frac{2\left\{ \mathcal{G}^{12,\left(R\right)}\left(\omega^{\prime}=0\right)\right\} ^{-4}}{\mathcal{Z}_{0}}\nonumber \\
 &  & \quad\times\sum_{n=0}^{\infty}e^{-\beta\left(\mathcal{E}_{n}-\mathcal{E}_{n_{0}}\right)}\left\{ \frac{\left(n+1\right)\left(n+2\right)}{\left(\mathcal{E}_{n+2}-\mathcal{E}_{n}\right)\left(\mathcal{E}_{n+1}-\mathcal{E}_{n}\right)^{2}}+\frac{n\left(n-1\right)}{\left(\mathcal{E}_{n-2}-\mathcal{E}_{n}\right)\left(\mathcal{E}_{n-1}-\mathcal{E}_{n}\right)^{2}}\right.\nonumber \\
 &  & \left.\phantom{\quad\times\sum_{n=0}^{\infty}e^{-\beta\left(\mathcal{E}_{n}-\mathcal{E}_{n_{0}}\right)}}\quad-\frac{\left(n+1\right)^{2}}{\left(\mathcal{E}_{n+1}-\mathcal{E}_{n}\right)^{3}}-\frac{n^{2}}{\left(\mathcal{E}_{n-1}-\mathcal{E}_{n}\right)^{3}}\right.\nonumber \\
 &  & \left.\phantom{\quad\times\sum_{n=0}^{\infty}e^{-\beta\left(\mathcal{E}_{n}-\mathcal{E}_{n_{0}}\right)}}\quad-\frac{n\left(n+1\right)}{\left(\mathcal{E}_{n+1}-\mathcal{E}_{n}\right)\left(\mathcal{E}_{n-1}-\mathcal{E}_{n}\right)^{2}}-\frac{n\left(n+1\right)}{\left(\mathcal{E}_{n+1}-\mathcal{E}_{n}\right)^{2}\left(\mathcal{E}_{n-1}-\mathcal{E}_{n}\right)}\right\} ,\label{eq:u_1 defined - 1}
\end{eqnarray}

\noindent and

\begin{eqnarray}
u_{2}^{2} & = & \frac{\left\{ \mathcal{G}^{12,\left(R\right)}\left(\omega^{\prime}=0\right)\right\} ^{-4}}{\mathcal{Z}_{0}}\sum_{n=0}^{\infty}e^{-\beta\left(\mathcal{E}_{n}-\mathcal{E}_{n_{0}}\right)}\left(\frac{n+1}{\mathcal{E}_{n+1}-\mathcal{E}_{n}}+\frac{n}{\mathcal{E}_{n-1}-\mathcal{E}_{n}}\right)^{2}\nonumber \\
 &  & -\frac{\left\{ \mathcal{G}^{12,\left(R\right)}\left(\omega^{\prime}=0\right)\right\} ^{-4}}{\mathcal{Z}_{0}^{2}}\sum_{n=0}^{\infty}\sum_{n^{\prime}=0}^{\infty}e^{-\beta\left\{ \left(\mathcal{E}_{n}-\mathcal{E}_{n_{0}}\right)+\left(\mathcal{E}_{n^{\prime}}-\mathcal{E}_{n_{0}}\right)\right\} }\left(\frac{n+1}{\mathcal{E}_{n+1}-\mathcal{E}_{n}}+\frac{n}{\mathcal{E}_{n-1}-\mathcal{E}_{n}}\right)\nonumber \\
 &  & \phantom{+\frac{i\left\{ \mathcal{G}^{12,\left(R\right)}\left(\omega^{\prime}=0\right)\right\} ^{-4}}{\mathcal{Z}_{0}^{2}}\sum_{n=0}^{\infty}\sum_{n^{\prime}=0}^{\infty}}\quad\times\left(\frac{n^{\prime}+1}{\mathcal{E}_{n^{\prime}+1}-\mathcal{E}_{n^{\prime}}}+\frac{n^{\prime}}{\mathcal{E}_{n^{\prime}-1}-\mathcal{E}_{n^{\prime}}}\right).\label{eq:u_2 defined - 1}
\end{eqnarray}

\noindent Note that $u_{1}$ corresponds to the coefficient $u$ introduced
in Ref.~\citep{Kennett}, but $u_{2}^{2}$ is a coefficient that did not
enter in that work, but is required to describe correlation function
dynamics. Note also that in the limit $\beta U\to\infty$, $u_{2}^{2}\to0$.

\section{Gapless spectrum in the HFBP approximation\label{app:gapless spectrum in the HFBP approx - 1}}

In this appendix we show that in the full HFB approximation the excitation
spectrum is not gapless in the SF phase.  We then show that the
HFBP approximation yields a gapless spectrum. In the SF phase, in order for the excitation spectrum to be gapless,
we require that 

\begin{eqnarray}
\tilde{C}_{\vec{k}=0} & = & 0,\label{eq:SF gapless cond - 1}
\end{eqnarray}

\noindent where $\tilde{C}_{\vec{k}}$ was defined in Eq.~(\ref{eq:C_k tilde defined - 1}).
To show this, first we substitute Eq.~(\ref{eq:E_MI - 1}) into Eq.
(\ref{eq:C_k tilde defined - 1}) to get

\begin{eqnarray}
\tilde{C}_{\vec{k}} & = & \left(C_{\vec{k}}\right)^{2}-\left(U+\mu\right)^{2}\left|\Sigma_{\vec{k}}^{22,\left(R\right)}\right|^{2},\label{eq:C_k tilde manipulated - 1}
\end{eqnarray}

\noindent where $C_{\vec{k}}$ was defined in Eq.~(\ref{eq:C_k - 1}).
In the full HFB approximation, the self-energy is given by Eqs.~(\ref{eq:Sigma_12^(R) - 1})
and (\ref{eq:Sigma_22^(R) - 1}). Using Eq.~(\ref{eq:phi equilibrium eqn of motion - 2})
one can rewrite $\Sigma_{\vec{k}}^{12,\left(R\right)}$ in the HFB
approximation as

\begin{eqnarray}
\Sigma_{\vec{k}}^{12,\left(R\right)} & = & \left(2dJ+\epsilon_{\vec{k}}\right)+\left\{ \mathcal{G}^{12,\left(R\right)}\left(\omega^{\prime}=0\right)\right\} ^{-1}-\frac{1}{2}u_{1}\left\{ iG_{\vec{r}=\mathbf{0}}^{22,\left(K\right)}\left(s^{\prime}=0\right)\right\} +u_{1}\phi^{2},\label{eq:HFB Sigma_12^(R) manipulated - 1}
\end{eqnarray}

\noindent where we assumed without loss of generality that $\phi$
is real, which implies that $iG_{\vec{r}=\mathbf{0}}^{22,\left(K\right)}\left(s^{\prime}=0\right)$
is real as well. Substituting Eq.~(\ref{eq:HFB Sigma_12^(R) manipulated - 1})
into Eq.~(\ref{eq:C_k - 1}) for $\vec{k}=0$ yields

\begin{eqnarray}
C_{\vec{k}=0} & = & -\frac{1}{2}u_{1}\left(U+\mu\right)\left\{ 2\phi^{2}-\left\{ iG_{\vec{r}=\mathbf{0}}^{22,\left(K\right)}\left(s^{\prime}=0\right)\right\} \right\} .\label{eq:HFB C_k=00003D0 - 1}
\end{eqnarray}

\noindent Lastly, we substitute Eqs.~(\ref{eq:HFB C_k=00003D0 - 1})
and (\ref{eq:Sigma_22^(R) - 1}) into Eq.~(\ref{eq:C_k tilde manipulated - 1})
to get

\begin{eqnarray}
\tilde{C}_{\vec{k}=0} & = & -2u_{1}^{2}\left(U+\mu\right)^{2}\phi^{2}\left\{ iG_{\vec{r}=\mathbf{0}}^{22,\left(K\right)}\left(s^{\prime}=0\right)\right\} .\label{eq:HFB C_k=00003D0 tilde - 1}
\end{eqnarray}

\noindent As we can see, Eq.~(\ref{eq:SF gapless cond - 1}) is not
satisfied in the full HFB approximation. However, in the HFBP approximation
-- which is equivalent to setting $iG_{\vec{r}=\mathbf{0}}^{11,\left(K\right)}\left(s^{\prime}=0\right)=iG_{\vec{r}=\mathbf{0}}^{22,\left(K\right)}\left(s^{\prime}=0\right)=0$
-- we clearly have a gapless spectrum.

\section{Static limit of $G^{\left(K\right)}$\label{app:static limit of G^(K) - 1}}

In this appendix, we show that 

\begin{eqnarray}
G_{\vec{k}}^{a_{1}a_{2},\left(K\right)}\left(\omega=0\right) & = & 0,\label{eq:static limit of G^(K) - 1}
\end{eqnarray}

\noindent for equilibrium systems. We start with Eq.~(\ref{eq:G^(K) eqn of motion - 1}),
which for equilibrium systems reduces to \citep{Stefanucci}

\begin{eqnarray}
G_{\vec{k}}^{a_{1}a_{2},\left(K\right)}\left(\omega\right) & = & \mathcal{G}^{a_{1}a_{2},\left(K\right)}\left(\omega\right)\nonumber \\
 &  & +\sum_{a_{3}a_{4}}\mathcal{G}^{a_{1}a_{3},\left(R\right)}\left(\omega\right)\Sigma_{\vec{k}}^{\overline{a_{3}}\overline{a_{4}},\left(R\right)}G_{\vec{k}}^{a_{4}a_{2},\left(K\right)}\left(\omega\right)\nonumber \\
 &  & +\sum_{a_{3}a_{4}}\mathcal{G}^{a_{1}a_{3},\left(K\right)}\left(\omega\right)\Sigma_{\vec{k}}^{\overline{a_{3}}\overline{a_{4}},\left(A\right)}G_{\vec{k}}^{a_{4}a_{2},\left(A\right)}\left(\omega\right).\label{eq:G^(K) equilibrium equations of motion - 1}
\end{eqnarray}

\noindent From Eq.~(\ref{eq:G0^(K)(w) - 1}), we have 

\begin{eqnarray}
\mathcal{G}^{a_{1}a_{2},\left(K\right)}\left(\omega=0\right) & = & 0,\label{eq:static limit of G0^(K) - 1}
\end{eqnarray}

\noindent which implies that

\begin{eqnarray}
G_{\vec{k}}^{a_{1}a_{2},\left(K\right)}\left(\omega=0\right) & = & \sum_{a_{3}a_{4}}\mathcal{G}^{a_{1}a_{3},\left(R\right)}\left(\omega\right)\Sigma_{\vec{k}}^{\overline{a_{3}}\overline{a_{4}},\left(R\right)}G_{\vec{k}}^{a_{4}a_{2},\left(K\right)}\left(\omega\right).\label{eq:G^(K) equilibrium equations of motion - 2}
\end{eqnarray}

\noindent The $G^{12,\left(K\right)}$ equation yields

\begin{eqnarray}
G_{\vec{k}}^{12,\left(K\right)}\left(\omega=0\right) & = & \mathcal{G}^{12,\left(R\right)}\left(\omega\right)\Sigma_{\vec{k}}^{12,\left(R\right)}G_{\vec{k}}^{12,\left(K\right)}\left(\omega\right)+\mathcal{G}^{12,\left(R\right)}\left(\omega\right)\Sigma_{\vec{k}}^{11,\left(R\right)}G_{\vec{k}}^{22,\left(K\right)}\left(\omega\right),\label{eq:G12^(K)(w=00003D0) eqn - 1}
\end{eqnarray}

\noindent whereas the $G^{22,\left(K\right)}$ equation can be rearranged as
follows

\begin{eqnarray}
G_{\vec{k}}^{22,\left(K\right)}\left(\omega=0\right) & = & \frac{\Sigma_{\vec{k}}^{22,\left(R\right)}}{\left\{ \mathcal{G}^{12,\left(R\right)}\left(\omega=0\right)\right\} ^{-1}-\Sigma_{\vec{k}}^{12,\left(R\right)}}G_{\vec{k}}^{12,\left(K\right)}\left(\omega=0\right).\label{eq:G22^(K)(w=00003D0) eqn - 1}
\end{eqnarray}

\noindent Substituting Eq.~(\ref{eq:G22^(K)(w=00003D0) eqn - 1})
back into Eq.~(\ref{eq:G12^(K)(w=00003D0) eqn - 1}) yields

\begin{eqnarray}
0 & = & \left[1-\mathcal{G}^{12,\left(R\right)}\left(\omega=0\right)\Sigma_{\vec{k}}^{12,\left(R\right)}\right.\nonumber \\
 &  & \left.\quad-\mathcal{G}^{12,\left(R\right)}\left(\omega=0\right)\frac{\left|\Sigma_{\vec{k}}^{22,\left(R\right)}\right|^{2}}{\left\{ \mathcal{G}^{12,\left(R\right)}\left(\omega=0\right)\right\} ^{-1}-\Sigma_{\vec{k}}^{12,\left(R\right)}}\right]G_{\vec{k}}^{12,\left(K\right)}\left(\omega=0\right).\label{eq:G12^(K)(w=00003D0) eqn - 2}
\end{eqnarray}

\noindent Since in general the expression inside the square brackets
is not zero, it must be the case that $G_{\vec{k}}^{12,\left(K\right)}\left(\omega=0\right)$
is zero, which also implies that $G_{\vec{k}}^{22,\left(K\right)}\left(\omega=0\right)$
is zero.

\bibliographystyle{elsarticle-num}
\bibliography{bibfile}

\end{document}